\documentclass[fleqn,usenatbib]{mnras}

\usepackage{newtxtext,newtxmath}
\usepackage[T1]{fontenc}

\DeclareRobustCommand{\VAN}[3]{#2}
\let\VANthebibliography\thebibliography
\def\thebibliography{\DeclareRobustCommand{\VAN}[3]{##3}\VANthebibliography}

\usepackage{graphicx}	% Including figure files
\usepackage{amsmath}	% Advanced maths commands
\usepackage{arydshln}
 
\usepackage{verbatim}
\usepackage{breqn}
\usepackage[export]{adjustbox}
\usepackage{longtable}
\usepackage{supertabular,booktabs}
\usepackage{threeparttable}
\newcommand{\lya}{Ly$\alpha$}
\newcommand{\lyb}{Ly$\beta$}

\newcommand{\kms}{$km s^{-1}$}

\newcommand{\HI}{\mbox{H\,{\sc i}}}

\newcommand{\OVI}{\mbox{O\,{\sc vi}}}

\newcommand{\CIV}{\mbox{C\,{\sc iv}}}

\newcommand{\SiIII}{\mbox{Si\,{\sc iii}}}

\newcommand{\NHI} {$N_{\rm HI}$}

\title[Low-redshift three-point correlation]{Redshift space three-point correlation function of IGM at $z<0.48$}

\author[Maitra et al.]{Soumak Maitra$^{1}$\thanks{E-mail: soumak@iucaa.in},
	Raghunathan Srianand$^{1}$, Prakash Gaikwad$^{2,3}$,  Nishikanta Khandai$^{4}$
	\\
	\\
	$^{1}$ IUCAA, Postbag 4, Ganeshkhind, Pune - 411007, India\\
	$^{2}$ Institute of Astronomy, University of Cambridge, Madingley Road, Cambridge, CB3 0HA, UK\\
	$^{3}$ Kavli Institute for Cosmology, University of Cambridge, Madingley Road, Cambridge, CB3 0HA, UK\\
	$^{4}$  School of Physical Sciences, National Institute of Science Education and Research, HBNI, Jatni - 752050, India\\
}

\date{Accepted XXX. Received YYY; in original form ZZZ}

\pubyear{2015}

\begin{document}
\label{firstpage}
\pagerange{\pageref{firstpage}--\pageref{lastpage}}
\maketitle

\begin{abstract}
The \lya\ forest decomposed into Voigt profile components allow us to study clustering properties of the intergalactic medium and its dependence on various physical quantities.
Here, we report the first detections of  probability excess of low-$z$ (i.e $z<0.48$) \lya\ absorber triplets over a scale of $r_\parallel\leq 8$ pMpc with a maximum amplitude of $8.76^{+1.96}_{-1.65}$ at a longitudinal separation of 1-2 pMpc.  {
We measure non-zero three-point correlation  ($\zeta = 4.76^{+1.98}_{-1.67}$) only at this scale with reduced three-point correlation value of Q = $0.95^{+0.39}_{-0.38}
$.
The measured $\zeta$ shows an increasing trend with increasing H~{\sc i} column density (\NHI) while Q does not show any \NHI\ dependence.}
%We find the radial profile of $\zeta$ to be different for high-$b$ (i.e $b>$40 \kms) and low-$b$  absorbers, { probably driven by the signal-to-noise ratio of the spectra used}.  
About 88\% of the triplets contributing to $\zeta$  {(at $z\le 0.2$)} have nearby galaxies ({ whose distribution is known to be complete for 0.1 L$_*$ at $z <0.1$ and for L$_*$ at $z\sim0.25$ and within 20' to the quasar sightlines})
within a velocity separation of 500 \kms\ and a median impact parameter of 405 pkpc. The measured impact parameters are consistent with appreciable number of triplets at {$z\le 0.2$} not originating from individual galaxies but tracing the underlying galaxy distribution. Frequency of occurrence of high-$b$
absorbers in triplets ($\sim$85\%) is a factor $\sim3$ higher than that found among the full sample ($\sim$32\%) . Using four different cosmological simulations, we  quantify the effect of peculiar velocities, feedback effects and show that most of the observed trends are broadly reproduced. However, $\zeta$ at small scales ($r_\parallel<1$ pMpc) and $b$-dependence of $\zeta$ in simulations are found inconsistent with the observations. This could either be related to the fact that none of these simulations reproduce the observed $b$-distribution and \NHI\ distribution for \NHI$>10^{14}$ cm$^{-2}$ self-consistently or to the widespread of signal-to-noise ratio in the observed data.

%, i.e., a value of $6.9\pm 1.3$ at scales of 1-2Mpc. The reduced three-point correlation corresponding to this scale is found to be $1.75\pm0.45$. We detect three-point correlation upto a length scale of 3Mpc beyond which it approaches 0. 
%Its radial profile is affected by exclusion at scales below 1Mpc similar to two-point correlation. The amplitude of longitudinal three-point correlation seems to be larger for higher \HI\ column density { absorbers}. Based on simulations, we find that redshift space distortion seems to play a major role in deciding the amplitude. We find stellar winds, AGN feedback and metal ion species to not play a significant role. We also find \lya\ triplets \NEW{in longitudinal direction}, especially those with higher \HI\ column densities to be strongly associated with nearby galaxies within an impact parameter of 1Mpc.
\end{abstract}


%%%%%%%%%%%%%%%%%%%%%%%%%%%%%%%%%%%%%%%%%%%%%%%%%%

%%%%%%%%%%%%%%%%% BODY OF PAPER %%%%%%%%%%%%%%%%%%

\section{Introduction}

The 
%forest of 
\lya\ absorption seen in the spectra of high-$z$ quasars are frequently used to probe the physics of the intergalactic medium (IGM) and parameters of the background cosmology \citep[see][]{rauch1998,meiksin2009}. 
While a vast majority of research using ground based observations have been focussing on higher redshifts  \lya\ forest (i.e $z>1.8$) owing to the atmospheric cut-off, observations using Hubble Space Telescope (HST) allow us to probe \lya\ forest at low-$z$ (i.e $z<1.3$) \citep[see for example,][]{bahcall1991,bahcall1993,bahcall1996,penton2000, tilton2016, danforth2016}.
%\lya\ forest at $z\sim 0$ traces much higher over-densties in comparison to \lya\ forest at high-$z$. Therefore, one is expected to detect strong signals in the two-point and higher order clustering statistics at low redshifts. While two-point correlation function of \lya\ absorption at low-$z$ is well probed \citep{ulmer1996,danforth2016}, higher order correlations are not explored in detail.

At high redshifts, it is believed that most of the baryons are located in the photoionized low density IGM that traces the underlying dark matter distribution for scales above the pressure smoothing scales \citep[see for example,][]{bi1997}. At low-$z$, the \lya\ absorption with a given {  neutral hydrogen column density}, \NHI, originates from higher over-densities {(i.e $\Delta \sim 35$ for log~\NHI=14.0 at $z=0$)} compared to that {(i.e  $\Delta \sim 10$ for log~\NHI=14.0 at $z = 2.0$)} at high-$z$
\citep{dave2010,smith2011,gaikwad2017a}. Frequent presence of broad \lya\ absorbers { \citep[BLAs are defined as systems with { Doppler parameter $b \ge 40$~\kms. If thermally broadened, the absorbing gas will have a temperature of $mb^2/2k \ge 10^5$ K}, see][]{Richter2006, Lehner2007}} and ionization modelling of high ionization absorbers \citep[like Ne~{\sc viii} and O~{\sc vi}, see ][]{Savage2005,tripp2008,hussain2017} suggests that collisional ionization {(for example, due to structure formation shocks)} may also be important for some of the low-$z$ \lya\ absorbers. 

{ In principle,} it is { possible} to associate the \lya\ absorbers with individual galaxies (or distribution of galaxies) at low-$z$.
%
%However, at low redshifts (i.e $z\le0.5$) the \lya\ 
Such studies have revealed that the
low-$z$ \lya\ forest absorption originates from different locations such as cool-dense circumgalactic medium \citep[CGM,][]{Werk2014}, dense hot intra-cluster medium \citep[ICM; see][]{Muzahid2017, Burchett2018} in addition to the filaments and voids defined by the distribution of galaxies \citep[][]{Stocke1995,Penton2002,Tejos2014}. 
It is also found that \NHI, $b$-parameter and metallicity of the gas depend on their location.
Thus, it is an usual procedure to study the properties of the low-$z$ IGM as a function of \NHI, $b$-parameter and metal abundance.

%\lya\ forest absorption in distant QSO spectra arises from \HI\ density fluctuations in the Intergalactic Medium (IGM) and is known to trace the underlying dark matter density fields at scales larger than the pressure smoothing scales. 

%

%
%Hence, one expects to find a much stronger correlation for these low-$z$ \lya\ absorbers . This means that low-$z$ \lya\ forest should probe much stronger gravitational instabilities arising from higher overdensities and should have a signature of strong non-gaussianity in the three-point correlation statistics. While detecting non-gaussianity at higher redshifts is difficult due to smaller signals, it is expected that the low-$z$ QSO sightlines from HST should give a stronger longitudinal three-point correlation signal.
%
At low-$z$ it is possible to correlate the spatial distribution and other properties of \lya\ forest with the cosmic web (i.e filaments, voids and clusters { in the galaxy distribution}) defined by galaxies. \citet{Penton2002}, by correlating the \lya\ absorbers at 0.003$\le z\le$0.069 with the galaxy distribution, arrived at the following conclusions. Apart from a few very strong \lya\ absorbers (i.e with \NHI$\ge10^{15}$ cm$^{-2}$) the strong \lya\ absorbers (i.e with \NHI = $10^{13.2} - 10^{15.5}$ cm$^{-2}$) are found to be aligned with the large scale distribution of galaxies. A small fraction  (i.e $22\pm8$\%) of \lya\ absorbers  are found to be distributed in cosmic voids \citep[see also][]{Stocke1995}. In general these void absorbers are found to have low \NHI\ (i.e  \NHI $\leq10^{13.2}$ cm$^{-2}$) . 

\citet{Wakker2015} studied \lya\ absorption towards 24 quasar sightlines that are close to two large local filaments. They find a strong correlation between \lya\ equivalent width (as well as $b$-parameter) and filament impact parameter. All the \lya\ absorption with 
\NHI $\geq10^{13}$ cm$^{-2}$ are found to have the filament impact parameter less than 2.1 Mpc. Interestingly the four BLAs found in their sample are all found to be located within 400 kpc to the filament axis and all the absorbers showing multiple velocity components are located within 1 Mpc to the filament axis. While the trends found in this study are very interesting, these are based on small number of sightlines (and systems) and it { is important} to expand such an analysis to large number of sightlines.
%\citep[for example,][]{Stocke1995, Penton2002, wakker2015, Burchett2020}.

\citet{Tejos2016} have studied absorption towards filaments connecting cluster pairs at $z<0.5$ towards the quasar J141038.39+230447. They found  tentative excesses of \HI\ (broad as well as narrow) and O~{\sc vi} absorption lines within rest-frame velocities of $\le 1000$ \kms from the cluster-pairs redshifts. They suggested that while O~{\sc vi} absorption may be associated with individual galaxies, narrow and broad \HI\ absorption are intergalactic in origin.  They also found the relative excess of BLAs to be larger than that of narrow \lya\ absorbers and used this to argue that BLAs may be originating from collisionally ionized gas in the filaments.

The clustering properties of the \lya\ absorbers can be used to probe the matter distribution in the universe. {  Majority of such studies in the literature focus mainly on high redshifts ($z>1.8$).}
%owing to the atmospheric cut-off from the ground based telescopes. } 
%At smaller scales, the physics is additionally governed by local temperature and photoionizing background, the thermal history of the Universe as well as various feedback processes. Over the years, high-redshift \lya\ forest ($z>1.7$) has been widely explored. 
Usually, these clustering studies are carried out in the redshift space using longitudinal (line of sight) correlation or conversely 1D flux power spectrum in the fourier space \citep{mcdonald2000,mcdonald2006,croft2002,seljak2006}. The 1D \lya\ forest flux power spectrum has been used to constrain the background cosmology  \citep{mcdonald2005,palanque2013}, mass of warm dark matter particles \citep{narayanan2000,viel2013w}, neutrino mass \citep{palanque2015a,palanque2015b,yeche2017,walther2020}, ionization state \citep{gaikwad2017a,khaire2019a}  as well as the thermal history of the IGM \citep{walther2019,gaikwad2019,gaikwad2020,gaikwad2020b}. 
Clustering studies of \lya\ forest can also { be carried out} in the transverse direction using closely spaced projected quasar pairs  or gravitationally lensed quasars \citep{Smette1995,Rauch1995,petitjean1998,aracil2002, Rollinde2005, Coppolani2006, Dodorico2006, hennawi2010}. 
It is found that correlation in the transverse direction is more sensitive to the 3D matter distribution in comparison to longitudinal direction which is dominated by thermal broadening effects \citep{peeples2010a,peeples2010b}. 
%However, the rarity of closely spaced projected quasar pairs makes correlation studies in the transverse direction much more difficult. This is even more true for investigating non-gaussianity in \lya\ clustering using higher order statistics in the transverse direction. 

{  The next order beyond the two-point correlation statistics (or power spectrum in fourier space) is the three-point correlation statistics (or bispectrum in fourier space). Higher order statistics are useful in studying the non-gaussianity in clustering imparted by the non-linear gravitational evolution as well as any primordial non-gaussianity in the density fields \citep{peebles1980}. They can act as an independent tool complementing the two-point statistics in constraining cosmological parameters and remove degeneracies between different cosmological parameters like bias and $\sigma_8$ \citep{fry1994,verde2002,Bernardeau2002}. While considerable work has been done studying three-point clustering statistics using galaxies \citep[see for example,][]{mcbride2011b,guo2016}, it remains largely unexplored in the case of clustering in \lya\ forest.

While the \lya\ forest is a good probe of underlying dark matter distribution the exact connection between \lya\ optical depth or column density to the dark matter density is not straightforward. In particular ionization and thermal inhomogeneities can have strong influence on this relation \citep[see][]{tie2019,maitra2020}. 
All this make probing two- and three-point correlation function in the \lya\ forest an important exercise. While all the theoretical explorations using simulations till date focus on transverse correlations, observationally it is possible to study only few triplets sightlines at high-$z$ \citep[e.g][]{cappetta2010, maitra2019}. However, enough spectra are available in the literature to probe the longitudinal (i.e redshift space) three-point correlation function \citep[for example,][]{viel2004a} at low and high-$z$. 
As discussed above, unlike high-$z$, the low-$z$ \lya\ absorbers can originate from varied environments like ICM, CGM and IGM. This makes it interesting to probe clustering at different scales.
Low-$z$ also provide additional advantage that we will be able to relate the observed \lya\ clustering properties with the underlying galaxy distribution and various feedback processes. This forms the main motivation of this work.
}

%This is simply because \lya\ forest probes lower overdensities in comparison to galaxies and the non-gaussianity is expected to be weaker for such overdensities. Nevertheless, it is essential for studying clustering at smaller scales (of the order of ~1 Mpc) and at higher redshifts.}
%The observational detection of 1D trasmitted flux bispectrum (or conversely longitudinal three-point correlation) at $z\sim 2$ using a large sample of QSO sightlines in \cite{viel2004a} has been shown to be consistent with the bispectrum obtained from a random distribution of absorbers within the errorbars. {  This could be because of insufficient redshift path length covered by the observed sightlines}.
%Hence it cannot be used to determine the non-gaussianity in underlying matter distribution. 
%Studies involving transverse three-point correlation are currently at explorative stages \citep{tie2019,maitra2019,maitra2020}  mostly using simulated data and we are expecting to detect signals in future large surveys. 

While line of sight two-point correlation of \lya\ forest at low-$z$ is studied in the past \citep[for example,][]{ulmer1996,Impey1999, Penton2002, danforth2016} higher order clustering is not explored. %This forms the main motivation of this paper.
Here, we
%The main motivation of this paper is to 
measure the redshift space (or longitudinal) 3-point ($\zeta$) and reduced 3-point correlation ($Q$) function of the IGM at $z\le0.48$. For this purpose, we use the  
%
%
%In this paper, we use the 
Voigt profile fitted \lya\ absorption components of IGM towards 82 UV-bright QSOs ($z_{em}$<0.85) observed using Hubble Space Telescope-Cosmic Origins Spectrograph (HST-COS) presented in \cite{danforth2016}. We report, for the first time, the detection of longitudinal three-point correlation of low-$z$ \lya\ absorbers ($z$<0.48) at scales $\le 4$ pMpc {  (Mpc in proper units)}.  
%We also find the reduced three-point correlation (Q) to be scale dependent, increasing with increasing scale.
%$=2\pm 1$, at a scale of 1.5 pMpc.
%with no strong evidence of evolution with scale or $N_{\rm HI}$. 
We study the dependence of $\zeta$ on \NHI, $b$-parameter and the presence of {\rm different metal ion species}. We also study the relationship between {regions showing triplet absorption} and galaxy distribution for $z\le 0.2$.

{ In the past, simulations have been used to study two-point correlation function of low-$z$ \lya\ absorbers \citep[see][for example]{pierleoni2008}.} In this work, we present the analysis of simulated IGM data at $z\sim0.1$ using four available cosmological hydrodynamical simulations.
{ We use these simulations to (i) check whether the observed dependence of clustering on \NHI\ and $b$-parameters are readily reproduced in the simulations; (ii) quantify the effect of peculiar velocities on the line of sight clustering and (iii) probe the effect of feedback on the line of sight clustering. 
 We show the peculiar velocities tend to enhance the two- and three-point correlation signals (by about 40-60\%) over the distance scale probed in this study. Presence of wind and AGN feedback (as implemented in the simulations { used here}) are shown to produce minor effect in the measured two- and three-point correlation functions. As these simulations have problems in reproducing the $b$-distributions and high \NHI\ end of the column density distribution function,  we do not make any serious attempt to exactly match our observations with simulations.}
%We also use these simulations to quantify the effect of peculiar velocities on the amplitude and scale of the measured three point functions. 
%We use these distinct absorption components or "{ absorbers}" to estimate the longitudinal three-point correlation as a probability excess of { absorber }triplet in comparison to a random distribution of { absorbers}. The three-point correlation defined this way directly probes the non-gaussianity in the \lya\ absorbers. 
%
%We report a $\sim 2.5\sigma$ detection of longitudinal three-point correlation of low-$z$ \lya\ absorbers ($z$<0.48) at scales< 3pMpc.  We also find the reduced three-point correlation function to be about $2\pm 1$ at a scale of 1.5pMpc with no strong evidence of evolution with scale or $N_{\rm HI}$. %
%\Anand{Text related to the simulations need to be updated in the end.}

This paper is organized as follows. In section 2 we provide the details of data used in our study. Section 3 summarises the results of two-, three- and reduced three-point correlation of low-$z$ \lya\ absorbers measured from observations. { In this section, we} also present the dependence of clustering on \NHI, $b$, $z$ and the presence of {\rm different metal ion species like \CIV, \OVI, and \SiIII}. In section 4, we investigate the connection between \lya\ clustering and galaxy distribution. In section~\ref{sec:simulations}, we present our analysis based on a set of $z\sim0.1$ hydrodynamical simulations with and without feedback. We discuss our main results in section~6. %We present the clustering results based on transmitted flux in the Appendix-\ref{Corr_flux_section}.
%
%\Anand{check the following}.
{
In this work we use the flat $\Lambda$CDM universe with the following cosmological parameter ($\Omega_\Lambda$, $\Omega_m$, $\Omega_b$, $h$, $n_s$, $\sigma_8$, Y) $\equiv$ (0.69, 0.31, 0.0486, 0.674, 0.96, 0.83, 0.24) based on \citep[][]{planck2014}. Cosmologicial parameters used in our simulations are slightly different and are summarised in section~\ref{sec:simulations}.}

\section{Data Sample}
\label{sample}
\begin{figure}
	\includegraphics[viewport=10 0 420 250,width=8cm, clip=true]{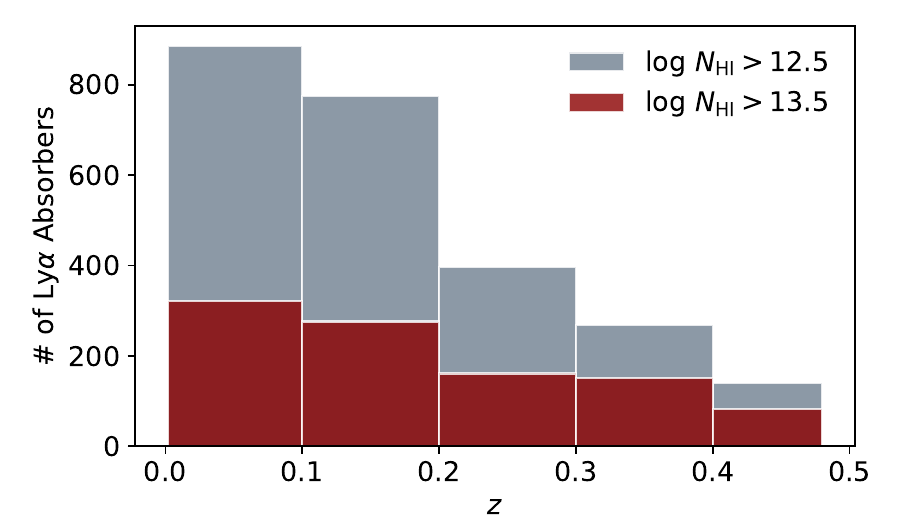}%
	
	\includegraphics[viewport=0 0 340 270,width=8.8cm, clip=true]{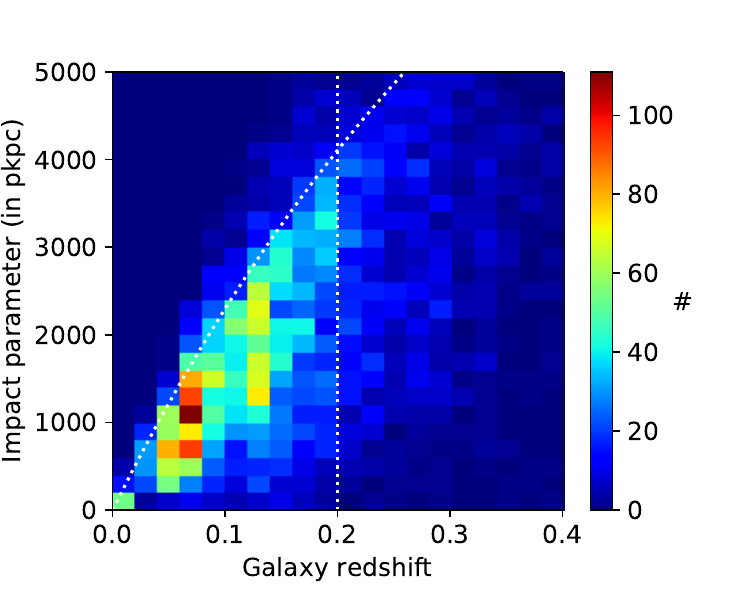}%

	\caption{{  \textit{Top:} Redshift distribution of number of  \lya\ absorbers in the sample used in this work.
	%Redshift intervals used are $z=0.05\pm 0.05,\ 0.15\pm 0.05,\ 0.25 \pm 0.05$ and $\ 0.39 \pm 0.09$. 
	We present this for  two \HI\ column density thresholds.}
	{ \textit{Bottom:} 2D histogram plot of redshift vs. impact parameter of the complete galaxy sample along the quasar sightlines used in this study. The vertical dashed line denotes the maximum galaxy redshift used for our study. The dashed curve shows the projected length scale as a function of redshift for an angular scale of 20'.}}
\label{Redshift_distribution}
\end{figure}

We use the publicly available data sample\footnote{\url{http://archive.stsci.edu/prepds/igm/.}} of low-redshift \lya\ spectra towards 82 UV-bright QSOs ($z_{em}$<0.85) observed using Hubble Space Telescope-Cosmic Origins Spectrograph (HST-COS) by \cite{danforth2016}. The sample covers \lya\ forest at $z\leq 0.48$ and { the spectra were obtained at a} resolution of $\sim$17\kms. In \cite{danforth2016}, the spectra were continuum fitted and 5138 absorption line features arising from the intervening IGM were identified. 
%This comprises of 4234 \lya\ lines, 606 \lyb\ lines and 1633 metal lines for 25 metal ion species. %\PG{The numbers are not adding to match 5138 why is that? What about other Lyman series lines (Ly-gamma, Ly-delta etc) identified in COS? } 
The redshift, column density, Doppler parameter $b$, equivalent width and the significance level of detection corresponding to each of these absorption features were tabulated. {  We use publicly available parameters of the Voigt profile components obtained by \citet{danforth2016} for the clustering study in this work.} {  The redshift distribution of the \lya\ absorbers used in this work is given in { the top panel of} Fig.~\ref{Redshift_distribution}.}

%\Anand{Let us define the biased sample here itself and identify the sightlines used for this. In the case of Stocke you set min z by taking galaxy v+300 km/s}.

For this work, we consider \lya\ absorption lines having $z \leq 0.48$, %While considering \lya\ forest for this redshift range, we avoid 
avoiding the proximity regions blue-wards to the quasar redshift within 1500 \kms\  {(corresponding to a proper distance of 20.65 pMpc at $z\sim0.1$)} and within 500 \kms\ red-ward of $z=0$ {   \citep[similar to][]{danforth2016}}. 
%\PG{Does this avoids geo-coronal line emission and \lya emission from milkyway?} %Also, in this sample of 82 QSOs, 6 of them  
{ The redshift range $1500<\Delta v<5000$ \kms bluewards of the quasar redshift may also be affected by high-velocity outflows from the quasar. So, absorption systems having strong absorption in high ion species but weak \HI\ , strongly non-gaussian absorption profiles or doublet ratio close to 1:1 indicating possible partial coverage of the source are removed \citep[see Sec.2.4.2][]{danforth2016}. }
Six quasars in this sample have originally been targeted to study the CGM near $z\leq 0.02$ galaxies \citep{stocke2013}\footnote{1ES1028+511 ($cz_{gal}$ = 649 and 934 \kms); 1SAXJ1032.3+5051 ($cz_{gal}$ = 649  \kms); HE0435-5304 ($cz_{gal}$ = 1673 \kms); PG0832+251 ($cz_{gal}$ = 5226 \kms); RXJ0439.6-5311 ($cz_{gal}$ = 1673 \kms) and SBS1108+560 ($cz_{gal}$ = 696 \kms)}. In order to remove any bias, %coming from these targeted samples of QSOs, 
we set a lower redshift limit for these sightlines to 300~\kms\ redwards of the redshift of the { target galaxy}. The redshift path length coverage of the \lya\ forest after removing these biased regions is about 19.9. 

%For checking the metallicity dependence of \lya\ clustering, we consider \CIV, \OVI\ and \SiIII\ metal line transitions in the absorbed spectra. 
%
%Depending on the rest wavelength of these metal ion species, the redshift coverage of these metal ion species corresponding to the \lya\ forest redshift range are: $z<0.16$ for \CIV, $0.1<z<0.48$ for \OVI\ and $z<0.48$ for \SiIII. 

%We also look for the connection of these observed \lya\ triplets with nearby galaxies.  For that,
We use a deep and wide galaxy redshift survey along 47 of HST-COS sightlines presented by \citet{keeney2018} to probe the connection between \lya\ clustering and galaxy distributions. We supplemented these data with the galaxies detected within 20' to the quasar sightline from SDSS and \citet{prochaska2011}. 34 sightlines in \citet{danforth2016} have galaxy information in \citet{keeney2018}. Five of these sightlines were also covered by \citet{prochaska2011}. We use the galaxy distribution around 8 other sightlines from \citet{prochaska2011}. Thus we have galaxy information around 41 sightlines in the sample of \citet{danforth2016}. In total we have 6174 galaxies close to these sightlines. { In the bottom panel of of Fig.~\ref{Redshift_distribution}, the redshift and impact parameter distribution of these galaxies is shown. Vertical dotted line marks $z=0.2$. We also show the impact parameter corresponding to 20'.}
We use these data to find the properties of the nearest galaxies to the isolated absorbers and the  absorbers showing strong two- and three-point redshift space (or velocity) correlation at $z\le 0.2$. At $z\sim0.1$ an angular scale of 20' corresponds to a projected length scale of $\sim$2.3 pMpc. {Apart from few cases the galaxy distribution from \citet{keeney2018} is known to be complete for $\sim0.1L*$ at $z<0.1$ and $L_*$ at z$\simeq$0.25. The data from \citet{prochaska2011} also reach similar depth by over half the angular scale (i.e up to 10' from the quasar sightline)}.

\section{Absorber-based statistics}

\cite{maitra2019,maitra2020} presented two- and three-point correlation studies of \lya\ forest
using 
%spectra decomposed into multiple 
Voigt profile components as a suitable probe of the clustering properties of the IGM  at $z>2$. {  These Voigt profile components can also be used to explore the connection between galaxies and intergalactic gas \citep[see][for example]{rudie2012a}.}
%{Even though it is well recognised that an individual Voigt profile component does not represent a physical { absorber}, they have been referred to as "{ absorbers}" for simplicity \citep[see for example][]{Stocke1995,penton2004}. In this work we follow the same terminology for the sake of continuity.}
%As we have used 
%With 
%techinques similar to what is done in the case of galaxies,
{ Throughout this paper, we will refer to these individual Voigt profile components as "absorbers". We will use the term absorption "system" to refer to the whole absorption profile.}
The  \lya\ { absorber} based approach allows { us} to study the two- and three-point correlations as a function of %neutral hydrogen column density 
$N_{\rm HI}$, {  $b$ and presence of metals}. In this section, we study the clustering properties of low-$z$ IGM by measuring longitudinal (i.e redshift space) three-point correlation of the \lya\ { absorbers}. For doing so, we first estimate the \NHI\ distribution of the \lya\ { absorbers} for the full sample and various sub-samples. This is an important first step to generate a set of mock sightlines having random distribution of \lya\ { absorbers} { that are used} as a comparison to estimate the longitudinal three-point correlation. We also { measure} the longitudinal two-point correlation in order to estimate the reduced three-point correlation, Q. %{For completeness sake, we present  the two- and three-point correlations obtained using statistics of transmitted flux in individual pixels in Appendix~\ref{Corr_flux_section}.}
%\PG{Please add a statement that you have decomposed the spectra using VIPER. You mention this in the text but very late. Reader may be confused weather you use VPFIT or any other code.}
%along these observed sightlines. 

\subsection{Neutral hydrogen column density distribution}
\label{sec_fnh}

\begin{figure}
	\includegraphics[viewport=0 0 370 230,width=8cm, clip=true]{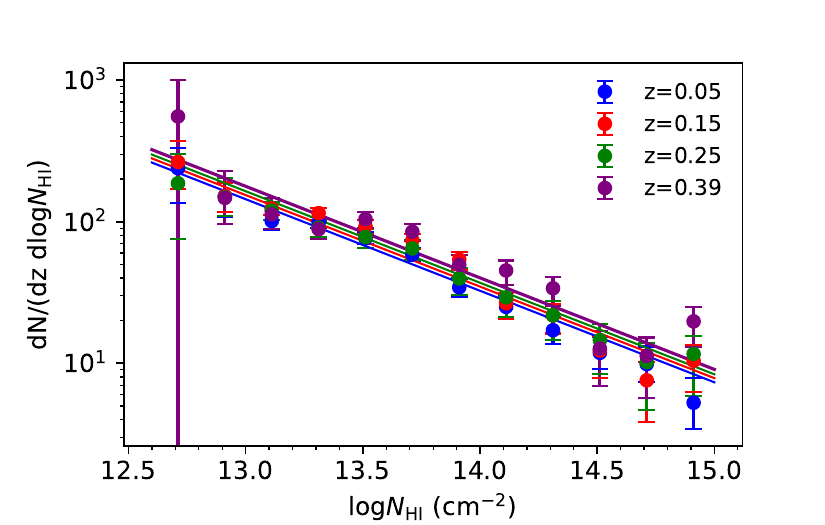}%
	
		\includegraphics[viewport=0 0 370 230,width=8cm, clip=true]{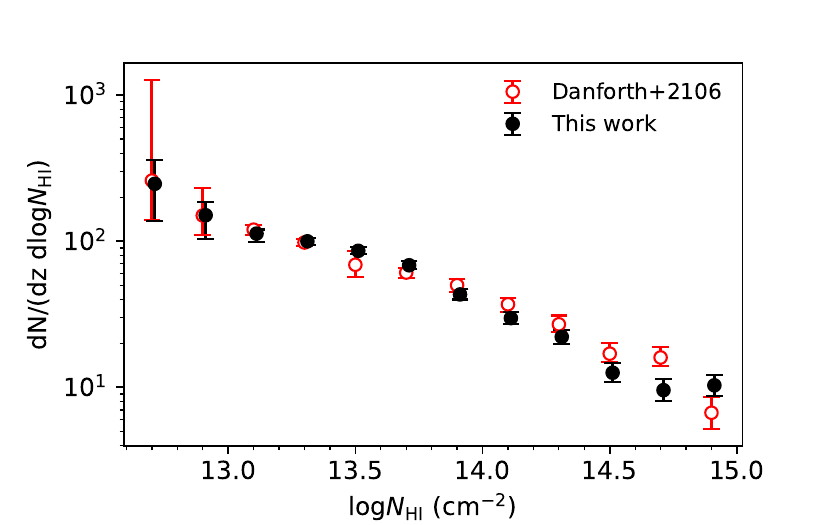}%
		
	\includegraphics[viewport=0 0 370 230,width=8cm, clip=true]{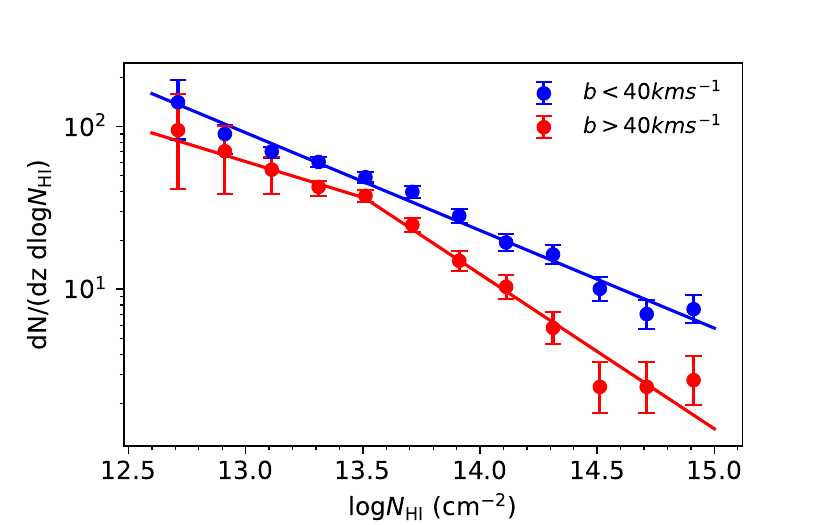}%

	\caption{ {\it Top panel}: \NHI\ distribution of \lya\ absorbers for four different redshift intervals ($z=0.05\pm 0.05,\ 0.15\pm 0.05,\ 0.25 \pm 0.05,\ 0.39 \pm 0.09$). This plot confirms a weak redshift evolution of $f(N_{HI},z)$. {\it Middle panel}: comparison of the \NHI\ distribution obtained by us for the entire sample with that of \citet{danforth2016}. {\it Bottom panel}: The \NHI\ distribution for 
	%two sub-sample: one having the line-width parameter 
	the low-$b$ ($b<40$ \kms) and the high-$b$ ($b>40$ \kms) sub-samples. {In the case of high-$b$ sub-sample the distribution is well fitted by a double power-law. The double power-law fit to high-$b$ and single power-law fit to the low-$b$ samples are also shown.}
	%\Anand{It may be good to show the double powerlaw fit overlayed..}
	}
\label{NHI_distribution}
\end{figure}

In HST-COS spectra, the Signal-to-Noise Ratio (SNR)  varies substantially across the observed wavelength range for a given sightline.
%depends on the wavelength $\lambda$ and varies substantially across the observed sightline. 
So, the detectibility of any absorption feature has  a wavelength (or $z$) dependence. We need the intrinsic distribution of \lya\ absorbers having different column densities (\NHI) to construct the random distribution of the { absorbers}
after appropriately taking into account
%and thus account for 
the $z$-dependence of detectibility along each sightline. 
%For our work, we take the Voigt profile fitting results for the \lya\ absorption features from \citet{danforth2016}.

Following the standard practice \citep[see for example,][]{Petitjean1993}, we define neutral hydrogen column density ($N_{\rm HI}$) distribution $f(N_{\rm HI},z)$ as the number of \lya\ absorbers having log$N_{\rm HI}$ in the range of log$N_{\rm HI}\pm d$log$N_{\rm HI}/2$ and lying within redshift interval of $z\pm dz/2$. Traditionally, this distribution is approximated by a power law of $N_{\rm HI}$ and $z$, i.e,
\begin{equation}
    f(N_{\rm HI},z)=\frac{d^2N}{d{\rm log} N_{\rm HI} dz}={\rm C_0}(1+z)^{\gamma}{N_{14}}^{\beta}.
\label{NHI_distribution_eqn}
\end{equation}
Here, $N_{14}$ is $N_{\rm HI}$ expressed in units of $10^{14}cm^{-2}$. 

{To calculate the intrinsic distribution of the absorbers, we  account for the "incompleteness" of the data sample. 
For a given column density we calculate the required spectral SNR so that the absorption line produced can be detected at $>4\sigma$  level. We do so by using the curve of growth for a given \NHI\ and assuming a median b value (i.e 34 \kms\ for the full sample). So for a given \NHI\ bin, only those pixels having { the observed SNR} greater than the SNR limit where the absorption can be detected { above 4$\sigma$ level} are identified.
%
%For an absorber having a column density, $N_{\rm HI}$, %value within a certain $N_{\rm HI}$ bin, 
%we estimate the spectral SNR {  limit} required to detect its \lya\ absorption  at  a 4$\sigma$
%level. 
%We estimate this SNR limit from the limiting 
%This is estimated from the
%equivalent width, required for detecting an absorption, calculated using curve of growth for a given $N_{\rm HI}$ and the median $b$ value of the sample (which is $\sim 34$ \kms\ for the full sample) .
%detection of \lya\ absorption. 
Then  we consider only the wavelength range covered by such pixels
%observed SNR is higher than this estimated SNR {  limit}
%required to detect absorbers in that $N_{\rm HI}$ bin 
\citep[as demonstrated in Fig.4 of][]{gaikwad2017b}
 %
%Only these wavelength ranges are considered 
for the calculation of redshift path length $dz$ in Eq.~\ref{NHI_distribution_eqn} . We calculate the total redshift path length corresponding to a given $N_{\rm HI}$ bin by integrating over all the quasar sightlines.
%\Anand{Give bit more details on how this is done. For example, what b values used etc should be given}.
%and use it to define the number of absorbers identified within a given redshift interval. 
}

The redshift path length calculated 
in this way 
takes care of incompleteness coming from regions having { the observed} SNR lower than what is required for detecting an absorption line in a certain $N_{\rm HI}$ bin\footnote{ Note that we use the median b values and not the full distribution of b for these calculations.}. We find that 25\%, 50\% and 75\% of the observed redshift path length is sensitive enough to detect absorbers having  log~$ N_{\rm HI}=$12.68, 12.84 and 13.0, respectively. 
%This is similar to the 25\%, 50\% and 75\% confidence in redshift path length obtained in
The corresponding values obtained by \citet{danforth2016} are log~$ N_{\rm HI}=$12.77, 12.93 and 13.09. The minor differences come from the fact that while \citet{danforth2016} considers all the $N_{\rm HI}$ measurements (including systems with measurements based on Ly$\beta$ absorption for which \lya\ is not covered), we consider only systems where \lya\ absorption is covered in the HST-COS spectra. {Also we avoid regions around 6 known galaxies that were searched for CGM absorption (see Section.~\ref{sample} for details).}

In the top panel of Fig.~\ref{NHI_distribution}, we plot $f(N_{\rm HI},z)$ in different ${\rm log}N_{\rm HI}$ bins 
%of [12.6,12.8,13.0,....,14.8 and 15.0] 
for 4 different redshift intervals ($z=0.05\pm 0.05$, $0.15\pm 0.05$, $0.25 \pm 0.05$ and  $0.39 \pm 0.09$). {  The error in the distribution is one-sided poissonian uncertainty in the number of absorbers corresponding to $\pm 1\sigma$} computed over all the sightlines in a redshift bin. {In the error, we also account for the uncertainty in $dz$ sourcing from the variation of completeness limit for a finite $N_{\rm HI}$ bin width.} We then fit $f(N_{\rm HI},z)$ according to Eq.~\ref{NHI_distribution_eqn}. The fitted parameter values are $\rm C_0=31\pm 2$, $\beta=-0.65 \pm 0.03$ and $\gamma=0.7\pm 0.3$. The $C_0$ and $\gamma$ values are similar to $f(N_{\rm HI})=(167)N_{13}^{-0.65\pm 0.02}\approx(37)N_{14}^{-0.65\pm 0.02}$ reported in \citet{shull2015}, $f(N_{\rm HI})=(23\pm 1)N_{14}^{-0.67\pm 0.01}$ reported in \citet{danforth2008} and {  $f(N_{\rm HI})=(25\pm 1)N_{14}^{-0.65\pm 0.02}$ reported in \citet{danforth2016} } considering no redshift evolution in $N_{\rm HI}$ distribution. As seen in Fig.~\ref{NHI_distribution}, $f(N_{\rm HI},z)$ depends strongly on the $N_{\rm HI}$ of the absorber while having a weak dependence on $z$ in the redshift range probed. For a sanity check, we compare this distribution with the one obtained in \citet{danforth2016} in the middle panel of Fig.~\ref{NHI_distribution} and find them to be similar
%. We find the two distributions to be similar 
within measurement uncertainties. {  One caveat which needs to be mentioned is that while \citet{danforth2016} calculate the \NHI\ distribution for the entire \HI\ sample, we only do so for the \HI\ \lya\ absorbers which we use for this study. Our computed errors match well with those of \citet{danforth2016}. } 

In the bottom panel of Fig.~\ref{NHI_distribution}, we plot $N_{\rm HI}$ distribution in two different bins based on $b$-values (high-$b$ sample with $b>40$ \kms\ and low-$b$ sample with $b<40$ \kms) considering systems in the full sample. The cut-off $b$-value of 40 \kms\ was chosen to delineate the possible BLAs from rest of the \lya\ absorbers as defined in \citet{Lehner2007}. About 31.9\% of the total \lya\ absorbers in our sample have $b>40$ \kms. { In both cases, we recalculated the redshift path length considering the median $b$-values of the sub-samples. }
%The distributions shown in Fig.~\ref{NHI_distribution} are corrected for the incompleteness. 
It is evident that both distributions have similar slope at low \HI\ column density end (i.e., log~$N_{\rm HI}$ $< 10^{14}$ cm$^{-2}$) . However, we do notice a fall in the number of high-$b$ systems at high  $N_{\rm HI}$ end. We fit the individual distribution using the form given in Eq.~\ref{NHI_distribution_eqn} ignoring the redshift evolution. The best fit values of $C_0$ and $\beta$  for low-$b$ and 
high-$b$ sub-samples are $C_0=23.0\pm1.5,\ \beta=-0.60\pm0.02$ and $C_0=9.5\pm1.0,\ \beta=-0.83\pm0.04$ respectively. {  In case of the high-$b$ sub-sample, we also fit the distribution with a double power law about $N_{\rm HI}=10^{13.5}$cm$^{-2}$ and obtain $C_0=22.1\pm2.4,\ \beta=-0.44\pm0.07$ for $N_{\rm HI}<10^{13.5}$cm$^{-2}$ and $C_0=12.4\pm1.3,\ \beta=-0.95\pm0.08$ for $N_{\rm HI}>10^{13.5}$cm$^{-2}$. } We use these fitted distribution  {(i.e  in general a single power-law fit and double power-law in the case of high-$b$ sub-sample)} to generate the random distribution of absorbers. 
%{Additionally, we also plot $f(N_{\rm HI},z)$ for two sub-samples of the data: one having line-width parameter $b<40$\kms\ and the other having $b>40$\kms. About 31.8\% of the total \lya\ absorbers in our sample have $b>40$\kms. As evident from the distribution, $b>40$\kms\ { absorbers} have a relatively flat $N_{\rm HI}$ distribution upto $N_{\rm HI}=10^{13.5}$cm$^{-2}$ as compared to $b<40$\kms\ { absorbers}. We will use these two sub-samples to study dependence of clustering on $b$.}

\subsection{Longitudinal two-point correlation function}

%\PG{You use slightly different estimator in paper-I from Landy and Szalay 1993
%zeta = (<DD> - 2<DR> + <RR>) / <RR>
%Are the two estimator give same answers?
%I thought these estimators give slightly different %correlation function?}
We follow a standard procedure for finding the longitudinal two-point correlation function of \lya\ absorbers. We calculate the probability excess of finding a pair of { absorbers} in the observed data relative to finding them in a random distribution of { absorbers} at certain redshift space separation. We select { absorbers} above a given $N_{\rm HI}$ threshold along the sightlines and estimate the longitudinal two-point correlation using the estimator, 
	\begin{equation}
		\xi( r_{\parallel})=\frac{<DD>}{<RR>}-1 \ ,
	\end{equation}	 
where "DD" and "RR" are the data-data and random-random pair counts of { absorbers} respectively at a separation of $r_{\parallel}$ \citep[see][]{kerscher2000}. {Our choice of a normal estimator for two-point correlation is motivated by the fact that the clustering amplitudes are relatively independent of the choice of estimators at small scales, as shown in \citet{kerscher2000}. We checked and found that it holds for \lya\ absorbers at the scales of interest in this study. Same is true for three-point correlation too (see Eq.~\ref{zeta}). So, choose to go with the normal estimators as they save the computational time significantly (especially in the case of three-point correlation).  The total data-data pairs $DD$ are summed over all the sightlines in their respective $r_{\parallel}$ bins and then normalized with $n_D (n_D-1)$ (where $n_D$ is the total number of absorbers).} 
%The estimator calculates the correlation between the observed distribution of { absorbers} relative to a random distribution of { absorbers}. 
We use 100 random sightlines for every data sightline to minimize the variance in random. All the random pair counts for a given separation $r_{\parallel}$ are also normalized with total number of pair combinations ($n_R (n_R-1)$). 
First, the distribution of { absorbers} along the random sightlines are generated using the best fitted power-law given by Eq.~\ref{NHI_distribution_eqn} for an observed sightline of length $dz$. We consider the random sightlines to have the same redshift range and same wavelength dependent SNR as the observed data. We compute the expected intrinsic number of random { absorbers} to be populated along the sightline (i.e. $N_{\rm abs}(R)$) by integrating Eq.~\ref{NHI_distribution_eqn} along the redshift pathlength and over the $N_{\rm HI}$ range in consideration,
\begin{equation}
    N_{\rm abs}(R)=\int_{z_l}^{z_u} \int_{N_{\rm HI, l}}^{N_{\rm HI, u}}f(N_{\rm HI},z)~dz~ d{\rm log}N_{\rm HI} .
\end{equation}
Here, $z_l$ and  $z_u$ are the minimum and maximum $z$ along the sightline respectively. $N_{\rm HI, l}$ is the lower $N_{\rm HI}$ threshold for the { absorbers} to be considered and we fix the upper limit $N_{\rm HI, u}$ to be $10^{18}$cm$^{-2}$. %\PG{Shoulden't it be $\sim 10^{15.2}$, maximum NHI you get in Fig. 2?} 
Next, we populate $N_{\rm abs}(R)$ number of random { absorbers} along the redshift path length by generating $N_{\rm abs}(R)$ randoms from the probability distribution of $(1+z)^{\gamma}$. For each of these randomly generated { absorbers}, we associate a random $N_{\rm HI}$ by drawing samples from the distribution $N_{\rm HI}^\beta$. { Using the observed SNR, at the location of each randomly generated { absorber}, we compute the detection significance of the absorption line it will produce { assuming the doppler width to be the median $b$ of the sample}.  Only lines with detection significance above 4$\sigma$ (similar to what has been used for the observations) are considered for the clustering analysis. In this way we account for the bias coming from non-uniform SNR across the spectrum.}
%{ \color{magenta} PG: We also need to specify what $b$ (I assume median $b$) values we used for these random samples as asked by referee.}
%
%We then assign the average SNR over a resolution element around the positions of these random { absorbers} from the observed spectrum as the SNR value for these { absorbers}. While calculating the correlations, we reject { absorbers} having SNR values less than the SNR limit (with 4$\sigma$ significance) set by the completeness requirement for  resolving the absorber with a chosen lower $N_{\rm HI}$ threshold.  %We do this for both observed as well as random sightlines.

{ We compute the two-point correlation logarithmically spaced $r_{\parallel}$ bins of [0.5-1, 1-2, 2-4, 4-8, 8-16, 16-32, 32-64] pMpc. We have taken this binning scheme specifically for the calculation of reduced three-point correlation function (see Eq.~\ref{Q_eqn}).  We compute three-point correlation for collinear triplet configurations, which we explain in the next subsection. For such configurations, the third arm of the triplet will be double the length of the other two arm lengths. So, we take the $r_{\parallel}$ bins such that the mean of the next bin value is exactly double that of the previous bin value. This makes calculation of the cyclic combination of two-point correlations, $\overline{\xi(r_1)}\times\overline{\xi(r_2)}+\overline{\xi(r_2)}\times\overline{\xi(r_3)}+\overline{\xi(r_1)}\times\overline{\xi(r_3)}$ (see Eq.~\ref{Q_eqn}), necessary for calculating the reduced three-point correlation at each bin easier.}

\begin{figure*}
	\includegraphics[viewport=0 8 300 290,width=5.9cm, clip=true]{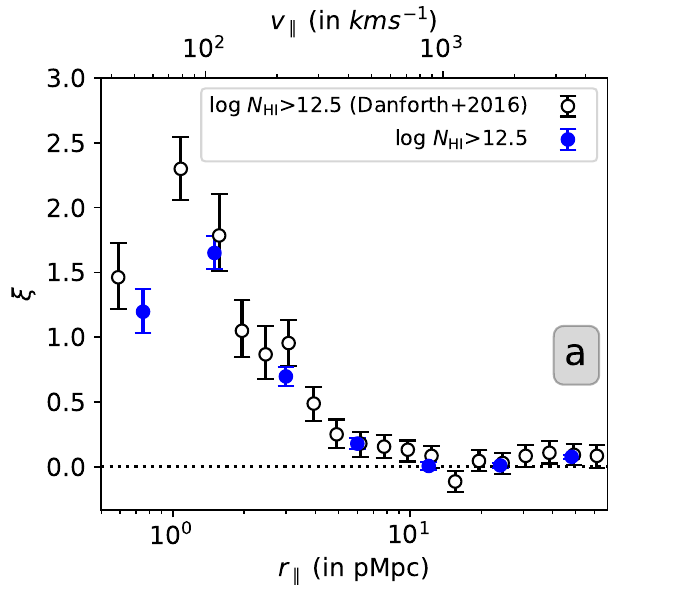}%
	\includegraphics[viewport=0 8 300 290,width=5.9cm, clip=true]{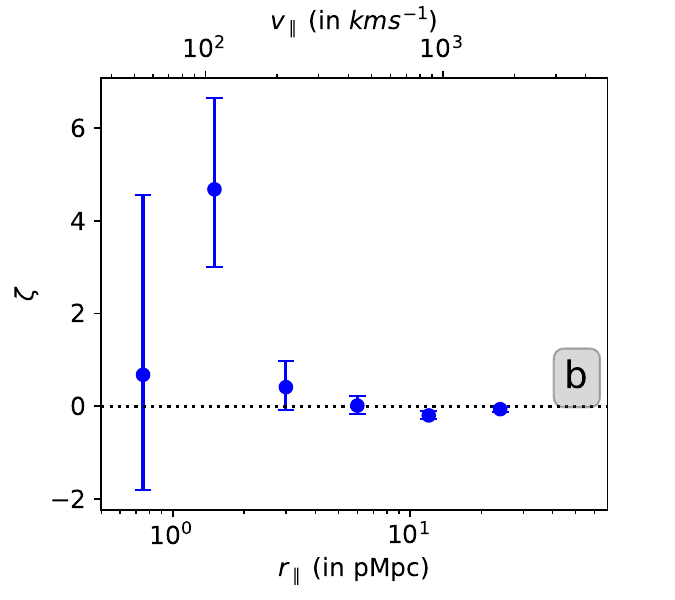}%
    \includegraphics[viewport=0 8 300 290,width=5.9cm, clip=true]{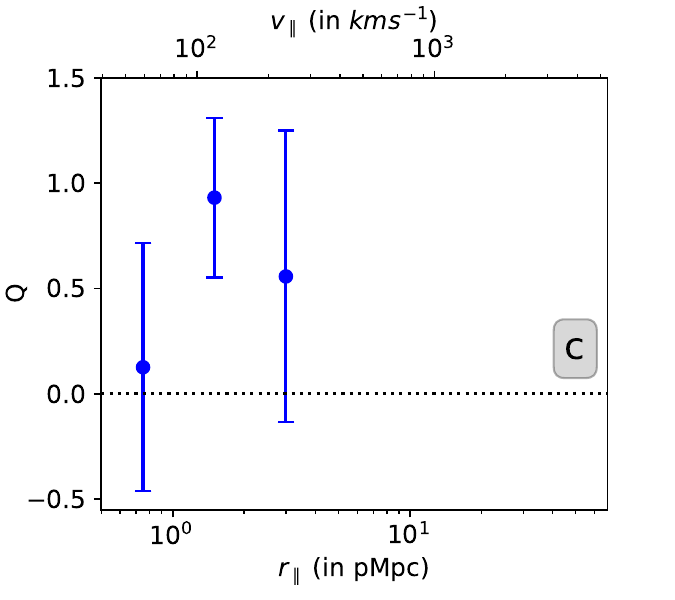}%
    
    \includegraphics[viewport=0 8 300 265,width=5.9cm, clip=true]{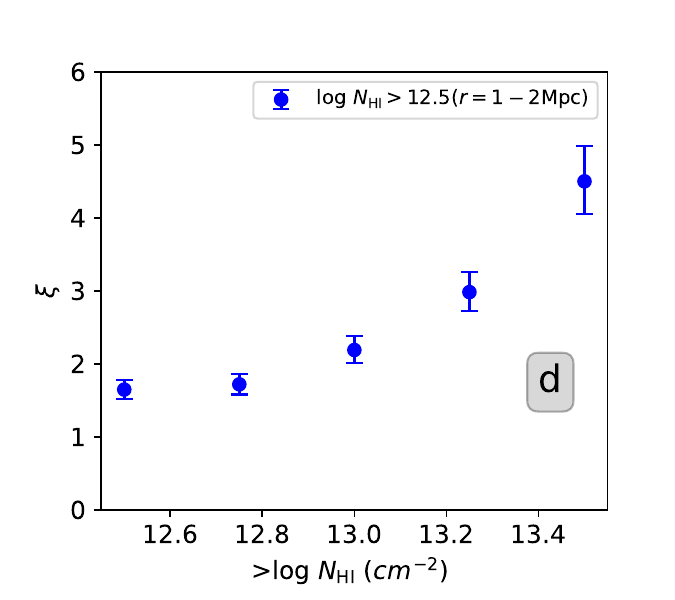}%
	\includegraphics[viewport=0 8 300 265,width=5.9cm, clip=true]{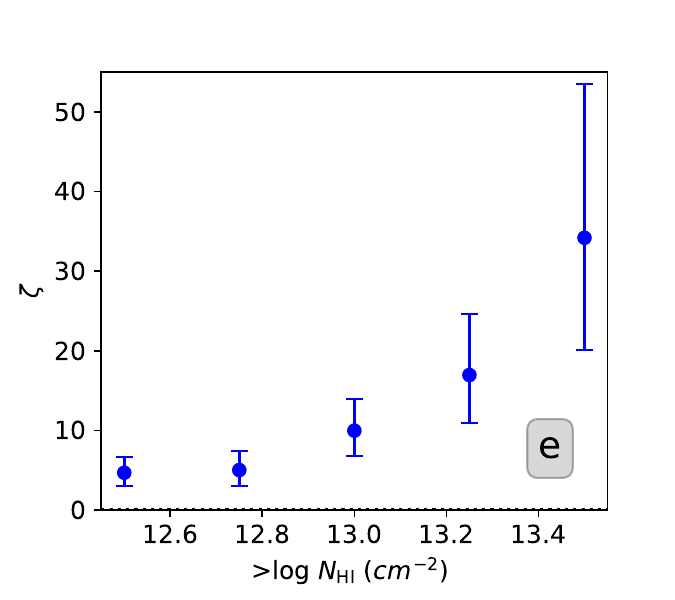}%
    \includegraphics[viewport=0 8 300 265,width=5.9cm, clip=true]{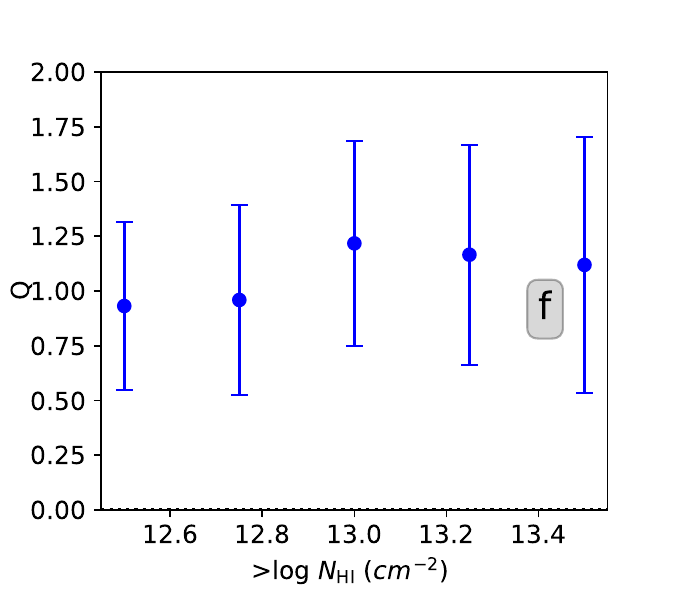}%
    
    \includegraphics[viewport=0 8 300 265,width=5.9cm, clip=true]{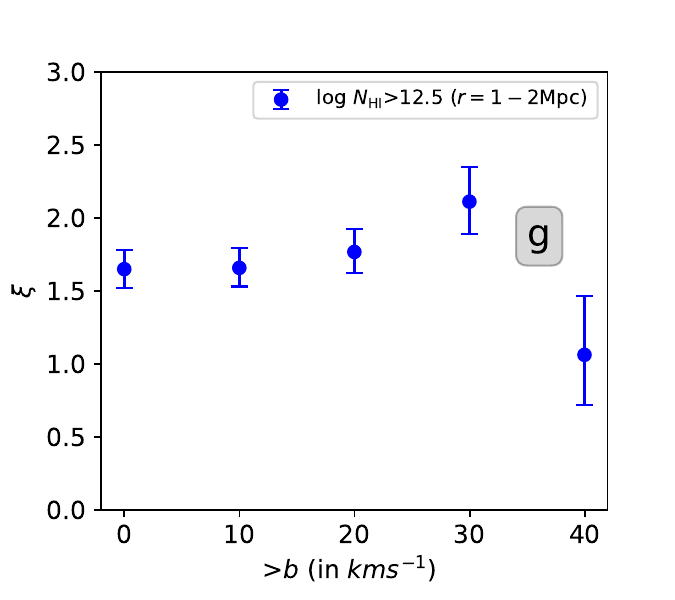}%
	\includegraphics[viewport=0 8 300 265,width=5.9cm, clip=true]{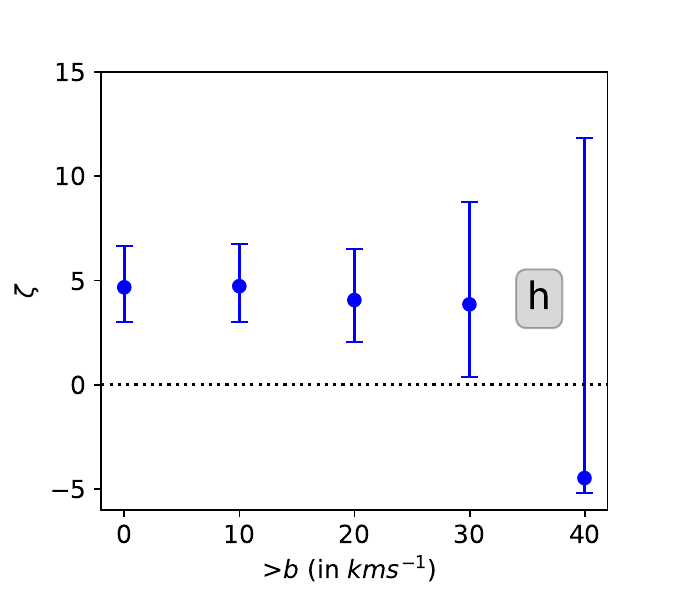}%
    \includegraphics[viewport=0 8 300 265,width=5.9cm, clip=true]{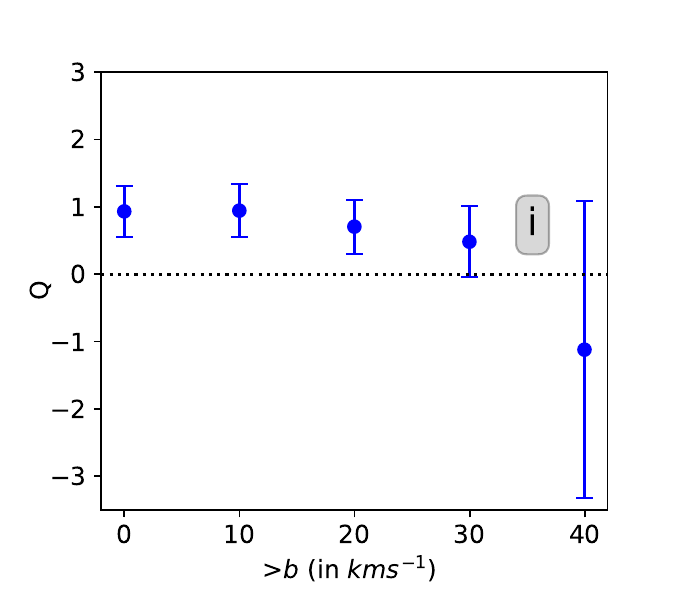}%
	\caption{Absorber-based longitudinal two-point, three-point and Q (left to right) of \lya\ absorbers as a function of longitudinal scale (top panels), $N_{\rm HI}$ thresholds (middle panels) and $b$ thresholds (bottom panels) . We consider $r_\parallel$ = 1-2 pMpc scale to probe the \NHI\ and $b$ dependence. The errors represent { one-sided poissonian uncertainty corresponding to $\pm 1\sigma$} about the mean value. In panel (a) we compare our measurements with those of
	%the two-point correlation measured in this study with those reported by 
	\citet{danforth2016} for consistency check. {  Our measurements of two-point and three-point correlation as a function of longitudinal scale is provided in Table~\ref{Corr_Table}.}
	%\PG{It would be good if you show size of r-parallel bins at least for one panel since you are using log scale. Also see comment in section 4.3}
	}
\label{Corr2}
\end{figure*}

In panel (a) of Fig.~\ref{Corr2}, we plot the average two-point correlation of { absorbers} with $N_{\rm HI}>10^{12.5}$cm$^{-2}$ for our full sample as a function of $r_{\parallel}$.
The error in the longitudinal two-point correlation is one-sided poissonian uncertainty corresponding to $\pm 1\sigma$ for all the data-data pairs. The uncertainty in the pairs for the large number of random absorbers taken is assumed to be relatively negligible. %As discussed before, we also assign weights to different sightlines based on the length of the sightlines for the different resampling of the sightlines. 
For a sanity check of our method, we compare our measurements with those of \citet{danforth2016}. The two-point correlation profile matches well within the errorbars (see panel (a) in  Fig.~\ref{Corr2}). {  Our two-point correlation measurements are given in Table~\ref{Corr_Table}.}

There are three scales of interest for longitudinal two-point correlation. At smaller scales ($r_{\parallel}< 1.0$ pMpc), we observe a suppression in the two-point correlation. This region of { suppression} for \lya\ absorbers is affected by thermal broadening along with instrumental resolution which sets a lower limit on scale for identification of multiple \lya\ absorption lines during Voigt profile decomposition. We also expect pressure broadening and small scale clustering (and turbulence) of the baryonic gas to play a part in { absorber } { suppression} at smaller scales.
At intermediate scales ($1.0$ pMpc$\leq r_{\parallel}\leq 6$ pMpc), longitudinal two-point correlation falls steadily and becomes consistent with zero beyond 10 pMpc.% {  The radial profile for the two-point correlation function obtained from the measured transmitted flux in individual pixels is shown in Appendix~\ref{Corr_flux_section}.}

Next we explore the $N_{\rm HI}$ dependence of $\xi$ in panel (d) of Fig~\ref{Corr2}. Here we mainly focus on the $r_\parallel$ bin of 1-2 pMpc. {Consistent with the past studies, the amplitude of the two-point correlation steadily increases with increasing $N_{\rm HI}$ threshold \citep[see for example,][]{Penton2002, danforth2016}}.
%\Anand{Add references}.
%
%
%{We also look for $N_{\rm HI}$ dependence at the scales of strongest clustering, which is shown in the bottom left panel of Fig.~\ref{Corr2}. At the scale of 1-2Mpc, the magnitude of two-point correlation steadily increases with increase in $N_{\rm HI}$ thresholds. 
%
%Higher $N_{\rm HI}$ absorbers typically originate from regions having larger underlying baryonic over-densities. 
As mentioned in the introduction, there is a strong correlation between \NHI\ and over-density. Therefore, \NHI\ dependence of clustering reflects the stronger clustering of more over-dense regions \citep[See][for discussions on this related to high-z IGM]{maitra2020}.
{  It is known that stronger \lya\ absorbers (i.e with $N_{\rm HI}>10^{14}$cm$^{-2}$) at low-$z$ are clustered strongly with the galaxies while the weaker absorbers are distributed more randomly or associated with galaxy voids or IGM \citep{Penton2002,Tejos2014}.} Therefore, increase in two-point correlation with $N_{\rm HI}$ threshold, could imply a stronger spatial clustering of \lya\ absorbers {  associated with environments of galaxies}.
%which probe larger underlying density fields. 
%Alternatively simulations suggests a strong connection between over densities and 
%{  in comparison to IGM} 
%\citep[refer to discussions in][]{maitra2020}. 

%\Anand{Refine this with reference: In the context of low-$z$ absorbers the \NHI\ dependent clustering could mean typically the low \NHI\ absorber tracing the IGM and high fraction of high \NHI\ { absorbers} are associated to galaxies. }
%
%}

\subsection{Longitudinal three-point correlation function}

{  The probability excess of finding a triplet of \lya\ { absorbers} in the observed data in comparison to a random distribution of { absorbers} can be used to estimate the longitudinal three-point correlation function. However, the two-point correlations associated with the three arms of the triplet need to be subtracted from this probability excess to get the true three-point correlation function (see Eq.~\ref{zeta}).} The triplets are defined as three collinear points along a sightline having separations $r_{\parallel, 1}$ between the first and second points and $r_{\parallel, 2}$ between the second and third points. In this work, we mainly consider the equal length configuration $r_{\parallel, 1}=r_{\parallel, 2}=r_{\parallel}$ (within the assumed bin-intervals) . 
%\PG{Is this criteria strictly true or do you allow for r-parallel to vary slightly in bins? If yes then what is the size of bin? It may be good to show this as an error in r-parallel in Fig. 4. This is because you may be taking r-parallel in linear bins but the correlation function is plotted in log scale on x-axis.} 
%Note, we measure the 
Our spatial separation measurements between the { absorbers} using the redshift difference
%. This will be 
is influenced by peculiar velocities. We address the effect of peculiar velocities on our measurements using simulated spectra in Section~
\ref{sec:simulations}.

 {The probability excess of \lya\ triplets with a separations of $r_\parallel$, $\rm PE_3$, is calculated for { absorbers} above a certain $N_{\rm HI}$ threshold using,
%and is estimated using, 
%\Anand{Change the text slightly to accommodate the following changes.} 
%\PG{Again you are using different expression here than used in Paper-I. There may be some bias introduce by this simple estimator?}
\begin{equation}
\rm PE_3 = \frac{<DDD>}{<RRR>}-1,
\end{equation}
where "DDD" and "RRR" are the data-data-data and random-random-random triplet counts. The $\rm PE_3$ is related to the three point correlation function, $\zeta( r_{\parallel})$, by
\begin{equation}\label{zeta}
		\zeta( r_{\parallel})={\rm PE_3}-\xi(r_1)-\xi(r_2)-\xi(r_3) \ ,
\end{equation}
where 
%"DDD" and "RRR" are the data-data-data and random-random-random triplet counts 
 $\xi(r_1),\xi(r_2)$ and $\xi(r_3)$ are the two-point correlations for arm lengths joining the three points considered for three-point correlation \citep[see Eq. 18 and 20 of][]{peebles1975}. 
 %The first term in the right hand side of Eq.~\ref{zeta} is the probability excess for the assumed triplet configuration.
 } 
 Similar to the two-point correlation, all the triplet counts at a certain separation $r_{\parallel}$ is normalized with total number of triplet combination. The $\rm PE_3$ and $\zeta$ for each data sightlines is calculated using 1000 random sightlines. The distribution of random { absorbers} along the sightlines and assignment of their $N_{\rm HI}$ values are done in a similar fashion as described earlier. { The three-point correlation is also computed in logarithmically spaced $r_{\parallel}$ bins as explained above.}
%
%Similar to two-point, the longitudinal three-point correlation is also calculated in logarithmically spaced bins of $r_{\parallel}=[0.5,~1,~2,~4,~8,~16,~32~{\rm and}~64]$ pMpc and is plotted in 
In panel (b) of Fig.~\ref{Corr2} we plot three-point correlation for $N_{\rm HI}>10^{12.5}$cm$^{-2}$ { absorbers}. 
%The relative contribution of each sightline to the average longitudinal three-point correlation function is weighted based on the length of the sightline. 
%The correlation is calculated for $N_{\rm HI}>10^{12.5}$cm$^{-2}$ { absorbers}. 

{We  measure positive three-point correlation (and probability excess) at scales below 4 pMpc (i.e line of sight velocity scales $<300$~\kms). At the scale of 1-2 pMpc, we have the strongest detection in three-point correlation (probability excess of $8.8^{+2.0}_{-1.7}$ and the corresponding three-point correlation of $\zeta = 4.8^{+2.0}_{-1.7}$  at  $\sim 2.6 \sigma$ significance level).} 
%We obtain a $\zeta$ value of $17.9\pm 5.5$. 
The amplitude of the three-point correlation in this $r_{\parallel}$ bin is higher than the corresponding two-point correlation. Similar to two-point correlation, we see the effects of suppression at scales below 1 pMpc. 
%However, compared to two-point, three-point correlation has a sharper spatial (i.e redshift space) profile. It falls down to zero beyond 4 pMpc for all $N_{\rm HI}$ thresholds. 
%
In panel (e) of Fig.~\ref{Corr2}, we plot $\zeta$ for different $N_{\rm HI}$ thresholds for the scale 1-2 pMpc. While a trend of increasing $\zeta$ with $N_{\rm HI}$ is evident, the measurement errors are large at high $N_{\rm HI}$ end due to small number of high \NHI\ absorbers involved. {  Our measurement of the probability excess and three-point correlation function as a function of longitudinal scale is provided in Table~\ref{Corr_Table}.
%In Appendix~\ref{Corr_flux_section}, we also present the three-point correlation function obtained based on transmitted flux in individual pixels.

}
%\PG{Is this due to small number of { absorbers}? then you can mention this.}  

%{We also see a strong $N_{\rm HI}$ dependence on three-point correlation, as shown in the bottom middle panel of Fig.~\ref{Corr2}.
%Though the error bars are large, we see a positive trend of longitudinal three-point correlation increasing with $N_{\rm HI}$ thresholds at the scale of 1-2 pMpc.}

Next, we estimate the reduced three-point correlation Q for \lya\ { absorbers}, which is defined as the three-point correlation normalized with cyclic product of two-point correlations (along the three arms connecting the three points considered for three-point correlation) as,
\begin{equation}
    Q=\frac{\overline{\zeta(r_1,r_2,r_3)}}{\overline{\xi(r_1)}\times\overline{\xi(r_2)}+\overline{\xi(r_2)}\times\overline{\xi(r_3)}+\overline{\xi(r_1)}\times\overline{\xi(r_3)}} \ ,
\label{Q_eqn}
\end{equation}
where $r_1,r_2$ and $r_3$ are the arm lengths joining the three points considered for three-point correlation. For the longitudinal three-point correlation considered in this study, $r_1=r_2=\Delta r_{\parallel}$ and $r_3=2\times \Delta r_{\parallel}$ \citep[see][]{groth1977,peebles1980}.
In panel (c) of Fig.~\ref{Corr2}, we plot the reduced three-point correlation Q as a function of longitudinal separations. We plot only up to a length scale of 4 pMpc because beyond that, we observe negligible three-point correlation. {We find a Q value of $0.95^{+0.39}_{-0.38}
$ at $r=1-2$ pMpc. In the $r=2-4$ pMpc bin, the Q value is found to be similar with larger errorbars. At $r<1$ pMpc, Q value is found to be negligible. We notice similar trend even when we consider only \lya\ { absorbers} with \NHI$>10^{13.5}$ cm$^{-2}$}.  %\PG{The three-point correlation function and cyclic three-point correlation function peaks at different r-parallel. Do you understand why is that happening?}
%Beyond that we observe a non-zero Q value of $2.2\pm 0.8$ at $r_{\parallel}=$ 1.5 pMpc. 
%Also, we do not see much scale evolution beyond the exclusion region within the errorbars. 
In panel (f) of Fig.~\ref{Corr2}, we plot Q as a function of $N_{\rm HI}$ thresholds (for $r_{\parallel}=$ 1-2 pMpc bin) .
The value of Q remains nearly constant with increasing \NHI\ thresholds albeit with large error bars. 
%{\color{blue} While it is seen that Q value seemingly depends on the $N_{\rm HI}$ threshold and decreases with increasing $N_{\rm HI}$ threshold value.  
%To get a better constrain on scale and $N_{\rm HI}$ dependence, we need a bigger dataset of low redshift QSO sightlines.
In section~\ref{sec:simulations}, we will compare the Q values predicted by the simulations with the observed values.

\begin{figure}
    \centering
    \includegraphics[viewport= 10 48 370 290,width=8.5cm, clip=true]{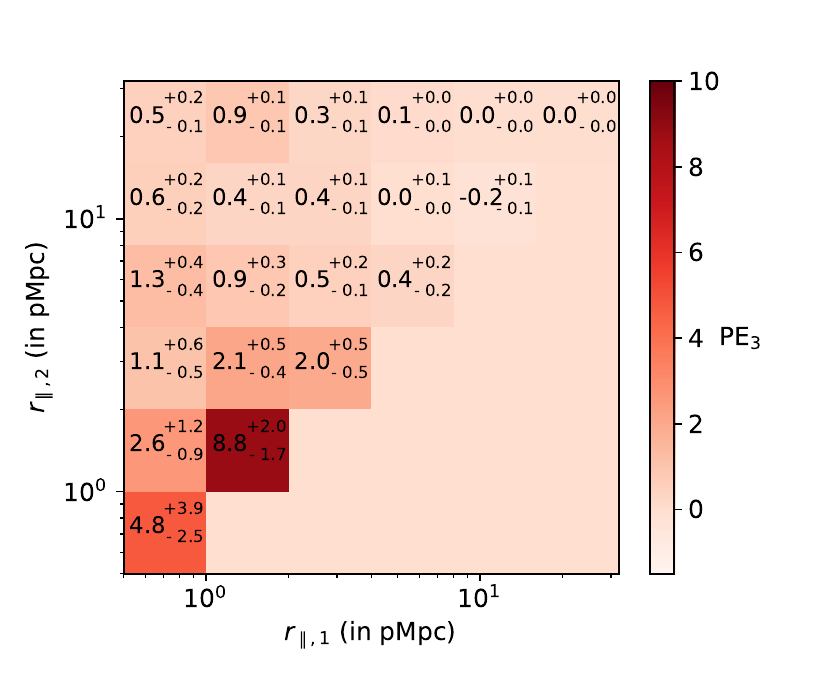}
    	
	\includegraphics[viewport= 10 13 370 290,width=8.5cm, clip=true]{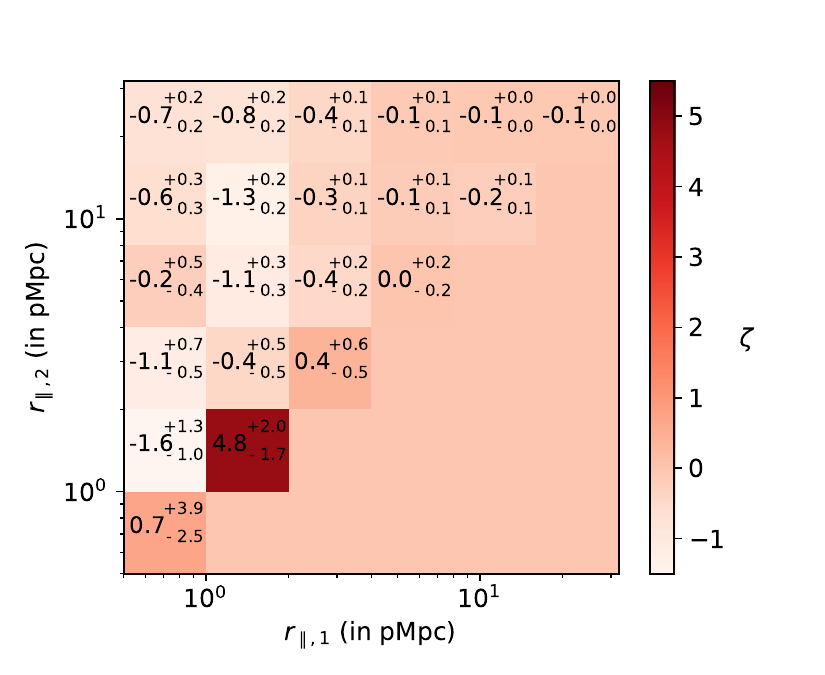}

	\caption{2D-plot of absorber-based longitudinal probability excess of \lya\ triplets (top) and three-point correlation (bottom) as a function of scales $r_{\parallel,1}$ and $r_{\parallel,2}$.  The measured values of $\rm PE_3$ and $\zeta$ with errors are provided in each bin. 
	%Strong three-point correlations are found for length scale less than 2 pMpc for the case of unequal arm.
	}
\label{Corr3_2d}
\end{figure}

Till now we have considered the case of equal length configuration (i.e $r_{\parallel, 1}=r_{\parallel, 2}=r_{\parallel}$) . In Fig~\ref{Corr3_2d}, we show the distribution of probability excess and three-point correlation for the case of
 $r_{\parallel, 1}\neq r_{\parallel, 2}$ for $N_{\rm HI}>10^{12.5}$cm$^{-2}$. 
{Among the off-diagonal elements the bins with $r_{\parallel, 1} = 0.5-1.0$ pMpc and $r_{\parallel, 2} = 1.0-2.0$ pMpc 
and  
% 
 %shows strong three-point correlation function (i.e $\zeta = 2.2\pm 0.9$) . For this bin we find the Q value to be {$0.5\pm 0.2$}. We find $\zeta = 2.0\pm0.5$ and
  %{$Q = 1.2\pm 0.4$}  for the bin 
  with $r_{\parallel, 1} = 1.0-2.0$ pMpc and $r_{\parallel, 2} = 2.0-4.0$ pMpc { show} a probability excess of $\sim2$ at more than 2.5$\sigma$ level. { However, none of these bins have} significant { non-zero} three point correlation. }
 %is also evident from the figure that we have significant (i.e $>3\sigma$ level) negative  $\zeta$ for $r_{\parallel,1} = 1-2$ pMpc with $r_{\parallel,2} = 4-4$ pMpc and $4-8$ pMpc.}
 %Even among bins with arm length ratio 1:2, Q increases with increasing scale albeit with large errors.
 {\citet{guo2016} have presented three point correlation function of galaxies for different configurations for 1:2 arm length ratio. Their result for $\theta\simeq 0$ (i.e squeezed configuration) will correspond to our equal arm configuration. 
 It is evident from their measurements 
 %(i.e for $\theta = \pi$) 
 that $\zeta$ of \lya\ forest is at least an order of magnitude smaller than what has been seen for galaxies for the same scales. In terms of velocity scale, the length over which one sees significant three point correlation function {(i.e $\simeq$ 1-2 pMpc or 146 \kms in the velocity scale at the median redshift i.e $z\sim0.15$ of our sample)} is consistent with the velocity dispersion of gas { clouds} in high mass galactic halos.} {  Conversely, this can also correspond to the large-scale structures in real space.}
 %\PG{Please give reference to work that measure velocity dispersion of gas { absorbers} in high mass halos.} 
 So it is important to explore how much contribution to the observed three point correlation comes from CGM (see section~\ref{con_gal}) . 
 %This we will explore in section~\ref{con_gal}.
 
% In what follows, while studying the dependence of clustering on the \lya\ parameters, we mainly focus on equal length configurations.
% \Anand{I think we should compare the Q values from low-z galaxy surveys}.
 
%

\begin{table*}
\caption{Observed values of longitudinal two-point ($\xi$), triplet probability excess ($\rm PE_3$) and three-point ($\zeta$) correlations.}
\begin{tabular}{cccccccc}
\hline
\multicolumn{2}{c}{$r$}  & \multicolumn{2}{c}{$\xi$} & \multicolumn{2}{c}{$\rm PE_3$} & \multicolumn{2}{c}{$\zeta$} \\
(in pMpc) & (in \kms) & $N_{\rm HI}>10^{12.5}$cm$^{-2}$ & $N_{\rm HI}>10^{13.5}$cm$^{-2}$ & $N_{\rm HI}>10^{12.5}$cm$^{-2}$ & $N_{\rm HI}>10^{13.5}$cm$^{-2}$ & $N_{\rm HI}>10^{12.5}$cm$^{-2}$ & $N_{\rm HI}>10^{13.5}$cm$^{-2}$ \\
\hline
\hline
\\
$0.5-1$ & 36.3-72.6 & $1.20^{+0.18}_{-0.16}$ & $3.65^{+0.68}_{-0.60}$ & $4.77^{+3.87}_{-2.47}$ & $-1.00^{+32.11}_{-0.00}$ & $0.72^{+3.89}_{-2.49}$ & $-12.81^{+32.14}_{-1.28}$ \\
\\
$1-2$ & 72.6-145.3 & $1.65^{+0.13}_{-0.13}$ & $4.50^{+0.49}_{-0.45}$ & $8.76^{+1.96}_{-1.65}$ & $44.33^{+19.34}_{-14.09}$ & $4.76^{+1.98}_{-1.67}$ & $34.18^{+19.36}_{-14.12}$ \\
\\
$2-4$ & 145.3-290.6 & $0.70^{+0.07}_{-0.07}$ & $1.14^{+0.22}_{-0.20}$ & $1.98^{+0.54}_{-0.46}$ & $7.76^{+4.32}_{-3.03}$ & $0.41^{+0.56}_{-0.48}$ & $4.88^{+4.34}_{-3.06}$ \\
\\
$4-8$ & 290.6-581.1 & $0.18^{+0.04}_{-0.04}$ & $0.59^{+0.13}_{-0.12}$ & $0.37^{+0.18}_{-0.16}$ & $2.66^{+1.32}_{-1.00}$ & $0.01^{+0.20}_{-0.18}$ & $1.31^{+1.35}_{-1.04}$ \\
\hline
\end{tabular}\label{Corr_Table}
\end{table*}

\subsection{Dependence of clustering on $b$-parameter}

\begin{figure}
 \includegraphics[viewport=0 0 305 270,width=8.0cm, clip=true]{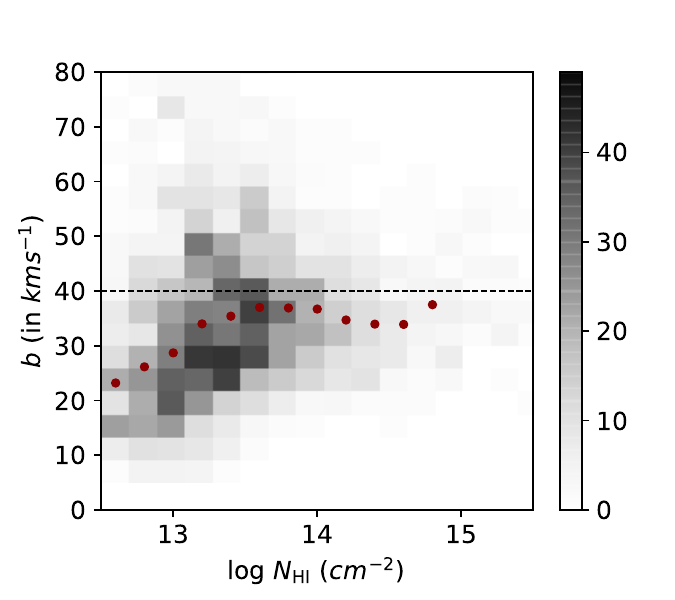}%
	\caption{2D histogram of $N_{\rm HI}$ vs. $b$ distribution. The horizontal dashed line at $b = 40$ \kms demarcates the high-$b$ and low-$b$ sub-samples. Dots provides the median $b$ value in each \NHI\ bin.
	A correlation is seen between \NHI\ and $b$ for small values of $b$ and median \NHI\ is flat at large-$b$ values. }
\label{NHI_b}
\end{figure}

We have seen that \lya\ clustering depends on \NHI\ and 
%{ absorbers} with different $N_{\rm HI}$ thresholds cluster differently and the clustering 
is stronger for higher $N_{\rm HI}$ { absorbers}. Here, we explore the dependence of \lya\ clustering on $b$-parameter. Purely based on the existence of $N_{\rm HI}$ $vs.$ $b$ correlations we expect the { absorbers} with high $b$-values to cluster more strongly. Additionally,
this exercise is motivated by the finding of \citet{Wakker2015}, that the BLAs tend to have low impact parameter (i.e $\rho \le 400$ pkpc) with respect to the filament axis compare to the narrow \lya\ absorbers that are found up to $\rho\sim 3$ pMpc. \citet{Tejos2016} have also found the number of BLA absorption associated with the filaments are $\sim6$ times in excess of random expectations. These absorbers may trace the warm ionized gas in the intracluster filaments.
%

%We divide the total sample of \lya\ { absorbers} into two sub-samples: one having $b<40$ \kms\ and one having 
%$b\ge40$ \kms. Remember for pure thermal broadening $b\sim40$ \kms corresponds to a temperature at which \HI\ can be collisionally ionized. Thus the above division will roughly separate the \lya\ into those in photoionization equilibrium and the ones where collisional ionization could be important. 

{Fig.~\ref{NHI_b} shows the scatter plot for the $N_{\rm HI}$ $vs.$ $b$ distribution in our sample.  In our  sample,
31.9\% of the \lya\ components have $b\ge40$ \kms. It is also evident from this figure that the distribution of $N_{\rm HI}$ seems different for high-$b$ and low-$b$ sub-samples. 
High-$b$ sub-sample seems to have less number of high $N_{\rm HI}$ (as also seen in Fig.~\ref{NHI_distribution}) . This is similar to the finding of \citet{Lehner2007} based on smaller number of sightlines. While one expects bias against detecting low \NHI\ high-$b$ absorbers when the SNR is low, lack of high-$b$ high \NHI\ absorption can be either physical or systematic bias introduced by the multiple component Voigt profile fits that tend to fit saturated lines with more narrow components.}

\begin{figure*}
 \includegraphics[viewport=0 8 310 290,width=8cm, clip=true]{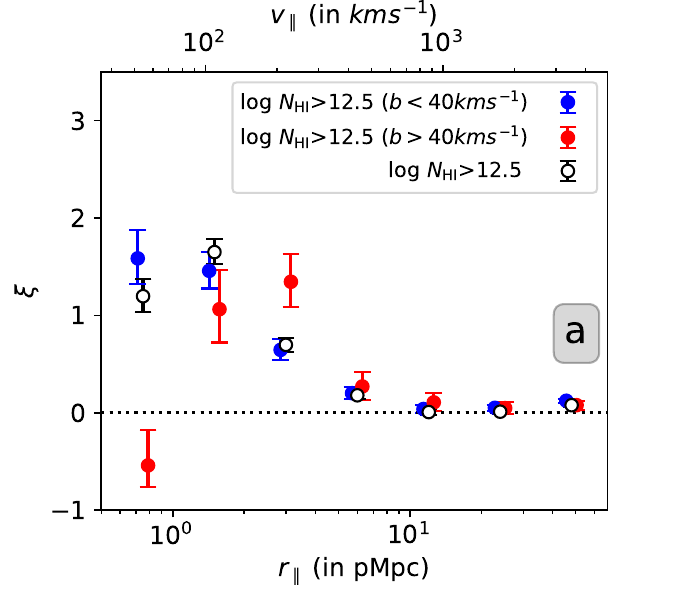}%
	\includegraphics[viewport=0 8 310 290,width=8cm, clip=true]{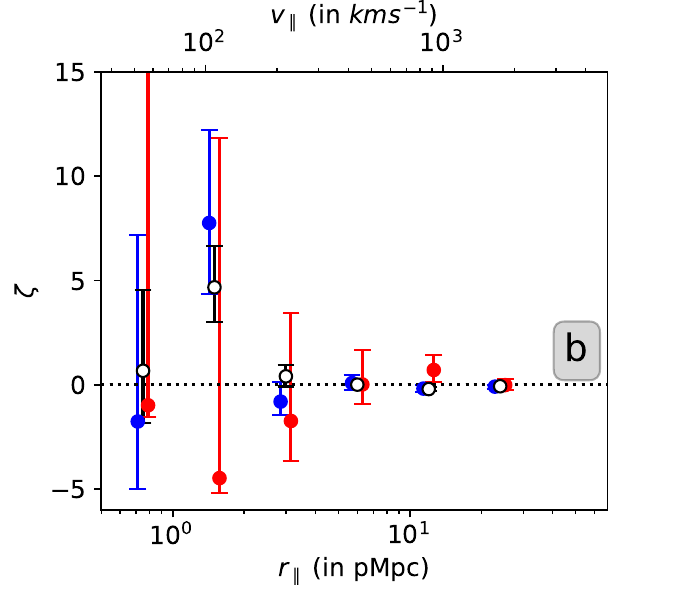}%

		\caption{Two-point (panel a) and three-point (panel b) correlations of \lya\ absorbers as a function of longitudinal scale for full sample as well as high-$b$ ($b>40$ \kms) and low-$b$ ($b<$ 40 \kms) sub-samples. The errors represent { one-sided poissonian uncertainty corresponding to $\pm 1\sigma$} about the mean value. The results for the low-$b$ sample roughly follow the that of the whole sample. { In the case of high-$b$ sub-sample, we do not find any triplets for $r_{\parallel}<4$ pMpc, this coupled with the low random probability to find high-$b$ triplets at these scale leads to large errors in $\zeta$.} 
		%It is evident from these figures that high-$b$ absorbers do not show strong clustering while low-$b$ absorbers show detectable clustering. We also notice that the clustering of low-$b$ absorbers are weaker than that of the whole sample. This is probably due to the fact that the cross-correlation of high-$b$ absorbers with low-$b$ has similar strength as the auto-correlation between the low-$b$ absorbers (see panel a) .
		}
\label{Corr_b}
\end{figure*}

In panel (g) of Fig.~\ref{Corr2}, we plot the two-point correlation function measured at $r_\parallel$ = 1-2 pMpc (equal arm configuration) as a function of different $b$-parameter thresholds. For generating the random distributions, as discussed in section~\ref{sec_fnh}, we compute the intrinsic \NHI\ distribution separately using appropriate median $b$ values for each sub-sample. Initially, we see a nearly constant two-point correlation with increase in $b$-parameter values. However, when we consider high-$b$ systems (i.e $b>40$ \kms) we notice that the two-point correlation function decreases. From panel (h) of Fig.~\ref{Corr2}, we find that the three-point correlation function also shows similar trend %at a lower $b$ value of 30 \kms. 
%However, the three-point correlation found for both $b>30$\kms\  and $b>40$\kms\ are consistent within 1$\sigma$ errorbar. Their is
with a sharp decline in the amplitude  for $b>40$ \kms\ case ({ with large errorbar}). The same trend is also shown by $Q$ that we plot in the panel (i) of Fig.~\ref{Corr2}. {  We do not find any triplets with all the components having $b>40$ \kms\ in $r_{\parallel}=1-2$ pMpc bin which results in a large negative mean three-point correlation.
%The fact that we find low two-point and three-point correlations for $b>40$ \kms\ \lya\ absorbers points to the bias against detecting high-$b$ absorbers.}  
%
%{ 
This could be real or artefact of some bias in the Voigt profile decomposition, in particular at small scales.
For example, presence of a broad absorber can conceal other broad components from being detected within the scales considered here (in particular when the SNR is not high), thereby lowering the two-point and three-point correlations.}
%

%This trend can be understood based on the scatter plot given in Fig.~\ref{NHI_b}. There is a strong correlation between \NHI\ and $b$ parameter for $b<40$ \kms. However, high \NHI\ components are missing when we limit ourselves to high-$b$ values. Therefore, the trend seen when we use low-$b$ cutoffs could be a mere reflection of increasing correlation with \NHI\ among low-$b$ absorbers.

To see whether this is a scale dependent result, in Fig.~\ref{Corr_b} we plot the longitudinal two- and three-point correlations as a function of scale separately for high-$b$ and low-$b$ sub-samples for $N_{\rm HI} > 10^{12.5}$ cm$^{-2}$. The random distribution of { absorbers} for each of these cases is drawn separately using the $N_{\rm HI}$ distributions shown in the bottom panel in Fig.~\ref{NHI_distribution}. 
As can be seen from this figure, at $r_\parallel<2$ pMpc, the two-point correlation between the low-$b$ absorber (while consistent with the full sample) is stronger than that of high-$b$ absorbers. Over the same scale the high-$b$ absorbers show largely negative three-point correlation. This confirms the lack of triplets with large-$b$ values at small scales. This is consistent with what we have seen in Fig~\ref{Corr2}.
%
%{\color{blue} The high-$b$ sub-sample has similar two-point correlation as the total sample at a scale of 1-2pMpc.
{ However for the scales of 2-4 pMpc, we see a stronger two-point correlation for the high-$b$ sub-sample.   
%Interestingly, this is also the scale at which we see a detectable three-point correlation in the high-$b$ sample (which is otherwise negative or zero in rest of the bins) . 
In the case of three-point correlation function, contribution comes from a single triplet of high-$b$ absorbers ($b=$ 42.4 \kms, 144.3 \kms\ and 51.2 \kms) seen along a single sight line (PKS~0405-123) at the redshift of 0.0251. For all other high $r_\parallel$ bins the low-$b$ and high-$b$ sub-samples have similar albeit with low values for correlation function.}
The trend seen for the two-point correlation is consistent with the { suppression} effects being severe in the case of high-$b$ absorbers for $r_\parallel<2$ pMpc and high-$b$ absorbers having higher $\xi$ at 2-4 pMpc. In the case of three-point, due to large errors we do not find any difference between high-$b$ absorbers and full sample for $r_\parallel > 4$~pMpc. 

\subsection{Effect of presence of metal ions on correlation}\label{Metal_Correlation}
\begin{figure*}
 { \large \qquad $z<0.16$ \quad\qquad\qquad\qquad\qquad\qquad\quad $0.1<z<0.48$ \qquad\qquad\qquad\qquad\qquad\qquad $z<0.48$}
\centering
%\vspace{-0.2cm}
    \includegraphics[viewport=0 40 295 290,width=5.9cm, clip=true]{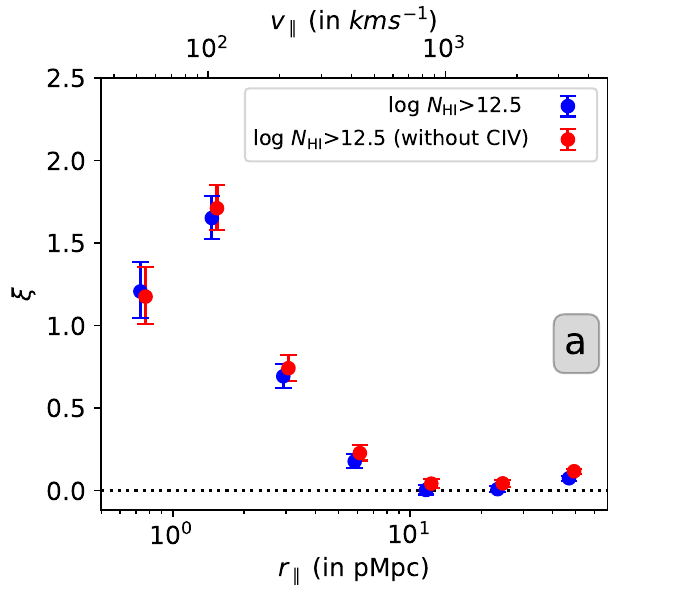}%
	\includegraphics[viewport=0 40 295 320,width=5.9cm, clip=true]{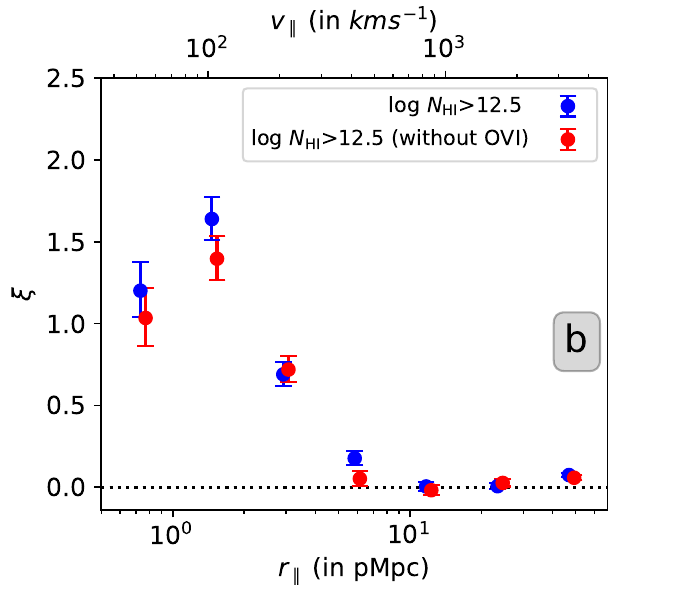}%
	\includegraphics[viewport=0 40 295 320,width=5.9cm, clip=true]{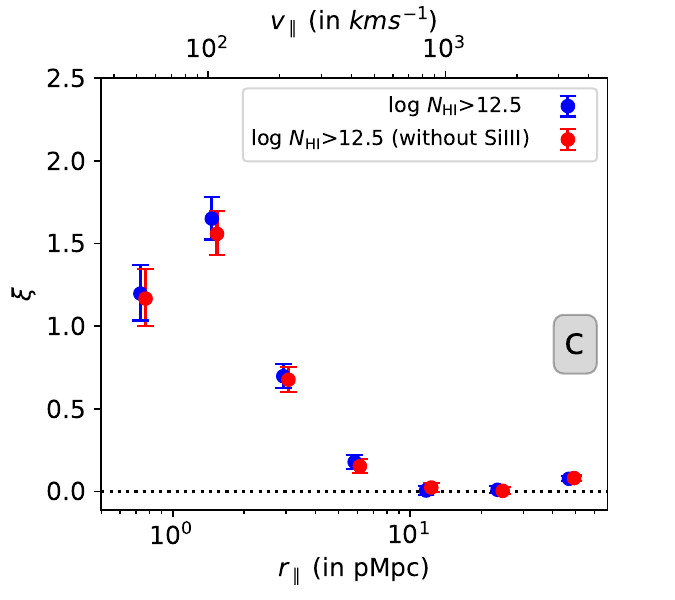}%
	%\vspace{0.2cm}

	\includegraphics[viewport=0 40 295 260,width=5.9cm, clip=true]{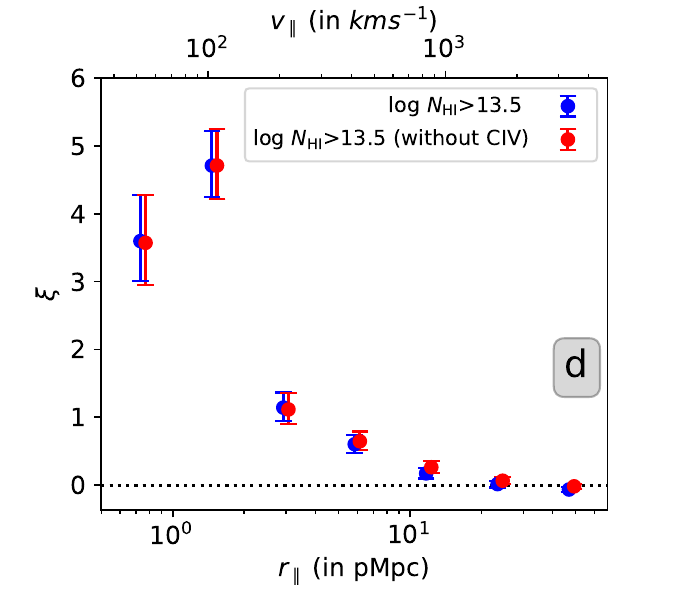}%
	\includegraphics[viewport=0 40 295 260,width=5.9cm, clip=true]{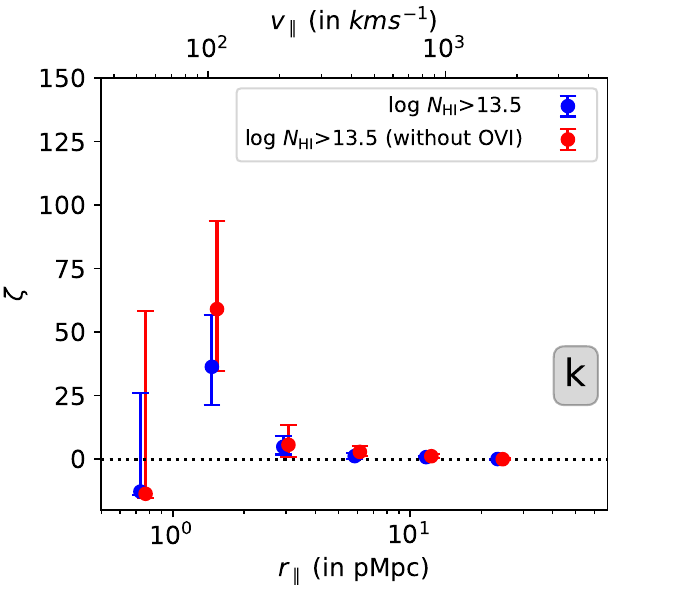}%
	\includegraphics[viewport=0 40 295 260,width=5.9cm, clip=true]{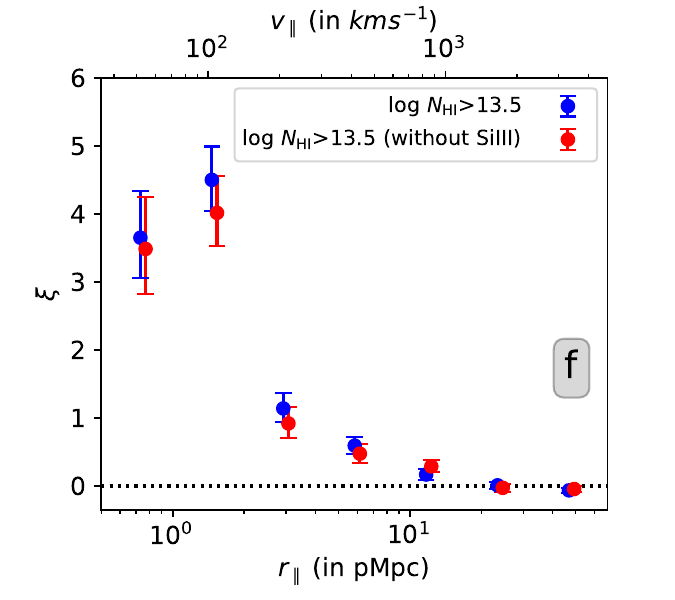}%
	%\vspace{0.2cm}
	%\includegraphics[viewport=0 50 350 330,width=5.8cm, clip=true]{Corr_2_CIV_N13_5.pdf}%
	%\includegraphics[viewport=0 50 350 330,width=5.8cm, clip=true]{Corr_2_OVI_N13_5.pdf}%
	%\includegraphics[viewport=0 50 350 330,width=5.8cm, clip=true]{Corr_2_SiIII_N13_5.pdf}%
	
	\includegraphics[viewport=0 40 295 260,width=5.9cm, clip=true]{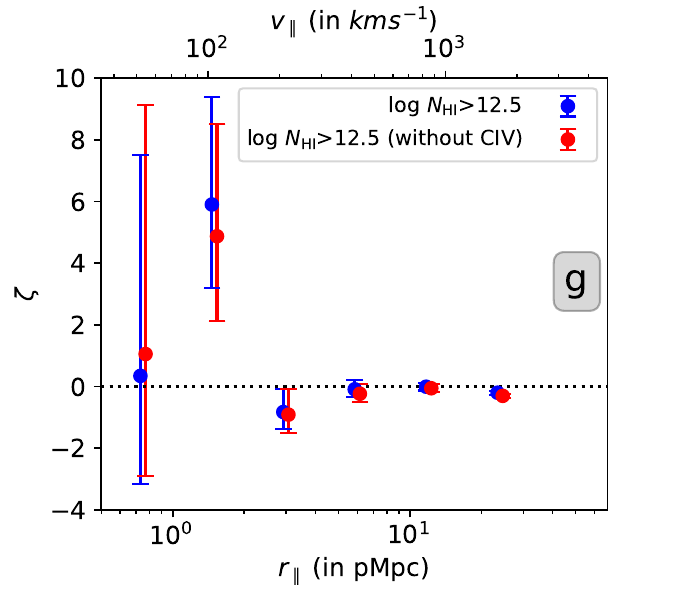}%
	\includegraphics[viewport=0 40 295 260,width=5.9cm, clip=true]{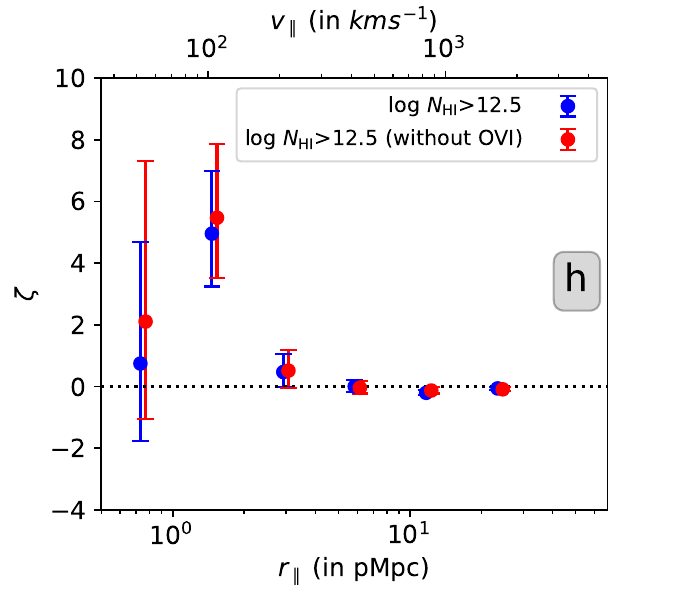}%
	\includegraphics[viewport=0 40 295 260,width=5.9cm, clip=true]{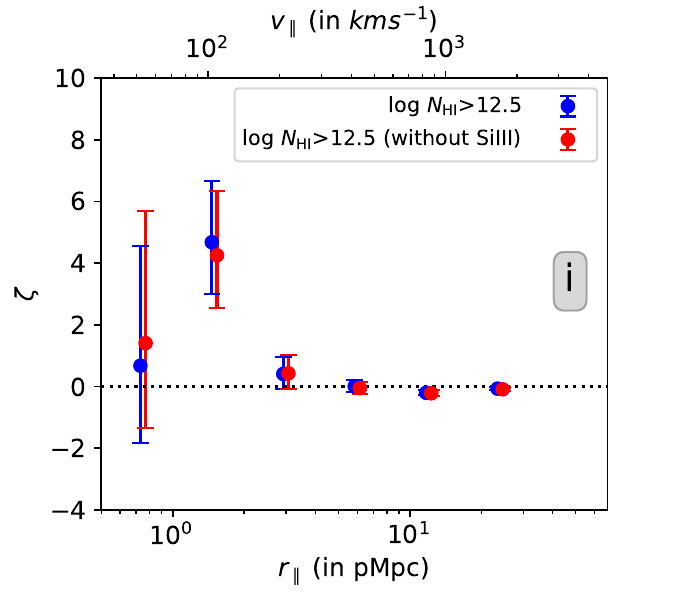}%
	%\vspace{0.2cm}
	
	%\includegraphics[viewport=0 50 350 330,width=5.8cm, clip=true]{Corr_2_CIV_N13_5.pdf}%
	%\includegraphics[viewport=0 50 350 330,width=5.8cm, clip=true]{Corr_2_OVI_N13_5.pdf}%
	%\includegraphics[viewport=0 50 350 330,width=5.8cm, clip=true]{Corr_2_SiIII_N13_5.pdf}%
	%\vspace{0.2cm}
	\includegraphics[viewport=0 0 295 260,width=5.9cm, clip=true]{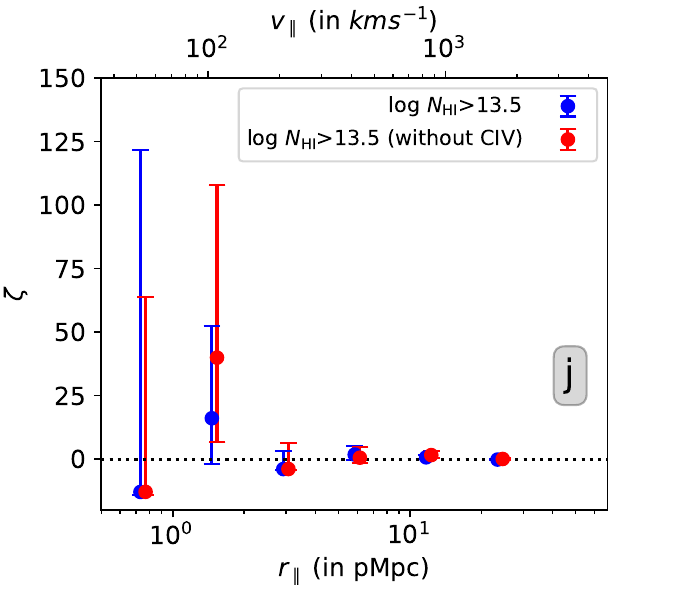}%
	\includegraphics[viewport=0 0 295 260,width=5.9cm, clip=true]{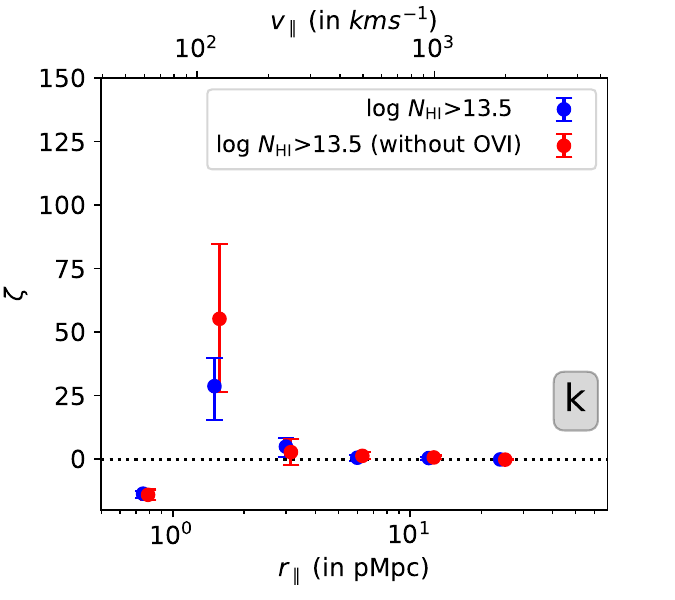}%
	\includegraphics[viewport=0 0 295 260,width=5.9cm, clip=true]{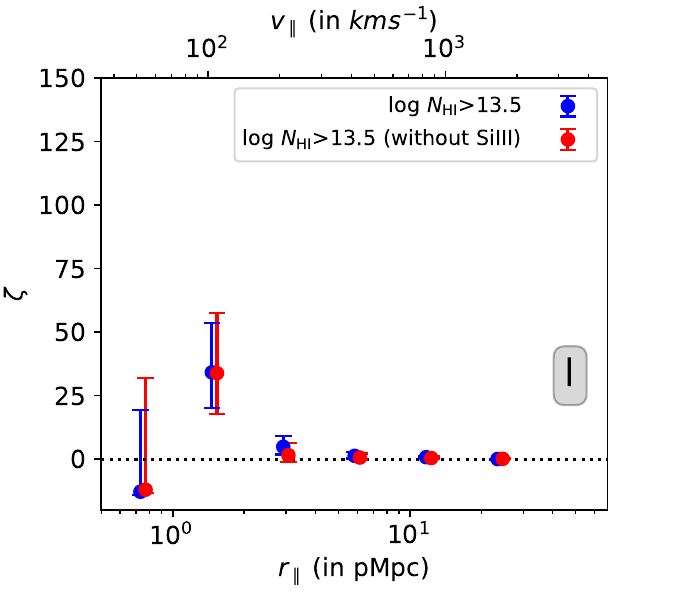}%

	\caption{The top two rows show two-point correlation of \lya\ absorbers with (blue) or without (red) associated metal lines such as \CIV\ (left panels), \OVI\ (middle panels) and \SiIII\ (right panels) as a function of longitudinal scale for $N_{\rm HI}>10^{12.5}$cm$^{-2}$ (top row) and $N_{\rm HI}>10^{12.5}$cm$^{-2}$ (second most top row) . The bottom two panels show longitudinal three-point correlation in the same fashion. The errors represent { one-sided poissonian uncertainty corresponding to $\pm 1\sigma$} around the mean value.}
\label{Corr_metals}
\end{figure*}

%\Anand{Note Danforth concludes that "This suggests that most of the radial-velocity clustering in the IGM can be attributed to strong, metal-bearing systems in the CGM, again consistent with the picture of strong, metal- enriched absorption being associated with galaxy halos".}

One important question which arises  is that whether the correlations we detect originate from the IGM or are dominated by a small population of absorbers 
originating from CGM of intervening galaxies.
\citet{danforth2016} have shown that the metal bearing \lya\ systems (based on the presence of \OVI) show stronger two-point correlation than the non-metal bearing systems (see their figure 18). Based on this they concluded that most of the radial velocity clustering of the \lya\ systems can be attributed to metal bearing systems originating from the CGM of intervening galaxies.
Here, we ask a slightly different question. { We would like to know whether the presence of different metal ion species influence the observed clustering properties of \lya\ absorbers.}
%are in any way related to the associated metal absorption. 

We base our study on \CIV, \OVI\ and \SiIII\ metal line transitions. Considering the wavelength range covered by the HST-COS medium resolution spectrum, 
%the \CIV\ 
%($\lambda \sim$1215.67\AA) 
%transition is covered typically for $z<0.48$. Associated 
%\CIV($\lambda\lambda$1548.20, 1550.77\AA ) can be covered only for $z<0.16$. 
redshift ranges over which \lya, \OVI\ and \SiIII\ associated with the \lya\ can be detectable are $z<0.16$, $0.1<z<0.48$  and $z<0.48$ respectively. 
So, for checking the dependence of \lya\ clustering on the presence of \CIV, %absorption, we consider a common redshift range of $z<0.16$. Similarly for
\OVI\ and \SiIII\, we consider redshift ranges of $z<0.16$, $0.1<z<0.48$ and $z<0.48$ respectively. We consider a metal line transition having redshift within the median $b$-parameter ($\sim 34$ \kms) of the redshift of \lya\ absorbers to be associated with it. %\PG{Why do you chose 34 km/s scale?}
{ We only consider components with metal { ion} line absorption having rest equivalent width above 30 m\AA\  \citep[as done by,][]{danforth2016}.}

Firstly, we consider \lya\ absorbers having $N_{\rm HI}>10^{12.5}$cm$^{-2}$. For $z<0.16$, 5.7\% of such \lya\ absorbers show detectable \CIV\ absorption. %\Anand{it will be good to define how we associated a \lya\ to \CIV\ absorption. You might have used the matching in redshift by some amont etc.}
For $0.1<z<0.48$, 19.8\% of the \lya\ absorbers show detectable \OVI\ absorption. For $z<0.48$,  5.9\% of the \lya\ absorbers are associated with \SiIII\ absorption. When we consider \lya\ absorbers having $N_{\rm HI}>10^{13.5}$cm$^{-2}$, these percentages increase to 16.8\% for \CIV, 33.0\% for \OVI\ and 16.1\% for \SiIII. Note that { for calculating these fractions,} we have considered all metal { ion} line detections without applying any column density cut-off based on detection sensitivity.

%\Anand{Shall we put a cut-off metal line detection to be rest equivalent width greater than 30 m\AA.}
%\PG{Above 13.5 Ly-alpha is complete for HST-COS sample but what about completeness of CIV, OVI and SiIII. The percentage you quote may be the lower limit on these species? Metal content could be higher than that as some lines may not be detected due to SNR. You may want to add this as footnote.} %\Anand{We should say some thing about their origin based on PI model works. I will do that in the next round.}

In top two rows of Fig.~\ref{Corr_metals}, we plot the two-point correlation of ``all \lya'' absorbers and those ``without associated metal { ion species}'' absorption for the three identified metal ion species and two \NHI\ cut-offs.
%We compare two-point and three-point correlation functions obtained for all the \lya\ absorbers with those obtained for \lya\ absorbers without metals.  
Note the ``all \lya" sample in each of these cases is different owing to different redshift ranges probed (redshift ranges are provided on top of each column in Fig.~\ref{Corr_metals}).
In the case of absorbers "without metal { ion species}'' we just remove only the Voigt profile components that have associated metal absorption.
%and do not remove the whole system.
%
As only a small fraction of \lya\ absorbers will be removed based on the presence of metal { ion species}, we naively expect their influence to be minimal. However, we still draw  appropriate random distributions of { absorbers} for each case separately while calculating $\xi$ and $\zeta$. 
%We fix a threshold of $N_{\rm HI}>10^{12.5}$cm$^{-2}$ for all the \lya\ absorbers. 
%
%In top two rows of Fig.~\ref{Corr_metals}, we plot the two-point correlation of ``all \lya'' absorbers and those ``without associated metal'' absorption for the three identified metal ion species and two \NHI\ cut-offs.
%
For $N_{\rm HI}>10^{12.5}~{\rm cm}^{-2}$, the $\xi$ measured for systems without \CIV, \OVI\ and Si~{\sc iii} are consistent with their respective ``all \lya'' samples.  This seem to be the case for $N_{\rm HI}>10^{13.5}~{\rm cm}^{-2}$  absorbers also.
%
%In the case of \OVI\ we do see reduction in $\xi$ by more than 3$\sigma$ \Anand{Check this and put correct value.} for the components without \OVI\ for 1-2 pMpc bin. However, results are consistent for other distance bins.
%Next we consider $N_{\rm HI}>10^{13.5}~{\rm cm}^{-2}$ case. 
%In the case of \CIV\ we find an indication of decrease in $\xi$ in $r_\parallel$ = 1-2 pMpc bin for without \CIV\ components. However, the difference is within 1.4$\sigma$ level. The measurements are consistent with one another in the remaining $r_\parallel$ bins.
%However,  we do not find any significant differences in the case of \OVI\ and \SiIII. 
Therefore, it appears that two-point correlation function we measure for the \lya\ absorption and its column density dependence  may not originate mainly from the metal line { bearing \lya} absorbers. 

In bottom two panels of Fig.~\ref{Corr_metals}, we plot the three-point correlations for two \NHI\ cut-offs. 
%As in the case of $\xi$,
The $\zeta$ does not show any significant difference between the full sample and the corresponding sample for \lya\ without \CIV, \OVI\ or \SiIII\ absorption. The three-point correlation measured for the \lya\ absorbers does not source primarily from a metal line detected components.  

\subsection{Redshift Evolution}
\begin{figure*}
	\includegraphics[viewport=0 5 300 290,width=6.cm, clip=true]{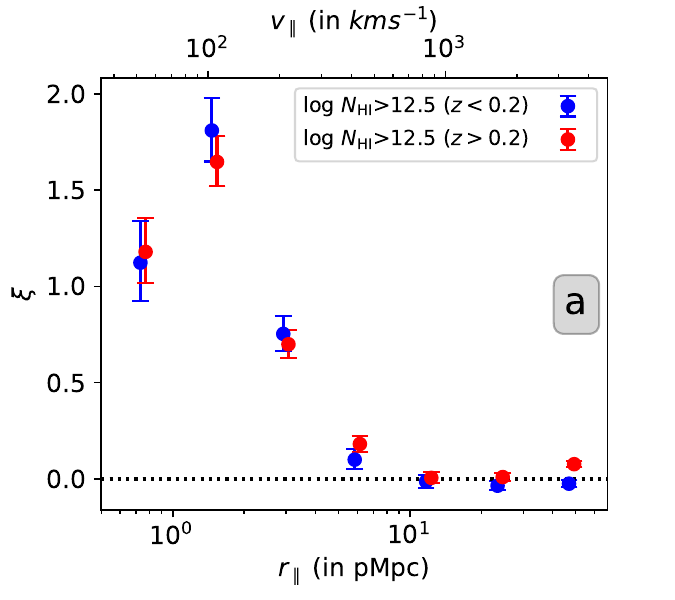}%
	\includegraphics[viewport=0 5 300 290,width=6.cm, clip=true]{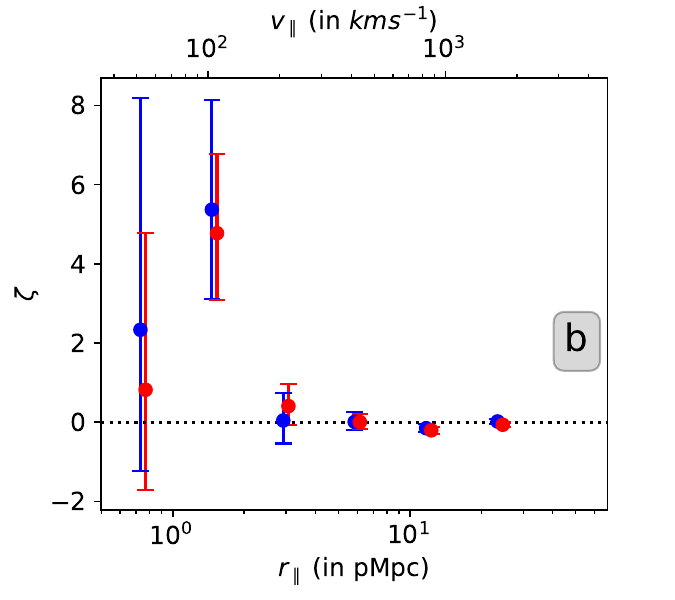}%
    \includegraphics[viewport=0 5 300 290,width=6.cm, clip=true]{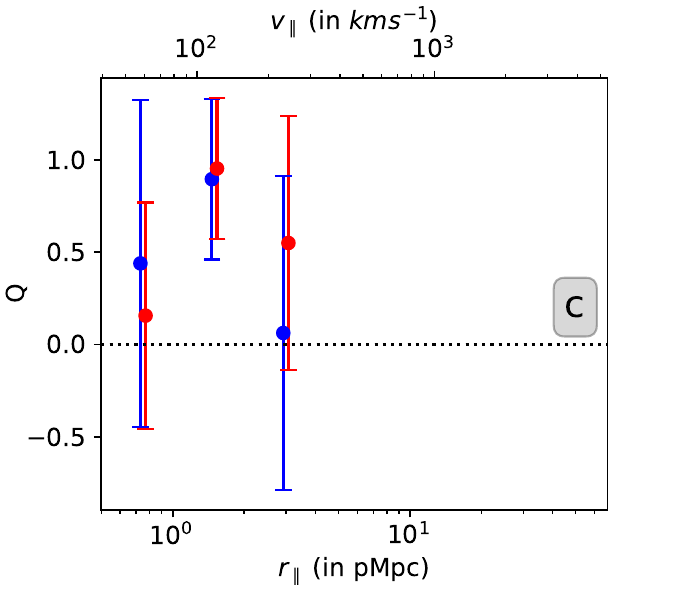}%
	\caption{Two-point, three-point and reduced three-point correlation (left to right) of \lya\ absorbers as a function of longitudinal scale for two different $z$-intervals. The errors represent { one-sided poissonian uncertainty corresponding to $\pm 1\sigma$} about the mean value of correlations.  }
\label{Corr_z}
\end{figure*}

In this section, we investigate the redshift evolution of the two- and three-point correlation %at $z\le0.45$. For this we consider 
by considering two redshift bins  $z<0.2$ and $0.2<z<0.48$ for $N_{\rm HI}>10^{12.5}~{\rm cm}^{-2}$. {  The choice of $z=0.2$ as a threshold between the two redshift bins is simply taken as the approximate midpoint of the  redshift range of our sample. }
The left and middle panels in Fig.~\ref{Corr_z} show the $\xi$ and $\zeta$ respectively as a function of distance { scale} for the two sub-samples. 
%It is evident that for $r_\parallel\le$ 3 pMpc the measured two-point correlation function for the low-$z$ sub-sample is systematically stronger compared to the high-$z$ sub-sample. 
%
%In the middle panel of Fig.~\ref{Corr_z} we show the results for the three-point correlation function. 
It is clear  from these two panels that { the measured values of} $\xi$ and $\zeta$ for the low-$z$ and
high-$z$ sub-samples are consistent with
each other
%those of the high-$z$ sub-sample 
within measurement uncertainties. 
%However, due to the small number of strongly clustered absorbers involved the error in the case of $\zeta$ measured for low-$z$ are larger.  
%
%As expected, 
We do not also find any difference in the Q profile for the two sub-samples (see right panel in Fig.~\ref{Corr_z}) .
%The measured Q values for the first three $r_\parallel$ bins are shown in the right panel of Fig.~\ref{Corr_z}. We do not find any difference as the associated errors are large. 
%
Thus over the redshift range considered here we do not find any evolution in the amplitude of the two- and three-point correlation function. 
%It will be good to increase the number of absorbers to confirm this trend at a higher statistical significance.  It particular it will be good to confirm the lack of evolution of Q as a function of $z$.
%
%\Anand{It will be good to check how much is the evolution one sees in the case of galaxy clustering over the same redshift range.Also comment on whether the sightline selection has any influence on the redshift evolution. Also respond to the bias discussed in section 2.5 of \citet{danforth2016}.} {Since we only deal with unbiased sightlines, do we need to discuss about bias here?}

%{In order to investigate the redshift evolution of clustering in these low-$z$ \lya\ { absorbers}, we divide the sample into two $z$ bins: $z<0.2$ and $0.2<z<0.48$. The longitudinal two-point and three-point correlation for $N_{\rm HI}>10^{12.5}$cm$^{-2}$ { absorbers} is calculated separately for these two samples similar to how it was generated for the complete sample. The random distribution of \lya\ { absorbers}, with respect to which the correlation is calculated, is generated separately for the two samples depending on the $N_{\rm HI}$ distribution of the { absorbers} in the two $z$ bins. We plot the two-point, three-point and reduced three-point correlation for these two $z$ bins in Fig.~\ref{Corr_z}. }
%\Anand{write some text here giving details.}

\section{Connection to galaxies}
\label{con_gal}
\begin{comment}
\begin{figure*}
\qquad Impact parameter < Virial Radius \qquad\qquad\qquad\qquad\qquad\quad Virial Radius < Impact parameter < 3Mpc
 \includegraphics[viewport=0 0 360 305,width=7cm, clip=true]{CDF_Impact_Factor_0_Rvir.pdf}%
	\includegraphics[viewport=0 0 360 305,width=7cm, clip=true]{CDF_Impact_Factor_Rvir_3Mpc.pdf}%

%	 \includegraphics[viewport=0 0 350 330,width=6cm, clip=true]{Corr_2_vs_r_feedback_dependence.pdf}%
%	\includegraphics[viewport=0 0 350 330,width=6cm, clip=true]{Corr_3_vs_r_feedback_dependence.pdf}%
%	\includegraphics[viewport=0 0 350 330,width=6cm, clip=true]{Q_vs_r_feedback_dependence.pdf}%

	\caption{Cumulative distribution function of galaxy impact parameter.}
\label{IF}
\end{figure*}
\end{comment}

\begin{figure*}

 \includegraphics[viewport=5 32 350 245,width=6.8cm, clip=true]{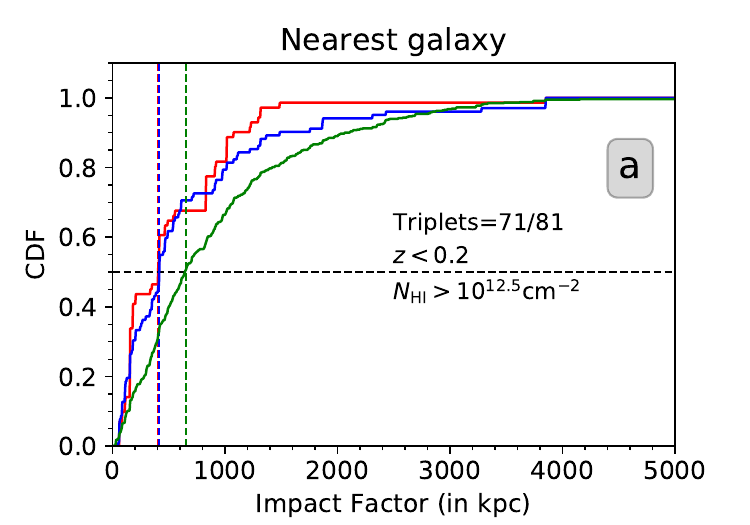}%
 \includegraphics[viewport=50 32 350 245,width=5.9cm, clip=true]{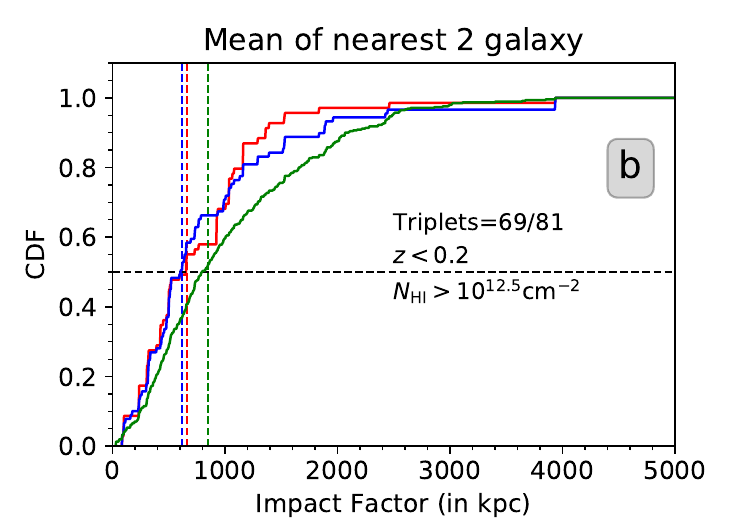}%
 \includegraphics[viewport=50 32 350 245,width=5.9cm, clip=true]{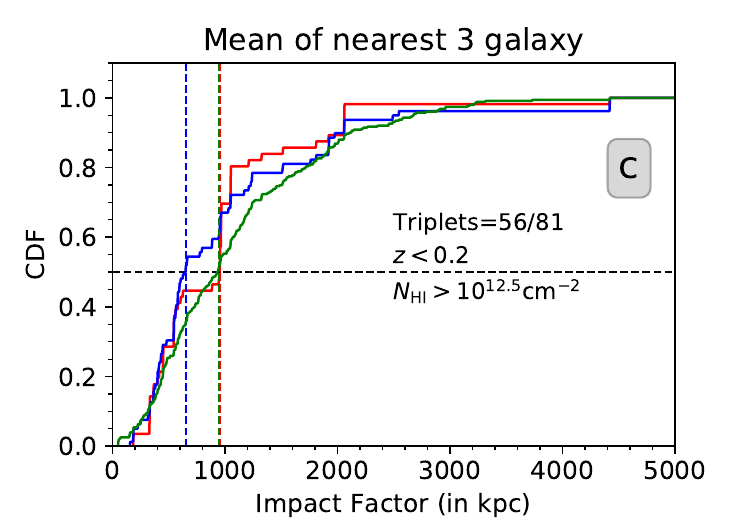}%
 
  \includegraphics[viewport=5 0 350 225,width=6.8cm, clip=true]{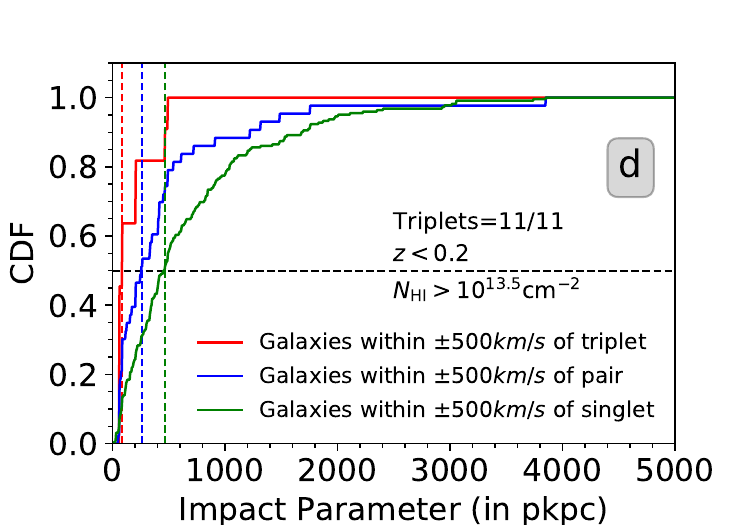}%
 \includegraphics[viewport=50 0 350 225,width=5.9cm, clip=true]{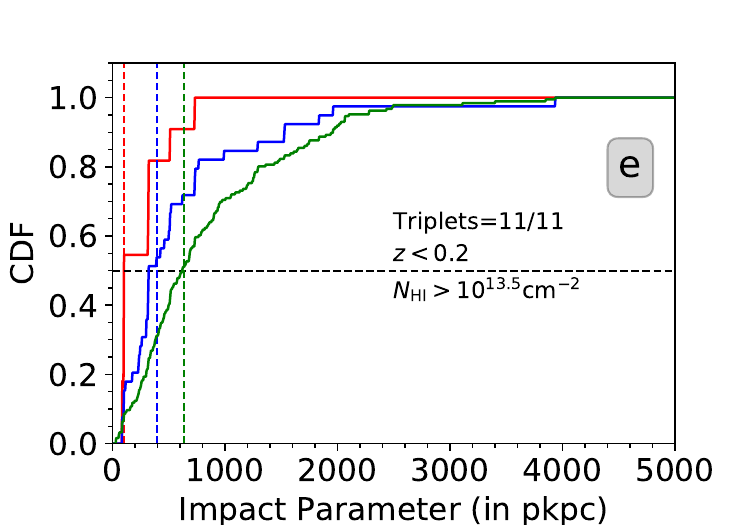}%
 \includegraphics[viewport=50 0 350 226,width=5.9cm, clip=true]{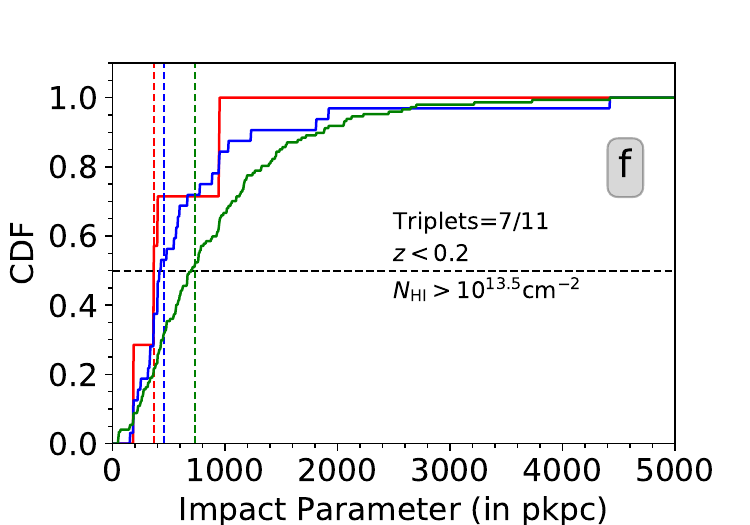}%

	\caption{Cumulative distribution function (CDF) of impact parameter of galaxies associated with isolated or "singlet", "pair" (excluding the ones that are part of the triplets) and "triplet" \lya\ absorbers. Left, middle and right columns show  results of impact parameter for the nearest, average of two nearest and three nearest galaxies respectively. Number of the identified triplets have associated galaxy (or galaxies) is also provided in each panel. Results for log~\NHI $>$12.5 and 13.5 are given in the top and bottom rows respectively. The vertical dashed lines give the median value of impact parameter for each sub-sample.}
\label{IF}
\end{figure*}

%\begin{figure*}
%
%  \includegraphics[viewport=13 20 340 250,width=6cm, clip=true]{PDF_L_1000kms_N12_5_near1.pdf}%
% \includegraphics[viewport=13 20 340 250,width=6cm, clip=true]{PDF_L_1000kms_N12_5_near2.pdf}%
% \includegraphics[viewport=13 20 340 250,width=6cm, clip=true]{PDF_L_1000kms_N12_5_near3.pdf}%
 
% \includegraphics[viewport=13 20 340 250,width=6cm, clip=true]{PDF_L_1000kms_N13_5_near1.pdf}%
% \includegraphics[viewport=13 20 340 250,width=6cm, clip=true]{PDF_L_1000kms_N13_5_near2.pdf}%
%\includegraphics[viewport=13 20 340 250,width=6cm, clip=true]{PDF_L_1000kms_N13_5_near3.pdf}%

%	 \includegraphics[viewport=0 0 350 330,width=6cm, clip=true]{Corr_2_vs_r_feedback_dependence.pdf}%
%	\includegraphics[viewport=0 0 350 330,width=6cm, clip=true]{Corr_3_vs_r_feedback_dependence.pdf}%
%	\includegraphics[viewport=0 0 350 330,width=6cm, clip=true]{Q_vs_r_feedback_dependence.pdf}%
%	\caption{Probability distribution function of galaxy Luminosity.}
%\label{L}
%\end{figure*}

\begin{table*}
    \caption{Galaxies nearby to identified \lya\ absorber triplets.   % \PG{Why are some Galaxies's velocity and impact parameter in red? Do you want to stress about these galaxies?}
    }
    \begin{threeparttable}
    \begin{tabular}{ccccccc}
    \hline
    QSO sightlines\tnote{1}  & $z$ (\lya\ triplet) & log $N_{\rm HI}$ of & $b$-parameter of &  $v_{\parallel}$ (Nearest Galaxy-\lya\ triplet )\tnote{2} & Impact parameter of Galaxy  \\
       &  & triplet system  & triplet systems (in \kms)   &  (in \kms) & from sightlines\tnote{2} (in pKpc)\\
    \hline
    \hline
    
  %  pg0003$^{\rm B,M}$ & 0.0911 & 12.5 & 0.5-1 & - & - \\
  \multicolumn{7}{c}{Equal arm length configurations: $r_{\parallel}=0.5-1$ pMpc}\\
    PG1116+215$^{\rm B,M}$ & 0.1656 & 12.63, 13.39, 13.06 & 15.8, 30.7, 44.9 & +98.3, -197.7, +312.0 & 156.0, 317.0, 531.4 \\
    PG1222+216$^{\rm B}$ & 0.1446 & 13.41, 13.32, 13.3 & 34.2, 18.6, 47.2 & {\color{red}  -956.9, -907.1, -941.2} & {\color{red}  1189.0, 1305.3, 1316.8} \\
    PKS2155-304$^{\rm B}$ & 0.1057 & 13.98, 13.33, 13.28 & 47.1, 21.6, 43.2 & -265.1, +36.1, -34.5, +33.4 & 924.0, 952.0, 1025.0, 1094.0 \\
    
    \\
     \multicolumn{7}{c}{Equal arm length configurations: $r_{\parallel}=1-2$ pMpc}\\
 %	3c66a & 0.1519 & 12.5 & 1-2 & - & - \\
 	H1821+643$^{\rm B}$ & 0.1217 & 14.21, 13.43, 13.14 & 38.1, 53.0, 38.1 & -15.8, +273.1, -328.7 & 156.3, 1160.5, 1550.4 \\
 	H1821+643$^{\rm B, M}$ & 0.1701 & 13.86, 13.68, 13.36 & 35.3, 58.0, 28.4 & +127.7, +209.7, -369.7 & 415.4, 1122.2, 1118.7 \\
 	H1821+643 & 0.1895 & 12.6, 12.72, 12.55 & 28.1, 26.1, 16.1 & -140.0, -155.1, -84.5  & 1019.8, 1054.5, 1084.3  \\
 	%p1103 & 0.0594 & 12.5 & 1-2 & - & - \\
 	%p1103 & 0.0597 & 12.5 & 1-2 & - & - \\
 	PG0953+414$^{\rm M}$ & 0.1423 & 12.73, 13.56, 13.48 & 15.3, 26.5, 30.9 & +237.7, +78.0, +51.2 & 405.0, 452.1, 506.7 \\
 	PG0953+414$^{\rm B,M}$ & 0.1426 & 13.56, 13.48, 13.2 & 26.5, 30.9, 52.7 & +155.4, -4.2, -31.0 & 405.0, 452.1, 506.7 \\
 	PG1048+342$^{\rm B,M}$ & 0.0061 & 14.8, 13.79, 13.99 & 42.7, 33.1, 86.9 & -6.0, -167.0 & 60.2, 149.2 \\
 	PG1048+342$^{\rm B,M}$ & 0.0057 & 14.07, 14.8, 13.79 & 30.4, 42.7, 33.1 & +84.1, -77.0 & 60.2, 149.2 \\
 	%pg1049$^{\rm B}$ & 0.1636 & 12.5 & 1-2 & - & - \\
 	PG1116+215$^{\rm B}$ & 0.1658 & 12.63, 13.06, 14.28 & 15.8, 44.9, 32.7 & +37.8 -258.1  +251.4 & 156.0,  317.0,  531.4 \\
 	%phl2525$^{\rm B}$ & 0.1696 & 12.5 & 1-2 & - & - \\
 	PKS0405-123$^{\rm B}$ & 0.0247 & 12.82, 12.56, 13.18 & 42.4, 13.4, 144.3 & {\color{red}  +1174.6, +1423.5, +1420.5} & {\color{red}  673.0,  799.0, 1105.0} \\
 	PKS0405-123$^{\rm B}$ & 0.1527 & 13.44, 12.54, 13.79 & 25.8, 13.5, 47.8 & 132.0,  223.0, -445.8 & 181.0, 2148.0, 3857.0 \\
	%pks0552$^{\rm B}$ & 0.1923 & 12.5 & 1-2 & - & - \\
 	%q0045$^{\rm B}$ & 0.1216 & 12.5 & 1-2 & - & - \\
 	Q1230+0115$^{\rm B,M}$ & 0.0948 & 12.99, 13.31, 14.33 & 69.7, 30.2, 46.3 & +46.8, -16.1, +16.7  & 113.4, 792.3, 917.2 \\
 	%rbs542$^{\rm M}$ & 0.0040 & 12.5 & 1-2 & - & - \\
 	%rbs542$^{\rm B}$ & 0.0636 & 12.5 & 1-2 & - & -\\
 	RXJ0439.6-5311$^{\rm B}$ & 0.1772 & 13.53, 13.6, 13.87 & 100.0, 19.7, 29.9 & -65.8, -183.0, +127.9 & 467.2, 562.5, 1848.2 \\
 	TON1187$^{\rm B}$ & 0.0354 & 13.38, 13.98, 13.59 & 18.9, 28.2, 56.4 & {\color{red} +1398.9, +1149.7, +1346.8} & {\color{red}  278.1, 276.3, 284.1} \\
 	
 	\\
 	 \multicolumn{7}{c}{Equal arm length configurations: $r_{\parallel}=2-4$ pMpc}\\
    H1821+643$^{\rm B}$ & 0.1899 & 12.6, 12.55, 12.73 & 28.1, 16.1, 41.6  & -235.5, -250.6, -180.0 & 1019.8,1054.5,1084.3 \\
     H1821+643$^{\rm B}$ & 0.1901 & 12.55, 12.61, 12.73 & 16.1, 22.4, 41.6  & +481.7,-287.1,-302.2 & 833.4,1019.8,1054.5 \\
    H1821+643$^{\rm B}$ & 0.1901 & 12.6, 12.61, 12.73 & 28.1, 22.4, 41.6  & +481.7,-287.1,-302.2 & 833.4,1019.8,1054.5 \\
    H1821+643$^{\rm B}$ & 0.1908 & 12.55, 12.73, 12.57 & 16.1, 41.6, 15.9 & +305.1, -463.3, -478.4 & 833.4,1019.8,1054.5 \\
    H1821+643$^{\rm B}$ & 0.1908 & 12.55, 12.73, 12.67 & 16.1, 41.6, 31.5, & +305.1, -463.3, -478.4 & 833.4,1019.8,1054.5 \\
    H1821+643$^{\rm B}$ & 0.1908 & 12.61, 12.73, 12.67 & 22.4, 41.6, 31.5 & +305.1, -463.3, -478.4 & 833.4,1019.8,1054.5 \\
    H1821+643$^{\rm B}$ & 0.1908 & 12.61, 12.73, 12.57 & 22.4, 41.6, 15.9 & +305.1, -463.3, -478.4 & 833.4,1019.8,1054.5 \\
    PG0953+414$^{\rm B}$ & 0.1914 & 12.91, 13.34, 13.27 & 25.2, 40.5, 55.3 & 0.0, -136.0 & 1232.6, 3692.8 \\
    PG1116+215$^{\rm B,M}$ & 0.1662 & 12.63, 14.28, 13.69 & 15.8, 32.7, 45.4 & -53.5, -349.4, +160.0 & 156.0, 317.0, 531.4 \\
    PG1116+215$^{\rm B,M}$ & 0.1662 & 12.63, 13.39, 13.69 & 15.8, 30.7, 45.4 & -53.5, -349.4, +160.0 & 156.0, 317.0, 531.4 \\
    PG1216+069$^{\rm B}$ & 0.1801 & 13.25, 13.52, 13.48 & 64.8, 53.3, 30.4 & +140.8,-67.6, 214.5 & 588.6, 415.4, 632.0 \\
    PG1216+069$^{\rm B}$ & 0.1801 & 13.25, 13.52, 12.77 & 64.8, 53.3, 12.4 & +140.8,-67.6, 214.5 & 588.6, 415.4, 632.0 \\
    %phl2525$^{\rm B}$ & 0.1700 & 12.5 & 1-2 & - & - \\
    PKS0405-123$^{\rm B}$ & 0.0251 & 12.82, 13.18, 13.11 & 42.4, 144.3, 51.2 & {\color{red}  +1051.2, +1300.0, +1297.0} & {\color{red} 673.0,  799.0, 1105.0 } \\
 	PKS0405-123$^{\rm B}$ & 0.1330 & 13.67, 13.22, 13.13 & 27.6, 24.2, 45.0 & +227.7, +235.7, +156.2 & 467.0, 501.0, 714.0 \\
    PKS0405-123$^{\rm B}$ & 0.1522 & 12.97, 13.44, 13.79 & 21.8, 25.8, 47.8 & +267.7, +358.8, -310.4 & 181.0, 2148.0, 3857.0 \\
    PKS0405-123$^{\rm B}$ & 0.1522 & 12.78, 13.44, 13.79 & 21.2, 25.8, 47.8 & +267.7, +358.8, -310.4 & 181.0, 2148.0, 3857.0 \\
    %pks0552 & 0.1926 & 12.5 & 2-4 & - & - \\
 	PKS0558-504$^{\rm B}$ & 0.0280 & 14.08, 13.13, 13.61 & 32.2, 47.3, 55.1 & - & - \\
    Q1230+0115$^{\rm B,M}$ & 0.0057 & 13.68, 15.25, 13.25 & 24.4, 37.3, 42.8 & {\color{red}  +505.6, +595.1} & {\color{red} 155.1, 168.0} \\
 	%rbs1892$^{\rm B}$ & 0.1582 & 12.5 & 2-4 & - & - \\
 	%rbs1892$^{\rm B,M}$ & 0.1582 & 12.5 & 2-4 & - & - \\
 	 \hline
	\end{tabular}
	\begin{tablenotes}
	\item[1] Superscript B and M denotes presence of BLA and any metal ion species, respectively in at least one of the absorber in the triplet system. 
	
	{   The triplets are organized according to the $r$ bin they belong to.}
	\item[2] {  The $v_{\parallel}$ and impact parameters of the nearest galaxies which have velocity separations larger than 500\kms\ from the \lya\ triplets have been highlighted in red. }
	\end{tablenotes}
	\end{threeparttable}
    \label{Table_IF}
    
\end{table*}

\begin{table*}
    \contcaption{
    %Galaxies nearby to identified \lya\ absorber triplets. Superscript B and M denotes presence of BLA and any metal ion species, respectively in atleast one of the absorber in the triplet system.
    }
    \begin{tabular}{ccccccc}
    \hline
    QSO sightlines & $z$ (\lya\ triplet) & log $N_{\rm HI}$ of & $b$-parameter of &  $v_{\parallel}$ (Nearest Galaxy-\lya\ triplet ) & Impact parameter of Galaxy  \\
       &  & triplet system  & triplet systems (in \kms)   &  (in \kms) & from sightlines (in pKpc)\\
    \hline
    \hline
    
  %  pg0003$^{\rm B,M}$ & 0.0911 & 12.5 & 0.5-1 & - & - \\
  
  \multicolumn{7}{c}{$r_1=r_{\parallel},\ r_2=2r_{\parallel}$ configurations: $r_{\parallel}=0.5-1$ pMpc }\\
 	3C263$^{\rm B, M}$ & 0.0633 & 13.92, 14.86, 15.24 & 27.5, 52.5, 41.0 & -15.5, -131.2, -119.9 & 62.4, 571.5, 577.9 \\
    H1821+643$^{\rm B, M}$ & 0.1215 & 14.21, 13.49, 13.43 & 38.1, 35.6, 53.0 & +40.9, +329.8, -272.0 & 156.3, 1160.5, 1550.4 \\
    H1821+643$^{\rm B,M}$ & 0.1217 & 13.49, 13.43, 13.14 & 35.6, 53.0, 38.1 & -15.8, +273.1, -328.7 & 156.3, 1160.5, 1550.4 \\
    H1821+643 & 0.1899 & 12.72, 12.55, 12.61 & 26.1, 16.1, 22.4 & -235.5, -250.6, -180.0 & 1019.8,1054.5,1084.3 \\
    PG0953+414$^{\rm B}$ & 0.0160 & 13.11, 13.48, 12.92 & 39.8, 54.2, 8.9 & +102.8, -92.4 & 159.2, 454.5 \\
    PG0953+414$^{\rm B}$ & 0.0161 & 13.48, 12.92, 13.52 & 54.2, 8.9, 24.8 & +64.4, -130.4 & 159.2, 454.5 \\
    PG1116+215$^{\rm B,M}$ & 0.1658 & 13.39, 13.06, 14.28 & 30.7, 44.9, 32.7 & +37.8, -258.1, +251.4 & 156.0, 317.0, 531.4 \\
    PG1216+069$^{\rm B}$ & 0.1799 & 13.25, 13.98, 13.52 & 64.8, 33.6, 53.3 & -7.4, +201.1, +274.9 & 415.4, 588.6, 632.0 \\
    PHL1811$^{\rm B,M}$ & 0.1205 & 13.01, 14.17, 13.8 & 77.4, 52.1, 19.6 & -91.3, +425.4 & 1222.3, 1840.0 \\
    PKS0405-123$^{\rm B,M}$ & 0.1666 & 12.6, 13.54, 15.01 & 21.8, 29.3, 50.7 & +116.2, -184.6, -364.7 & 115.0, 2472.0, 3177.0 \\
    PKS2155-304 & 0.0170 & 13.43, 13.53, 13.3 & 22.9, 24.5, 39.9 & -25.4 & 113.0 \\
    PKS2155-304$^{\rm B}$ & 0.0542 & 13.85, 13.61, 12.75 & 34.9, 52.5, 23.7 & -37.8 & 544.0 \\
    %rbs1892 & 0.1908 & 12.5 & 1:2 & - & - \\
    TONS210$^{\rm B}$ & 0.0858 & 13.05, 13.07, 12.91 & 73.0, 19.8, 19.5 & - & - \\
    
    \\
    \multicolumn{7}{c}{$r_1=r_{\parallel},\ r_2=2r_{\parallel}$ configurations: $r_{\parallel}=1-2$ pMpc }\\
 	3C263$^{\rm B, M}$ & 0.1137 & 14.06, 13.27, 13.87 & 46.4, 56.8, 22.1 & +13.7, +8.4, +258.9 & 351.0,  689.4, 708.9 \\
 	3C273$^{\rm B}$ & 0.0671 & 14.08, 12.6, 12.62 & 37.0, 73.6, 25.0 & {\color{red} +2168.5} &{\color{red}  661.0} \\
 %	b0117 & 0.1165 & 12.5 & 1-2 & - & - \\
 %	f1010 & 0.1138 & 12.5 & 1-2 & - & - \\
 	H1821+643$^{\rm M}$ & 0.1215 & 14.21, 13.49, 13.14 & 38.1, 35.6, 38.1 & +40.9, +329.8, -272.0 & 156.3, 1160.5, 1550.4 \\
 	H1821+643 & 0.1895 & 12.6, 12.72, 12.61 & 28.1, 26.1, 22.4 & -140.0,  -155.1,  -84.5 & 1019.8, 1054.5, 1084.3 \\
 	H1821+643$^{\rm B}$ & 0.1899 & 12.72, 12.55, 12.73 & 26.1, 16.1, 41.6 & -235.5, -250.6, -180.0 & 1019.8, 1054.5, 1084.3 \\
 	H1821+643$^{\rm B}$ & 0.1914 & 12.73, 12.57, 12.67 & 41.6, 15.9, 31.5 & +149.8, +200.2, +159.9 & 833.4, 1301.0, 1847.2 \\
 	HE0153-4520$^{\rm M}$ & 0.1489 & 13.34, 13.25, 12.98 & 35.6, 29.3, 26.0 & -133.4, +229.5, -62.9 & 1079.3, 1085.2, 1477.9 \\
 	HE0153-4520$^{\rm B,M}$ & 0.1706 & 12.66, 13.71, 14.33 & 32.9, 100.0, 39.6 & -435.7, -64.1,  -12.8 & 911.3, 2145.1, 2384.1 \\
 %	mrk876 & 0.0634 & 12.5 & 1-2 & - & - \\
 %	p1103 & 0.0594 & 12.5 & 1-2 & - & - \\
% 	p1103 & 0.0597 & 12.5 & 1-2 & - & - \\
 %	pg0838 & 0.1242 & 12.5 & 1-2 & - & - \\
 	PG0953+414$^{\rm B}$ & 0.0160 & 13.11, 13.48, 13.52 & 39.8, 54.2, 24.8 & +102.8, -92.4 & 159.2, 454.5 \\
 	PG0953+414 & 0.0161 & 13.11, 12.92, 13.52 & 39.8, 8.9, 24.8 & +64.4, -130.8 & 159.2, 454.5 \\
 	PG0953+414$^{\rm B,M}$ & 0.1423 & 12.73, 13.56, 13.2 & 15.3, 26.5, 52.7 & +237.7, +78.0, +51.2 & 405.0, 452.1, 506.7 \\
 	PG0953+414$^{\rm B}$ & 0.1426 & 12.73, 13.48, 13.2 & 15.3, 30.9, 52.7 & +155.4, -4.2, -31.0 & 405.0, 452.1, 506.7 \\
 	PG1048+342$^{\rm B,M}$ & 0.0057 & 14.07, 14.8, 13.99 & 30.4, 42.7, 86.9 & +84.1, -77.0 & 60.2, 149.2 \\
 	PG1048+342$^{\rm B}$ & 0.0061 & 14.07, 13.79, 13.99 & 30.4, 33.1, 86.9 & -6.0, -167.0 & 60.2, 149.2 \\
 	PG1116+215$^{\rm B,M}$ & 0.1662 & 13.06, 14.28, 13.69 & 44.9, 32.7, 45.4 & -53.5, -349.4, +160.0 & 156.0,  317.0,  531.4 \\
 	PG1216+069$^{\rm B,M}$ & 0.1239 & 14.65, 14.6, 14.54 & 24.5, 28.4, 44.6 & +56.1, -64.1, +184.2 & 88.0, 91.6, 389.0 \\
 	PG1216+069$^{\rm M}$ & 0.1239 & 14.65, 14.6, 14.1 & 24.5, 28.4, 23.3 & +56.1, -64.1, +184.2 & 88.0, 91.6, 389.0 \\
 	PG1216+069$^{\rm B}$ & 0.1799 & 13.25, 13.98, 13.48 & 64.8, 33.6, 30.4 & -7.4, +201.1, +274.9 & 415.4, 588.6, 632.0 \\
 	PG1216+069$^{\rm B}$ & 0.1799 & 13.25, 13.98, 12.77 & 64.8, 33.6, 12.4 & -7.4, +201.1, +274.9 & 415.4, 588.6, 632.0 \\
 	PG1307+085$^{\rm B,M}$ & 0.1413 & 12.91, 13.92, 13.28 & 49.7, 34.0, 28.5 & -245.5, -421.6 & 1318.0, 1408.0 \\
 	PG1307+085$^{\rm B,M}$ & 0.1413 & 12.91, 13.92, 14.02 & 49.7, 34.0, 40.2 & -245.5, -421.6 & 1318.0, 1408.0 \\
 	PG1424+240$^{\rm B,M}$ & 0.1471 & 14.66, 14.74, 13.51 & 49.9, 38.3, 55.5 & -121.6, +380.5, -464.2 & 493.2, 968.5, 1378.9 \\
 %	phl2525 & 0.1700 & 12.5 & 1-2 & - & - \\
 	PKS0405-123$^{\rm B}$ & 0.0251 & 12.56, 13.18, 13.11 & 13.4, 144.3, 51.2 & {\color{red} +1174.6, +1423.5, +1420.5} & {\color{red}  673.0,  799.0, 1105.0} \\
 	PKS0405-123 & 0.1522 & 12.97, 13.44, 12.54 & 21.8, 25.8, 13.5 & +267.7, +358.8, -310.4 & 181.0, 2148.0, 3857.0 \\
 	PKS0405-123 & 0.1522 & 12.78, 13.44, 12.54 & 21.2, 25.8, 13.5 & +267.7, +358.8, -310.4 & 181.0, 2148.0, 3857.0 \\
 	PKS0405-123$^{\rm B,M}$ & 0.1829 & 14.61, 13.98, 12.7 & 43.7, 33.8, 36.7 & -255.94, -149.4, -169.7 & 3854.0, 4017.0, 5395.0 \\
 %	pks0552 & 0.1107 & 12.5 & 1-2 & - & - \\
 %	pks0552 & 0.1489 & 12.5 & 1-2 & - & - \\
% 	pks0552 & 0.1498 & 12.5 & 1-2 & - & - \\
% 	pks0552 & 0.1926 & 12.5 & 1-2 & - & - \\
 %	pks0637 & 0.1437 & 12.5 & 1-2 & - & - \\
 	PKS1302-102$^{\rm B,M}$ & 0.1925 & 14.47, 13.95, 13.64 & 34.0, 39.5, 54.8 & -194.0, -22.9, +120.5 & 209.0, 434.0, 464.0 \\
 	PKS1302-102$^{\rm M}$ & 0.1925 & 14.47, 13.95, 13.6 & 34.0, 39.5, 23.2 & -194.0, -22.9, +120.5 & 209.0, 434.0, 464.0 \\
 	Q1230+0115 & 0.0485 & 12.75, 13.49, 13.16 & 35.9, 29.9, 20.8 & -95.3, -15.2, -158.2 & 912.3, 1104.2, 1148.3 \\
 	Q1230+0115$^{\rm B}$ & 0.0554 & 12.56, 12.65, 12.92 & 20.7, 38.5, 56.7 & {\color{red}  +1559.4,  +1673.1, +880.1} & {\color{red}  353.0, 517.0, 533.0 } \\
 %	rbs1892 & 0.0826 & 12.5 & 1-2 & - & - \\
 %	s080908 & 0.0230 & 12.5 & 1-2 & - & - \\
 	SBS1108+560$^{\rm B,M}$ & 0.1385 & 13.27, 15.25, 14.31 & 42.6, 23.9, 25.2 & -89.3,  -225.0, -429.3 & 336.3, 453.0, 465.2 \\
 	SBS1122+594$^{\rm B}$ & 0.1375 & 13.34, 13.14, 13.63 & 77.6, 14.9, 100.0 & -453.6, -97.6, -390.3 & 1299.2, 1488.3, 1759.1 \\
 	SBS1122+594$^{\rm B}$ & 0.1381 & 13.14, 13.63, 13.84 & 14.9, 100.0, 150.0 & -264.6,  -53.8 & 1488.3, 2189.5 \\
 	SBS1122+594 & 0.1578 & 13.4, 13.91, 13.27 & 33.4, 18.9, 21.3 & -28.5, -261.7, -303.2 & 556.8, 633.4, 687.4 \\
 	
\hline
	\end{tabular}
    \label{Table_IF_continued}
\end{table*}

In this section,we study the connection between 
%the redshift distribution of \lya\ absorbers with the cosmic structures traced by galaxies. We do so by associating 
the \lya\ absorbers that are isolated, pairs or triplets that contribute to the observed two- and three-point correlations with nearby galaxies
%. For this purpose, we use the galaxies 
in the sample discussed in Section~\ref{sample}.
%
 %\Anand{I think 90\% completeness is for $z<0.1$ so we should focus more on this.} 
{ We have not considered the galaxies around 3C57 for our analysis due to poor completeness of the galaxy sample \citep[see discussions in][]{keeney2018}.}
We identify all the \lya\ triplets with equal arm (for $r_{\parallel,1}\le4$ pMpc) and 1:2 configuration (for $r_{\parallel,1}\le2$ pMpc) at $z<0.2$ that are present along 41 sightlines having galaxy information.
%We restrict our discussions to $z\leq 0.2$ keeping in mind that the galaxy sample is more complete at the lower redshifts for fainter galaxies.
%We consider triplets at $z<0.2$ having $r_{\parallel,1}=r_{\parallel,2}=r_{\parallel}$ and $r_{\parallel}$ within 0.5-1, 1-2 and 2-4 pMpc (denoted as equal arm configuration),
%We also consider $r_{\parallel,1} = 2\times r_{\parallel,2}$ for $r_{\parallel,1} = 0.5$ and 1 pMpc (i.e 1:2 configuration), 
{  Triplets with such configurations are chosen because $\rm PE_3$ is detected significantly for these configurations (see Fig.~\ref{Corr3_2d}) .}

Details of the triplets and the associated galaxies are provided in Table~\ref{Table_IF}. First four columns in this table give QSO name, absorption redshift of the central component of the triplet, column densities of individual components in the triplet system and { the $b$-parameters of the components}. Fifth column of this table gives the line of sight velocity separation for up to 3 nearest {  catalogued} galaxies with respect to the absorption redshift { (redshift of the central absorber)}. {  We define nearest galaxies by their transverse distance from the sightlines.}
We consider only those galaxies that are within $\pm$500 \kms. {This velocity is chosen to account for the typical velocity dispersion in galaxies (i.e $\le$350 \kms) and that of the \lya\ triplets (i.e $\sim$300 \kms)}. Impact parameters of these galaxies ({ the distance between the absorber and  the galaxy measured using the angular separation between the QSO sightline and the galaxy}) are provided in the last column.
In eight cases we find nearest galaxies having velocities in the range 500 -2500 \kms (red colour entrees in Table~\ref{Table_IF}) . 
%These are presented in red colour in the table. In 3 cases, 
We do not detect any nearby galaxy within 2500 \kms\ for 2 triplets. {We discuss these specific cases in detail below.}
%{These are triplets at $z = 0.169$ towards 3C57, $z= 0.0280$ towards PKS~0558-504 and $z = 0.0858$ towards TON~S210. The lack of galaxies identification in the first case could be attributed to poor completeness achieved in the galaxy distribution towards 3C57.}

First we consider the "equal arm" configuration.
There are three triplets in the  $r_{\parallel}$ = 0.5-1.0 pMpc bin %for the "equal arm" configuration. 2 
and two of them show associated galaxies. 
%with a velocity separation $<$ 500 \kms with respect to the absorption redshift. 
The impact parameter of these galaxies are 156 and 924 pkpc. In the third case the nearest cataloged galaxy has a velocity difference of $\sim$ 957 \kms\ (and an impact parameter of $\sim$1.2 pMpc) with respect to the absorption redshift. All the three triplets have at least one component having $b>40$ \kms. Only one of these absorbers show detectable metals {  at $z=0.1656$ with the nearest cataloged galaxy at an impact parameter of 156~pkpc}.  Note that none of these systems satisfy the \NHI\ threshold of $10^{13.5}$cm$^{-2}$. 

In the case of $r_{\parallel}$ = 1.0-2.0 pMpc bin (where we detect $\zeta$ with best significance level), there are thirteen triplets contributing to the three-point correlation. Eleven of them have at least one component having $b>40$ \kms and only six 
%out of the 14 identified triplets do 
show detectable metal absorption {with impact parameters to the nearest cataloged galaxies in the range 60-467 pkpc (see column 6 of Table.~\ref{Table_IF})}.
There are two cases where the same absorber contributes twice to our triplet counts (i.e multiple component systems with two combinations consistent with our triplet definition).  
Therefore, there are eleven independent systems contributing to the triplet count. One of these systems also contribute to $r_{\parallel}$ = 0.5-1.0 pMpc. We find nine of these eleven independent systems identified here have nearby galaxies with velocity separations within $\pm$500 \kms.  The impact parameter of the nearby galaxies varies from 60 pkpc to 1 pMpc with a median impact parameter of 181 pkpc.
We notice that systems contributing multiple times to the three-point function tend to have low (i.e $<$500 pkpc) impact parameters. In the remaining two cases we do identify nearby galaxies but with large velocity separations (i.e between 1000-1500 \kms). Interestingly only two systems (three independent triplets) in this list satisfy \NHI$\ge10^{13.5}$ cm$^{-2}$. In both cases galaxies are found with velocity separations less than 200 \kms and impact parameter of the nearest galaxy $\le$ 500 pkpc.

%\Anand{Soumak you need to update the numbers in the following paragraph}.Next we consider 
In the case of $r_{\parallel}$ = 2.0 - 4.0 pMpc
bin there are eighteen triplets (originating from nine independent systems) contributing to the measured $\zeta$. In all these cases at least one of the components has $b>40$ \kms\ and only three of them show detectable metal lines. For six of these independent systems we find galaxies with velocity separations within $\pm$500 \kms. In one case we identify galaxies having velocity separation just above the cut-off.  In these six systems the measured impact parameters of the nearest galaxy in the range 0.156-1.102 pMpc  with a median value of 833.4 pkpc. In one case we identify the nearest galaxy to have velocity separations of 1050 \kms and impact parameter of 673 pkpc. In one case we do not have any galaxies nearest to the absorbers. In the case of $z=0.0280$ triplet towards PKS~0558-504 the galaxy observations are complete up to 0.1 L*. However, the maximum impact parameter probed is $\sim$ 400 pkpc.
Considering several of the associated galaxies for this configuration are at impact parameter $>400$ pkpc (see Table~\ref{Table_IF}), we need to search for galaxies at slightly higher impact parameters before confirming this triplet as a void absorber.
%In the case of $z= 0.169$ triplet towards 3C57 non-detection of galaxies can be attributed to poor completeness \citep[see discussions in][]{keeney2018}.} This is the only triplet that satisfies the \NHI\ threshold of 10$^{13.5}$ cm$^{-2}$ in this $r_{\parallel}$ bin. 
{ Interestingly, none of the triplets satisfy the \NHI\ threshold of 10$^{13.5}$ cm$^{-2}$ in this $r_{\parallel}$ bin.}
{As discussed before, in only one triplet (i.e $z = 0.0251$ towards PKS~0405-123), we find all the three components having $b>40$ \kms the nearest cataloged galaxies have velocity separations in excess of 1000 \kms and impact parameters in the range $673-1105$ pkpc.}

Next  we consider the configuration with an arm length ratio 1:2. In the $r_{\parallel,1}$ = 0.5-1 pMpc bin we identify thirteen triplets (and eleven independent systems). Eleven of these triplets show at least one of the components having $b>40$ \kms and only six of them show detectable metal absorption. We could identify at least one nearby galaxy in
%with velocity separation $<500$ \kms for 
ten out of eleven independent systems. In these systems the nearest impact parameter ranges from 62 pkpc to 1.2 pMpc with a median value of 158 pkpc. {For one triplet (i.e $z =0.08586$ towards TON S210) we do not detect any associated galaxy within 2500 \kms. The galaxy observations are deep enough to detect 0.1L* galaxy within an impact parameter of 960 pkpc. Most of the galaxies detected in other cases for this configuration are well within this impact parameter. Therefore, lack of galaxy identification could mean this triplet being a void absorber.}
Only one system satisfies the \NHI\ threshold of 10$^{13.5}$ cm$^{-2}$ and this happen to be the system with the lowest impact parameter in this
$r_{\parallel,1}$ bin.

%We also consider 
For the $r_{\parallel,1}$ = 1-2 pMpc bin we
%for 1:2 arm length ratio and 
identify thirty-four triplets (from twenty-five independent systems). Out of this, twenty-four triplets have at least one component with $b>40$ \kms\ and only sixteen  triplets show detectable metal lines. We find thirty-one triplets having at least one nearby galaxy with velocity separation  $<500$ \kms.
For these triplets, the nearest impact parameter range from 60 pkpc to 3.8 pMpc with a median value of 410 pkpc. For three triplets, we find nearest galaxies with large velocity separations (1100-2200\kms) with the nearest impact parameters in the range of 350-660 pkpc. Only seven triplets satisfy \NHI\ threshold of 10$^{13.5}$ cm$^{-2}$ and  all of these have nearby galaxies within a velocity separation of 200 \kms and with impact parameters ranging from 60 pkpc to 500 pkpc.

{\it In summary, for five configurations considered here we identify at least one associated galaxy within a velocity separation of 500 \kms\ for $\sim88$\% of the triplets. 
%{If we consider the the velocity separation to be $<300$ \kms then we have xx\% of the triplets have galaxy association}.
The measured impact parameters of the nearest galaxies range from 62-3800 pkpc with a median value of 405 pkpc. Therefore, a good fraction of triplets originate from impact parameters that are inconsistent with them being associated to a single galaxy. The occurrence of $b>40$ \kms\ absorbers are more frequent among the triplets ($\sim 85$\%, as opposed to the $31.9$\% for isolated absorbers). {However, in only one case we see all the three components having $b>40$ \kms. In only twenty-one cases ($\sim$25\%) we have two components having $b>40$ \kms.} 
Only 40\% of all the triplets are associated with metal line absorption. Most of the triplets are originating form low \NHI\ systems with only $\sim$14\% of the identified triplets having all components having \NHI$\ge10^{13.5}$~cm$^{-2}$.}
%We consider the redshift of the middle absorber as the redshift of the system. We then search for nearby galaxies in the galaxy redshift survey which are within $\pm 500$ \kms\ of the triplet systems.

Next we ask the question, is the impact parameter distribution for the triplets different from those of doublets and isolated absorption lines. For this we calculate the cumulative distribution function (CDF) of the impact parameters of respective sightlines  from galaxies in three ways: first we consider the impact parameter of the nearest galaxy, second the average impact parameter of the 2 nearest galaxies (nearest in transverse direction) and third, the average impact parameters of the 3 nearest galaxies. If we do not find the minimum required number of galaxies, then we do not consider those cases. As above, we consider only galaxies which are within $\pm 500$\kms\ of \lya\ pairs (and $r_{\parallel}$ within 0.5-1, 1-2 and 2-4 pMpc) that do not belong to a triplet system. Similarly, we also identify galaxies which are within $\pm 500$ \kms\ of isolated individual \lya\ absorbers (singlet), which do not belong to either a triplet or a pair. 
%We also calculate the probability distribution functions of galaxies associated with a \lya\ pair and \lya\ singlet separately in the three cases we mentioned before. 

In Fig.~\ref{IF}, we plot the CDF of  impact parameters of these galaxies associated with triplet, pair and singlet \lya\ systems. The left, middle and right columns denote impact parameters considered for the nearest galaxy, average of two nearest galaxies and average of three nearest galaxies, respectively. Top panels are for \NHI $> 10^{12.5}$ cm$^{-2}$ case and the bottom panels are for \NHI $> 10^{13.5}$ cm$^{-2}$.

%We also plot the same distribution function but with a larger redshift space bin of $\pm 1000$\kms\ in the bottom two rows. The first and third rows show the distribution for $N_{\rm HI}>10^{12.5}$cm$^{-2}$ { absorbers} while the second and fourth rows show it for $N_{\rm HI}>10^{13.5}$cm$^{-2}$ { absorbers}. All the information of the \lya\ triplets and the corresponding nearby galaxies used are given in Table.~\ref{Table_IF}. We also show the number of triplets associated with galaxy systems "n" to the total number of triplet systems "N" in the format "Triplets=n/N" for each plot.

%We identify 83 triplets in the redshift range of our interest with \NHI $>10^{12.5}$ cm$^{-2}$. 
%For 72 of these absorbers we do find galaxies with line of sight velocities within 500 \kms\ to the absorbers (panel (a) in Fig.~\ref{IF}) .  
As discussed before for the \NHI $> 10^{12.5}$ cm$^{-2}$ case, the median impact parameter with the nearest galaxies for the triplets in our sample is 405.0 pkpc and the largest separation found is 3.8 pMpc.
%confirming that $\sim$84\% of our triplets have at least one galaxy within an impact parameter of 3.8 Mpc and the velocity separation of 500 \kms. 
%
%The median impact parameter is higher than the typical virial radius of nearby galaxies. 
%Therefore, it is most likely that most of the triplets we find for \NHI$>10^{12.5}$ cm$^{-2}$ may not be originating from individual halos. 
%
In the case of doublets (respectively singlets) the median impact parameter is 415.4 pkpc (respectively  645.1 pkpc) and the largest separation found is 3.8 pMpc (respectively  5.4 Mpc) . From panel (b) and (c) of Fig~\ref{IF}, we notice that 69 (i.e 85\% triplets) and 56 (i.e 69\% triplets) out of 81 triplets have at least two and three nearby galaxies with velocity separation $<500$ \kms. 
The median separations of two nearest galaxies are 658.4 pkpc, 608.9 pkpc and 796.2 pkpc for triplets, doublets and singlets respectively. While there is clear distinction in the median impact parameter for the single and multiple (i.e doublets plus triplets) \lya\ absorbers, we do not find any significant difference between the median impact parameter of doublets and triplets.
The median separation for three nearest galaxies are 953.5 pkpc, 651.9 pkpc and 931.5 pkpc for triplets, doublets and singlets respectively. { Even though the median separation may seem similar between triplets and singlets, as can be seen below the KS test results suggest the two distributions are statistically different. }

We perform KS-test to compare different CDFs and results are summarized in Table~\ref{tab:ks}.  Here D is the maximum separation between the two distributions and Prob(D) is the probability that this D occurs by chance. This table confirms that there is no statistically significant difference between the distribution of impact parameters for triplets and doublets. While the impact parameter distribution of singlet is significantly different from that of triplets. 
%when we consider the nearest neighbour or two neighbour.

% This is also supported by the KS-test results presented in Table~\ref{tab:ks}.

%However, some doublets have associated galaxies at much larger impact parameters than the maximum value found for the triplets.
%

Next we consider \NHI $>$ $10^{13.5}$ cm$^{-2}$ case, we identify eleven triplets and all of them have nearby galaxies within line of sight velocity separation $< 500$ \kms. 
All eleven triplets show $b>40$ \kms components and nine of them show detectable metals.
In this case the median separations of these nearby galaxies are  88 pkpc, 256 pkpc and 
459 pkpc for the triplets, doublets and singlets respectively. In this case it is possible that some of the correlated absorption may originate from the halos of individual galaxies.
Impact parameters measured are systematically smaller than the corresponding values for 
\NHI $>$ $10^{12.5}$ cm$^{-2}$. We also confirm this with KS-test. {  The D and Prob(D) values
%for the impact factors of $N_{\rm HI}>10^{12.5}$cm$^{-2}$ and $N_{\rm %HI}>10^{13.5}$cm$^{-2}$ triplets from the %nearest galaxies 
are 0.54 and 0.004 respectively. } 
%
%It is also evident from panels (e) and (f) of Fig.~\ref{IF}, that 11 and 7 of these triplets have two or three nearby galaxies respectively. However, to draw any strong conclusions between triplets originating from different \NHI\ thresholds we need a much bigger sample of high column density absorbers.

As \citet{keeney2018} also provide stellar mass and luminosities of individual galaxies we also searched for any differences in the observed luminosities and inferred stellar masses of nearest galaxies {(as stated before, within a velocity separation of 500 \kms)} associated with triplets compared to doublets and singlets. We do not find any strong trend. This may once again hint towards the idea that most of the \lya\ absorbers studied here are not physically linked to individual galaxies.
%
%In summary, the three point correlation we found for \NHI $>$ $10^{12.5}$ cm$^{-2}$ absorbers are not predominantly from halos of individual galaxies and closely related to the distribution of galaxies compared to the isolated singlet \lya\ absorbers. This is consistent with our finding in the previous section that the \lya\ absorbers without associated metals also show three-point correlation.
%We do find high frequency for the association of high-$b$ absorption component with the triplets.
%Our study indicates that a good fraction of high \NHI\ triplets (i.e \NHI $\ge10^{13.5}$ cm$^{-2}$) may originate from galaxy halos. However, the number of such systems considered in this study is very low (i.e only 12) and we need  a significant number of such systems to draw any strong conclusions.
%
The results presented are consistent with the findings in previous studies that most \lya\ absorbers originate from the filamentary structures with BLAs and multiple absorbers having low impact parameters to the filament axis compared to the low \NHI\ isolated absorbers \citep[see][]{dave2001a,Penton2002,dave2010,Wakker2015, Tejos2016}. The volume probed and the completeness level reached in terms of luminosity  for the galaxy will not allow us to perform an investigation similar to \citet{Wakker2015}. However, it will be possible to explore for some of the very low-$z$ triplets using SDSS data base. 
We leave that exercise to a future work.

\begin{table*}
\caption{Results of KS-test}
\begin{tabular}{cccccccc}
\hline
Case       &    log $N_{\rm HI}$       & \multicolumn{2}{c}{Triplet vs. doublet} & \multicolumn{2}{c}{Triplet vs. singlet} & \multicolumn{2}{c}{Doublet vs. singlet}  \\
 & threshold & D & Prob(D) &  D & Prob(D) &  D & Prob(D) \\
\hline
\hline
Nearest neighbour & 12.5 & 0.11 & 0.69              & 0.27 & 0.0001           & 0.23 & 0.0002  \\
2 neighbours     & 12.5 & 0.10 & 0.77 & 0.23 & 0.002   & 0.18 &  0.013            \\
3 neighbours     & 12.5 & 0.12 & 0.74 & 0.23 & 0.01 & 0.16 & 0.05\\
\\
Nearest neighbour & 13.5 & 0.36 & 0.15              & 0.56 & 0.001    & 0.23 & 0.04         \\
2 neighbours     & 13.5 & 0.40 & 0.09 & 0.58 & 0.001  & 0.25 & 0.03             \\
3 neighbours     & 13.5 & 0.27 & 0.71 & 0.46 & 0.08 & 0.23 & 0.09\\
\hline
\end{tabular}
\label{tab:ks}
\end{table*}

\section{Simulations}
\label{sec:simulations}

In this section, we study the clustering of \lya\ absorbers using four hydrodynamical simulations for $z=0.1$. Main motivations for this exercise are: {(i) to measure the two- and three-point correlation functions for different scales; (ii) to quantify the effect of redshift space distortions in the measurement of line of sight correlation functions, (iii) to see whether the observed dependence of $\zeta$, $\xi$ and Q  on \NHI\ and $b$ are naturally realised in the simulations as well, and (iv) to see the effect of feedback (Wind and AGN) on the correlation functions. However, we do not make any attempt to fine tune the simulation parameters and/or various sub-grid physics used to reproduce the observations.}

%The motivation behind simulating the low-$z$ \lya\ forest is to quantify the effects of redshift space distortion. We would also like to compare the observational dependencies of simulated $\xi$, $\zeta$ and Q on $N_{\rm HI}$, $b$ and the metallicity with those seen in simulations. We also use simulations with and without wind and AGN feedback to explore their effects on clustering. 
%
%The details of the simulations used in this work is given in Table.~\ref{Sim_tab}. 

%We use t
The MassiveBlack-II  (MBII) hydrodynamic  simulation \citep{khandai2015} 
%for studying the effect of redshift space distortion and metallicity. The simulation had been 
was run in a 100$h^{-1}$ cMpc ({ Mpc in comoving units}) cubic periodic box with $2\times 1792^3$ particles using {\sc p-gadget} which is a hybrid version of {\sc gadget-3} upgraded to run on Petaflop-scale supercomputers.  This simulation used the UV background model of \citet{haardt1996} and incorporates feedback associated with star formation and black hole accretion and
%The 
cosmological parameters 
%used are
from
$(\Omega_m=0.275,\Omega_{\Lambda}=0.725, \Omega_b=0.046,h=0.702,\sigma_8=0.816,n_S=0.96)$ from WMAP7 \citep{komatsu2011}. 

We also use three hydrodynamical simulation boxes from publicly available Sherwood Simulation suite\footnote{\url{https://www.nottingham.ac.uk/astronomy/sherwood/}} \citep{Bolton2017} to explore the effect of wind and AGN feedback. All of them are performed in a 80 h$^{-1}$ cMpc cubic box with $2\times512^3$ particle using P-Gadget-3 \citep{springel2005}. The model ``80-512" is run with {\sc quick\_lyalpha} \citep[as described in][]{viel2005} command without any stellar or wind  feedback. 
The second simulation ``80-512-ps13" implements the star formation and energy driven wind model of \citet{Puchwein2013}, but without AGN feed back. The third simulation ``80-512-ps13+agn" implements the AGN feedback in addition to the star formation and energy driven wind models.
All three simulations have the same initial seed density field, utilize \citet{haardt2012} UV background and use the same set of cosmological parameters from \citet{planck2014}, where $\Omega_m=0.308,\Omega_{\Lambda}=0.692, \Omega_b=0.0482,h=0.678,\sigma_8=0.829,n_S=0.961$. \citet{Nasir2017} provides a detailed comparison of predictions of these three models with observations.
%Note that the cosmology used in the Sherwood simulation suite is different from that used in MBII.
%For both the MBII and Sherwood Simulation suite, we use the $z=0.1$ snapshot.

Recent studies \citep[see][]{kollmeier2014,shull2015, khaire2015b, gaikwad2017a,gaikwad2017b} have shown that the \HI\ ionization rate ($\Gamma_{\rm HI}$) at $z\sim0.1$ is higher than those predicted by \citet{haardt2012}.
%
%We would like to mention that our aim is not to match the simulated statistics with the observations since we still need to improve the simulations in terms of reproducing $N_{\rm HI}$ and $b$ distribution self-consistently. Rather we aim to explore the effect of redshift space distortion using MassiveBlack-II simulation and the effect of wind and AGN feedback from the three Sherwood simulations. 
%
Therefore, for calculating the \HI\ density fields from the simulations, we use the  $\Gamma_{\rm HI}$ at $z\sim0.1$ from \citet{khaire2019} uniformly for all the simulations. We do not adjust $\Gamma_{\rm HI}$ to match the mean flux with observations. We generate 4000 lines of sight through these boxes using standard procedure described in our earlier papers \citep[][]{gaikwad2017b, maitra2020, gaikwad2020b}. { We find the mean transmitted flux of 0.973, 0.979, 0.977 and 0.978 for MBII, 80-512, 80-512-PS13 and 80-512-PS13+AGN simulations respectively. These are close to what is observed, i.e, 0.961 for the full sample and 0.967 for the \lya\ absorption at $z<0.2$.}
We convolve the spectrum with a gaussian profile with FWHM=17 \kms (instead of using HST line spread function), add a gaussian noise { corresponding to SNR$=$50 per pixel} and use spectral sampling (i.e $\sim$5 \kms\ per pixel) similar to the HST-COS spectra. {The usage of Gaussian LSF is justified as observationally what we have is the deconvolved $b$-distribution. }
%
%simulate instrumental smoothing effects by convolving with a simple gaussian profile with . We also add a gaussian noise with a high SNR value of 50 uniformly across the wavelength for all the simulated sightlines (unlike HST-COS spectra where it is wavelength dependent and has a median SNR of 12) . 
We use high SNR spectra to get an insight into the intrinsic clustering of the \lya\ absorbers. 
%We also consider SNR$\sim$12 for few cases to understand the effect of SNR. 
Simulated spectra  are fitted with multiple Voigt profile components using {\sc VIPER} \citep[see][for details]{gaikwad2017b}. As in the observations, we consider only components that satisfy rigorous significant level in excess of 4 for our analysis.

\subsection{Distributions of $b$-parameter and $N_{\rm HI}$}

\begin{figure*}
\includegraphics[viewport=0 0 300 254,width=5.8cm, clip=true]{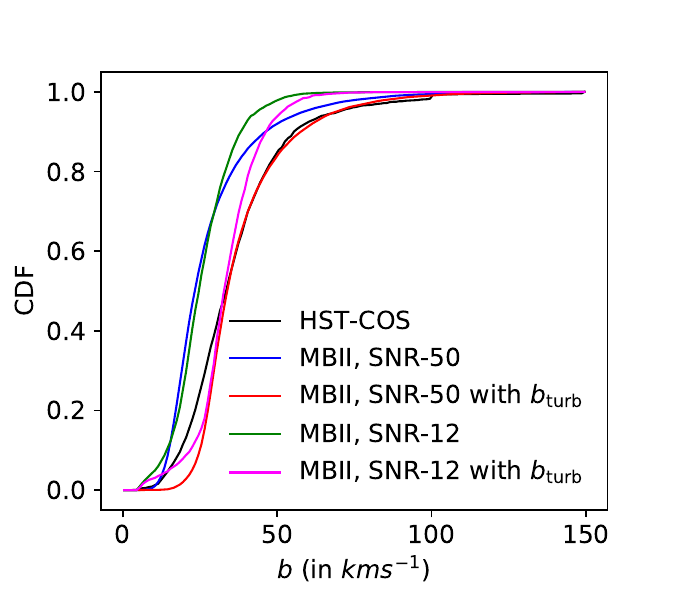}%
\includegraphics[viewport=0 0 300 254,width=5.8cm, clip=true]{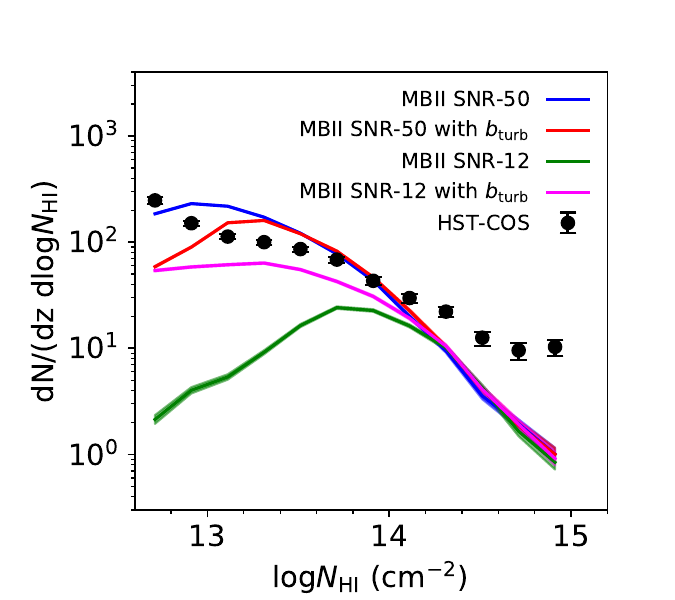}%
 \includegraphics[viewport=0 0 300 254,width=5.8cm, clip=true]{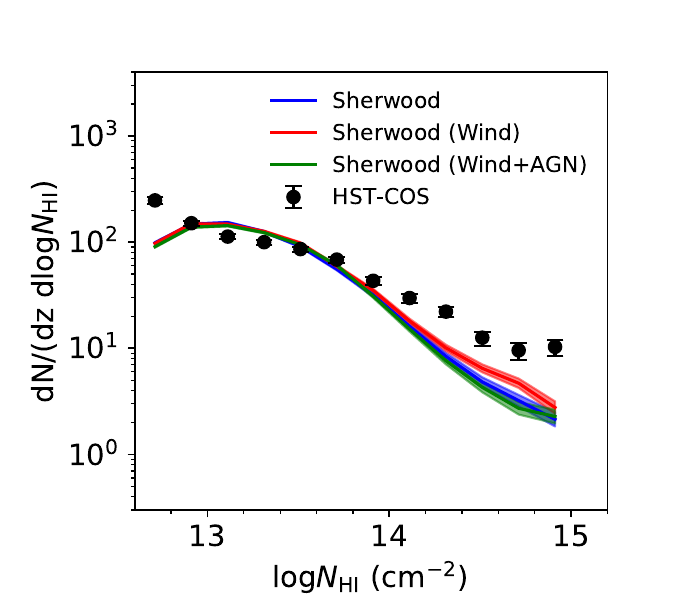}%
	\caption{{\it Left panel}: Cumulative distribution function for $b$-parameter for MBII simulations (for two SNR) are compared with the observed distribution. Simulations tend to produce low-$b$ values compared to the observed distribution. Addition of $b_{\rm turb}$ reduces this difference. Our high SNR model produces over all CDF closer to the observed distribution.
	{\it Middle panel}: \NHI\ distribution predicted by MBII simulations are compared with the observed distribution. {  The high SNR simulation predicts a larger number of low \NHI\ { absorbers} in comparison to observations. The number goes down with the addition of $b_{\rm turb}$ in the low \NHI\ end. The low SNR simulations produce significantly lesser low \NHI\ { absorbers} due to incompleteness. On the high \NHI\ ends, the simulations  produce significantly lesser { absorbers} than the observations.} {\it Right Panel}: 
	\NHI\ distribution predicted by Sherwood simulations.
	Like MBII simulations,  { absorbers} with \NHI$>10^{14}$ cm$^{-2}$ are under-predicted. {  However, the low \NHI\ end matches relatively better than MBII simulations.}}
\label{NHI_b_sim}
\end{figure*}

%\begin{figure}
% \includegraphics[viewport=0 0 300 300,width=8cm, clip=true]{NHI_dis_feedback.pdf}%%
%	\caption{Neutral hydrogen column density distribution for SHERWOOD simulation suite.}
%\label{NHI_dis_feedback}
%\end{figure}

\begin{figure}
 \includegraphics[viewport=0 0 400 260,width=10.8cm, clip=true]{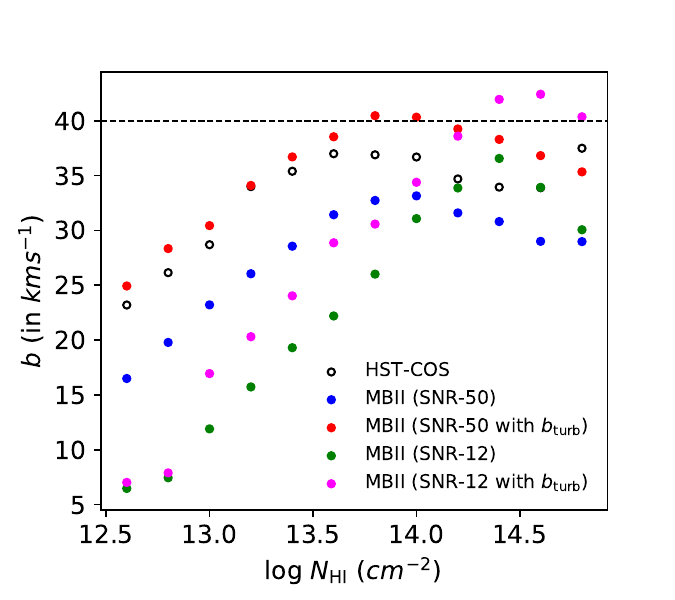}%
	\caption{
Median b-parameter in different \NHI\ bins 
{ obtained from} our simulations are compared with observations.
%found in observations is compared with predictions for different simulations.
%	Neutral hydrogen column density $N_{\rm HI}$ vs line-width parameter $b$ color plot for simulations. 
The horizontal dotted line at $b = 40$ \kms demarcates the high-$b$ and low-$b$ sub-samples. 
%\Anand{Plot this exacly as in Fig 6 and as suggested by the referee also include for SNR=12.}
%Point provide the median value of \NHI\ in each $b$ bins. 
}
\label{NHI_b_sim2}
\end{figure}

%Before measuring two- and three-point correlation functions 
To begin with, we explore how well the simulations reproduce the observed \NHI\ and $b$ distributions. 
In panel (a) of Fig.~\ref{NHI_b_sim}, we compare the cumulative distribution function of the $b$-parameter of \lya\ absorbers found in MBII simulation with the observed distribution. We find that the simulation produces lower $b$-values (median b of $\sim 25.0$ \kms) in comparison to the observations (median $b$ of $\sim 34.4$ \kms). \citet{Nasir2017} have found Sherwood simulations also to produce low-$b$ values. 
%Even with the modified the $\Gamma_{\HI}$ we find the same trend. 
The median $b$-values are 24.4, 24.7 and 26.1 \kms for models 80-512, 80-512-ps13
and 80-512-ps13+agn respectively
%The quoted median $b$-parameters are 
for { absorbers} with \NHI$>10^{12.5}$ cm$^{-2}$.
%Slightly higher value found for MBII simulations can be attributed to a better convergence due to its high resolution \citep[see][for discussions on convergence]{Nasir2017}.
%We generate SNR-12 sightlines along with SNR-50 sightlines to check whether SNR plays any role in the $b$-distribution. However, we find the $b$-values are lower irrespective of the SNR of the sightlines. 
%Things does not improve even when we use a low SNR of 12. 
%This seems to be the case for the three Sherwood simulations also.

This issue of hydrodynamical simulations at low-$z$ is discussed in the literature \citep{viel2016,gaikwad2017b,Nasir2017}. The solution can come from additional heating sources not considered in these simulations and/or from the inclusion of sub-grid micro-turbulence missing in the simulations \citep[as explored in][]{oppenheimer2009, gaikwad2017b}.
We consider the second case
%
%In order to match the $b$-distribution, one can artificially introduce additional heating \citep[as done in][]{viel2016}. However, this tends to suggest a lower value of $\Gamma_{\rm HI}$. One can also 
by introducing additional line broadening by adding a non-thermal micro-turbulence $b_{\rm turb}$ component to the Doppler parameter $b$ in quadrature  \citep[$b^2=b^2_{\rm thermal}+b^2_{\rm turb}$, see][]{gaikwad2017b}. {  This micro-turbulence term is added to the temperature field along the simulated sightlines before calculating the transmitted flux.} We perform Voigt profile decomposition and obtain \NHI\ and $b$-parameters using {\sc viper}.  
As can be seen in panel (a) of Fig.~\ref{NHI_b_sim}, by adding a constant non-thermal turbulence term ($b_{\rm turb}\sim 20$ \kms), we  match  the median of the observed $b$-distribution. 
%
%As shown by \citet{gaikwad2017b} this modification does not change the \NHI\ distribution dramatically and hence the derived $Gamma_{\HI}$ from observations.

In panel (b) of Fig.~\ref{NHI_b_sim}, we compare the $N_{\rm HI}$ distribution we obtained from MBII simulation with the observations. {As the observed distribution is corrected for incompleteness, it reflects the intrinsic distribution. This also justifies the usage of simulated spectra with higher SNR.}
It is evident from the figure that the simulations slightly over predict the {  weak \NHI\ absorbers ($10^{13}$cm$^{-2}<N_{\rm HI}<10^{14}$cm$^{-2}$)}. However, they tend to produce significantly lesser number of high $N_{\rm HI}$ systems (i.e \NHI $> 10^{14}$ cm$^{-2}$). While a small increase in $\Gamma_{\HI}$ can provide a better matching in the low \NHI\ end, the difference in the high \NHI\  end will be increased.
It is also clear from the figure that inclusion of $b_{\rm turb}$  significantly affects the distribution at the low \NHI\ end. {In particular, the under prediction of low \NHI\ { absorbers} can be attributed to the additional line broadening making some weak absorption lines to go below the significant level of detection (i.e for a given SNR and \NHI, lines with higher $b$-values are difficult to detect). However, as expected the additional broadening has not affected the \NHI\ distribution at higher column densities (in this case above $10^{13.3}$ cm$^{-2}$). 
Similarly, the simulated spectra at lower SNR under predicts the low \NHI\ systems (i.e \NHI $< 10^{14.3}$ cm$^{-2}$). Addition of $b_{\rm turb}$ moves the \NHI\ values, where the { simulated} data is complete, to a higher value even in  this case. Alternatively, addition of $b_{\rm turb}$ will push some low \NHI\ systems below our detection limit.}

In Fig~\ref{NHI_b_sim2} we plot median $b$
%\NHI\ 
for different \NHI-bins (similar to Fig.~\ref{NHI_b}). {As expected the MBII simulations with SNR=50 have lower median $b$-values  compared to observations for all the \NHI\ bins. Addition of turbulence broadening improves the matching. When we consider the lower SNR simulations  the difference between the observed and predicted median $b$-values for a given \NHI\ is larger than that seen for high SNR case. Also simple addition of $b_{\rm turb}$ did not help even when we match the overall median $b$ values with observations (see Fig~\ref{NHI_b_sim}). 
This is because at low SNR detection of broad absorption at low \NHI\ is difficult. Also  the median $b$ values at high \NHI\ in this is higher as Voigt profile fitting will will require less number of components to achieve statistically significant fit.
%This disparity between the observation and low SNR simulation can be attributed to a distribution of low and high SNR regions in the observed sightlines. 
%At low SNR 
%In the case of low SNR simulations the detection of low $N_{\rm HI}$ and high $b$ absorbers is difficult while also predicting the intrinsically blended high $N_{\rm HI}$ systems as a single high $N_{\rm HI}$ and high $b$ absorber.
}

%This also confirms that the median \NHI\ is slightly lower for a given $b$ in MBII simulations. \Anand{You should plot Fig 12 exaclty like Fig 6 and also include the results for SNR = 12.}

The \NHI\ distribution in the case of three Sherwood simulations are shown in panel (c) of Fig~\ref{NHI_b_sim}. Despite using slightly different $\Gamma_{\HI}$ and different Voigt profile fitting routines our results match well with that of \citet{Nasir2017}. While inclusion of wind feedback marginally increase the number of high \NHI\ components addition of AGN feedback nullifies this effect. 
%Even in this figure we can clearly see the lack of \NHI$\ge 10^{14}$ cm$^{-2}$ absorbers. 
%This figure also suggests that the lack of high \NHI\ absorber may not be linked to the presence or absence of feedback in the simulations.
%
Lack of high \NHI\ systems in the simulated spectra 
is again a well known result \citep[see also,][]{shull2015,viel2016,gurvich2017,Nasir2017}. 
%It is possible that one will be able to match the observed distribution if the distribution in SNR at a given redshift is taken careoff instead of using single SNR in simulated spectra and by using simulation boxes
%at different redshifts over the observed $z-$range \citep[see for example,][]{gaikwad2017b}. 
%\Anand{Prakash, confirm the above statement}. \PG{Anand, this is partly true. The other reason could be the evolution of $\Gamma_{\rm HI}$ with redshift at $z<0.5$ has to be accunted for while computing CDDF.}
%
{\it In summary, we find that there are some inherent short coming in simulations considered here {(probably coming from some missing sub-grid physics)} in reproducing the observed $N_{\rm HI}$ and $b$-distribution self-consistently. Keeping this in mind we shall proceed with the clustering analysis.}
%
%So, we will not make an effort in matching the simulated clustering statistics with observations. 
%Therefore, we restrict our aims to investigate the effects of redshift space distortion, feedback and metallicity on clustering without attempting to match the observed clustering. 
%For this, we will use high SNR sightlines to capture the intrinsic clustering properties of \lya\ absorbers.

%\Anand{ We can show the \NHI\ distribution and b-distribution and compare them with the observed distribtution. Basically you can say this is well known result and cite Viel 2017, Gurvich 2017 etc. Main point is to say simulation tend to produce less number of high \NHI systems. We try to match b distribution by adding turbulance (Gaikwad) or additional heating (Viel) . We will discuss the turbulence case } 

\subsection{\lya\ clustering}
%\subsection{Effect of redshift space distortion}
\begin{figure*}
 \includegraphics[viewport=0 38 300 290,width=6cm, clip=true]{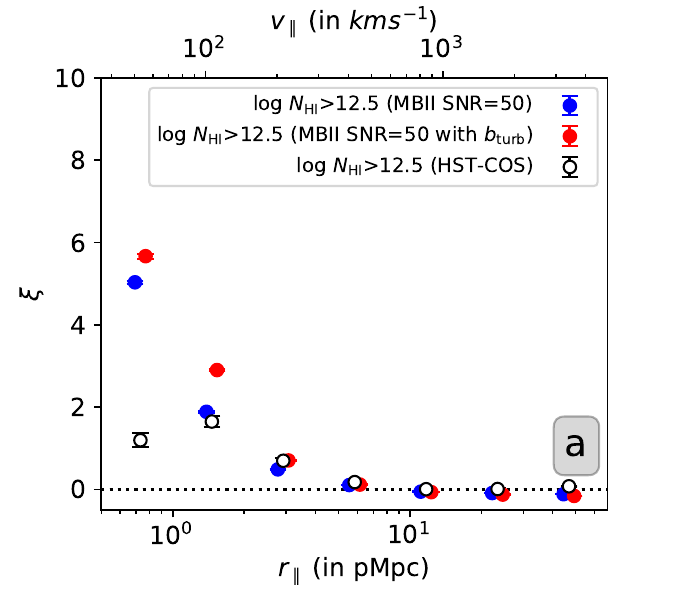}%
	\includegraphics[viewport=0 38 300 290,width=6cm, clip=true]{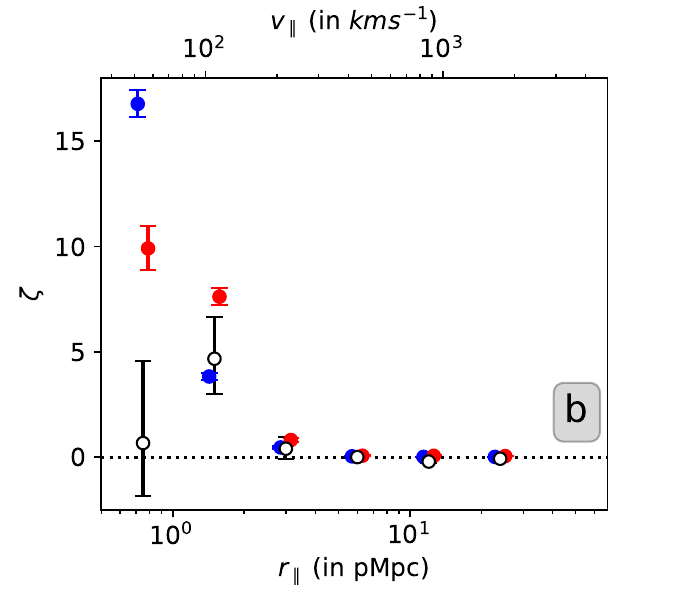}%
	\includegraphics[viewport=0 38 300 290,width=6cm, clip=true]{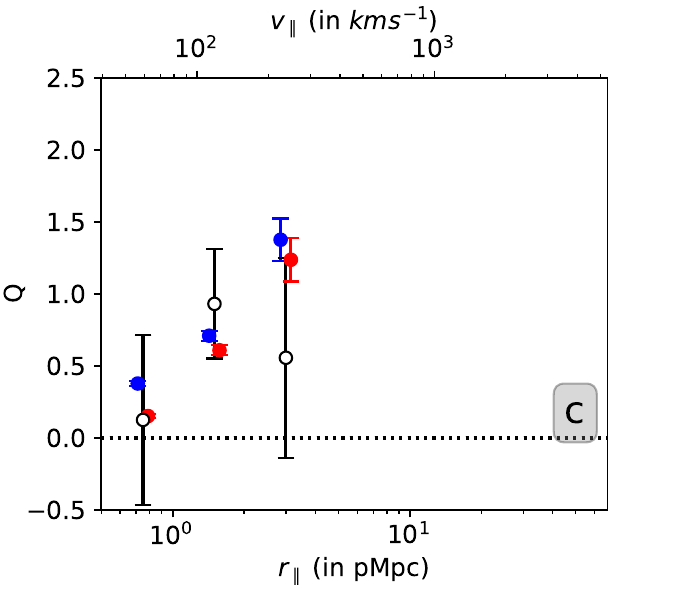}%
	
	\includegraphics[viewport=0 7 300 259,width=6cm, clip=true]{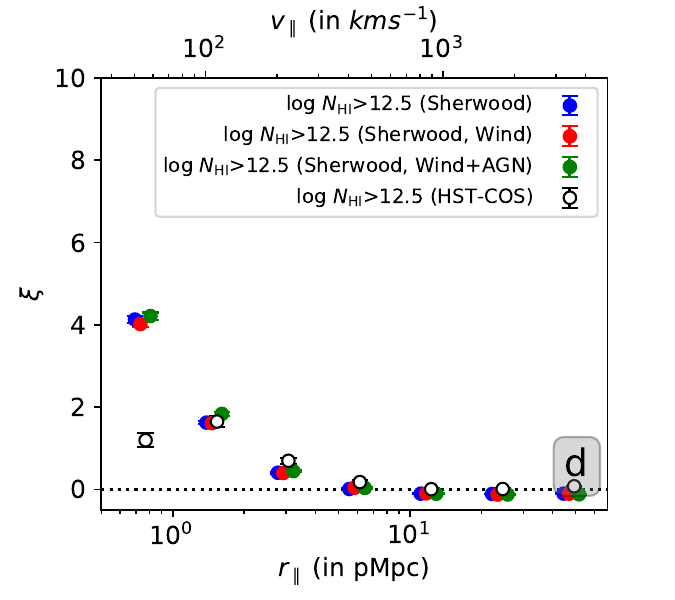}%
	\includegraphics[viewport=0 7 300 259,width=6cm, clip=true]{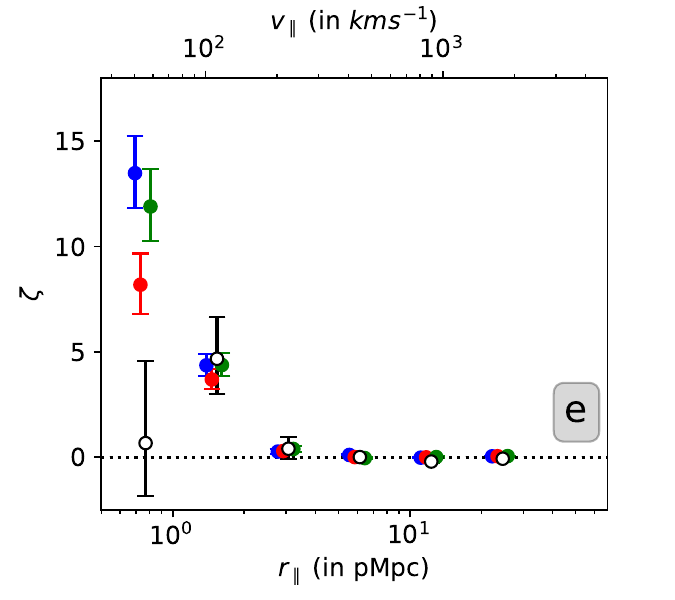}%
	\includegraphics[viewport=0 7 300 259,width=6cm, clip=true]{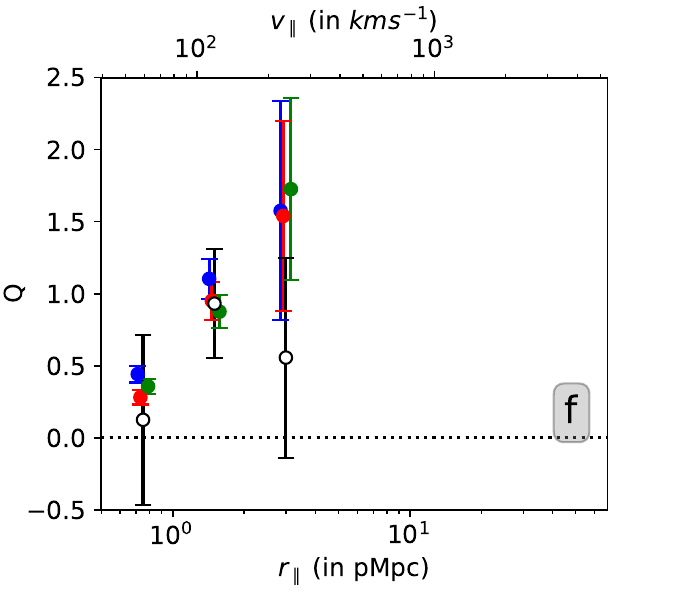}%
	\caption{\textit{Top panels}: Results from MBII simulations. Longitudinal two-point (left), three-point (middle) and reduced three-point (right) correlations of \lya\ absorbers as a function of longitudinal scale in simulations for $N_{\rm HI}>10^{12.5}$cm$^{-2}$.
	Results for simulated spectra with two different SNR and including $b_{\rm turb}$ in the case of SNR=50 are presented together with the observations.
	%
	%We calculate the correlations for two cases: one with non-thermal turbulence $b_{\rm turb}$ added to the line-width parameter $b$ and one without $b_{\rm  turb}$.
	\textit{Bottom panels}: Results from all the three simulations in the Sherwood suite studied here. 
	%is compared with the observations.
	%Effect of wind and AGN feedback on the absorber-based longitudinal two-point (left), three-point (middle) and reduced three-point (right) correlations of \lya\ absorbers as a function of longitudinal scale in simulations.
	%The errors represent 68\% confidence interval about the mean value of correlations calculated for different sightlines. The correlations have been calculated at $z=0.1$.
	}
\label{Corr_sim}
\end{figure*}

\begin{figure*}
 \includegraphics[viewport=0 8 300 290,width=6cm, clip=true]{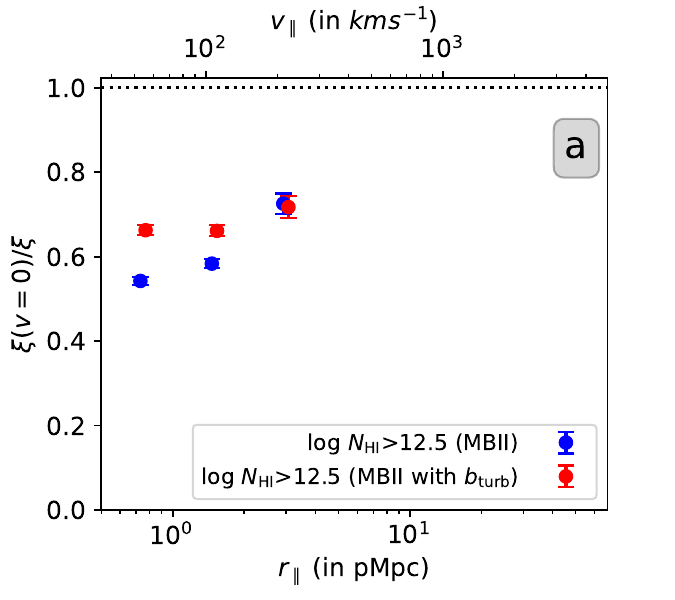}%
	\includegraphics[viewport=0 8 300 290,width=6cm, clip=true]{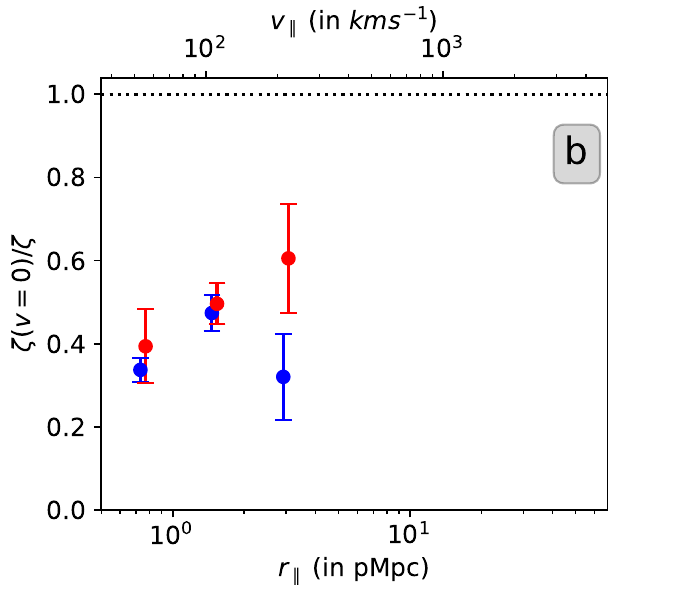}%
	\includegraphics[viewport=0 8 300 290,width=6cm, clip=true]{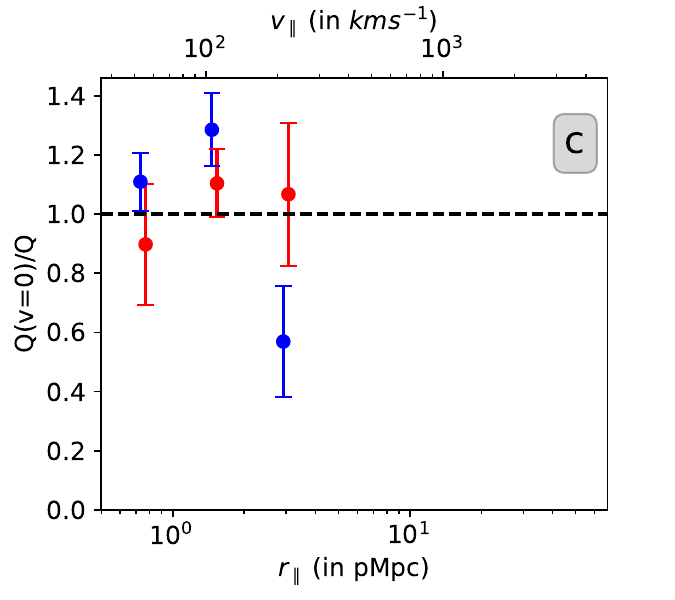}%
	
	\includegraphics[viewport=0 8 300 260,width=6cm, clip=true]{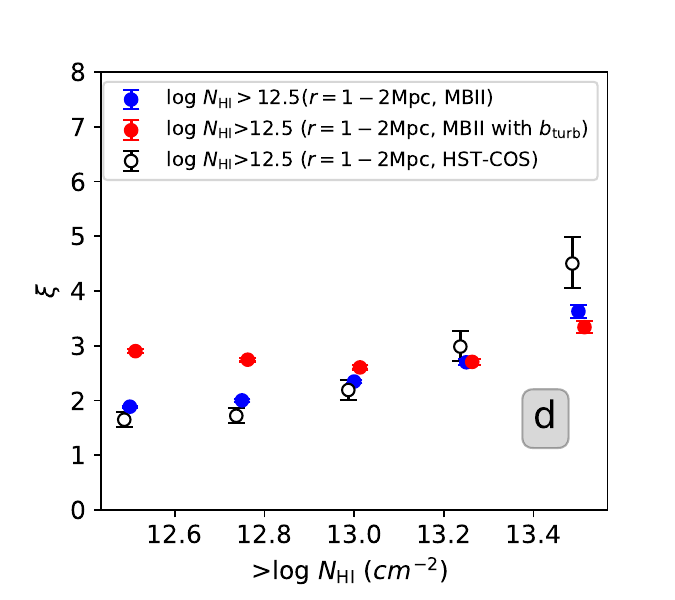}%
	\includegraphics[viewport=0 8 300 260,width=6cm, clip=true]{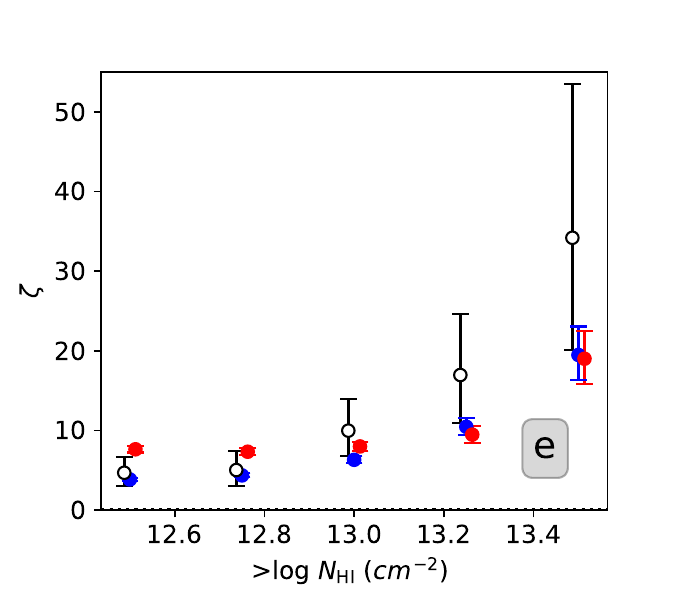}%
	\includegraphics[viewport=0 8 300 260,width=6cm, clip=true]{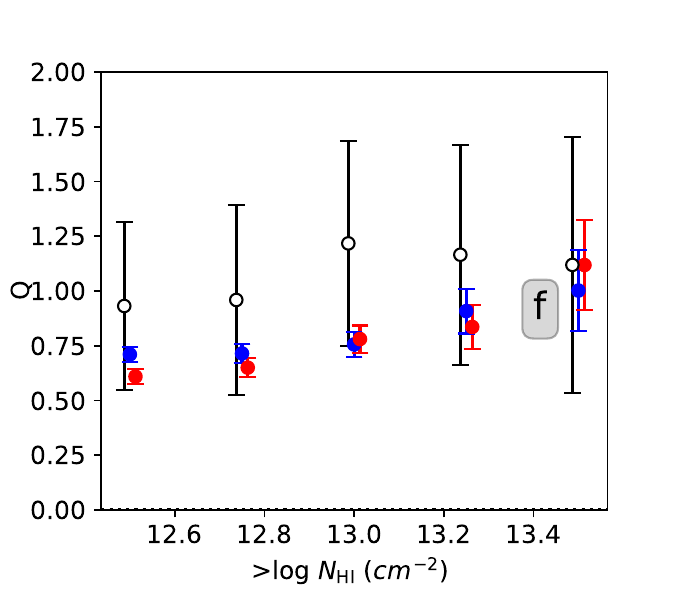}%
	
	\includegraphics[viewport=0 8 300 260,width=6cm, clip=true]{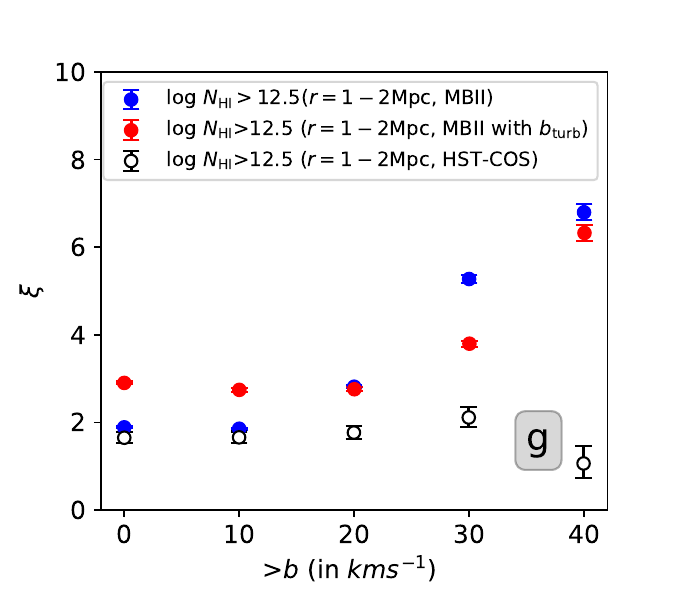}%
	\includegraphics[viewport=0 8 300 260,width=6cm, clip=true]{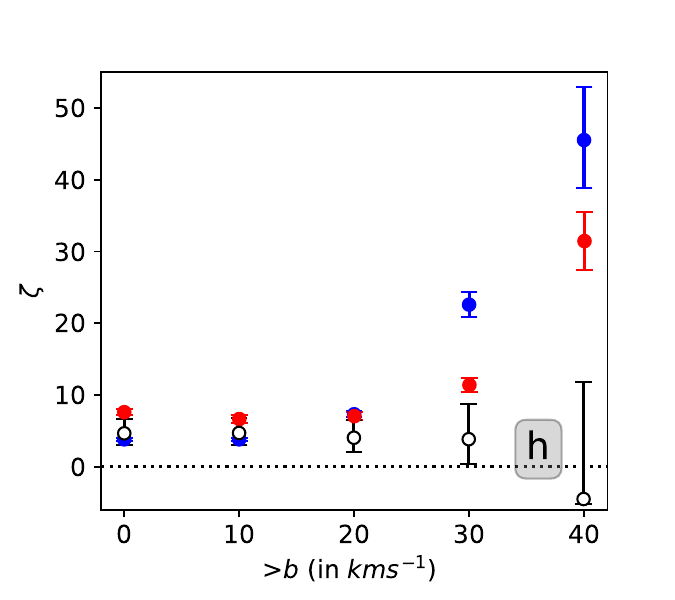}%
	\includegraphics[viewport=0 8 300 260,width=6cm, clip=true]{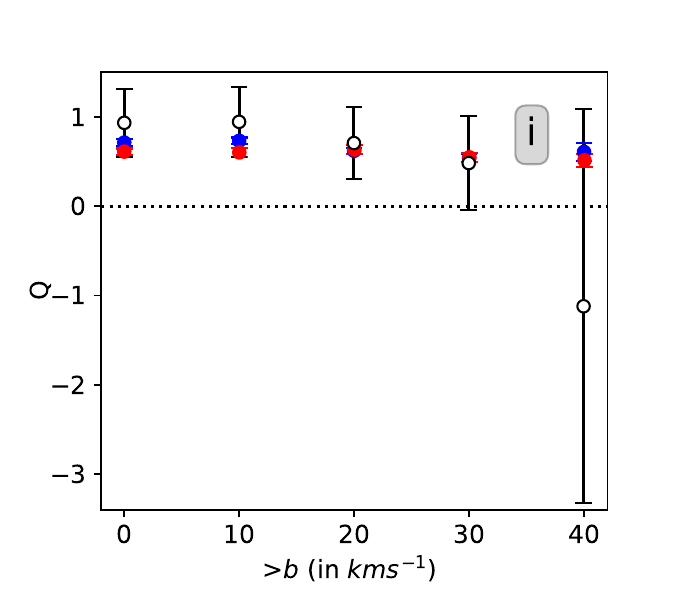}%

	\caption{Dependence of redshift space distortions (top panels), $N_{\rm HI}$ thresholds (middle panels) and $b$ thresholds (bottom panels) on the longitudinal two-point (left), three-point (middle) and reduced three-point (right) correlations of \lya\ absorbers { in MBII simulations}. 
	%as a function of longitudinal scale in simulations for two $N_{\rm HI}$ thresholds (red and blue) . 
	In the top row we show the results as a function of $r_{\parallel}$ for two \NHI\ thresholds. In the bottom two rows we show the results for $r_{\parallel} = 1-2$ pMpc, with and without including $b_{\rm turb}$. 
	%We use MBII simulation to explore redshift space distortion effect.  %The errors represent 68\% confidence interval about the mean value of correlations calculated for different sightlines. The correlations have been calculated at $z=0.1$.
	}
\label{Corr_v}
\end{figure*}

%We take the MBII simulation snapshot at $z=0.1$ and shoot 8000 sightlines through it in random directions. In the case of Sherwood simulations we have used 4000 sightlines.
%We generate \lya\ transmitted flux spectra for these sightlines with and without including $b_{\rm turb}$  for the MBII simulations.
%In the case of Sherwood simulations we do not consider simulations with $b_{\rm turb}$.
%
%
%We also consider the case without $b_{\rm turb}$ and also 
%swithing off any peculiar velocities. We use the spectra from these simulation box to study the effect of peculiar velocities on our clustering measurements.
%
%once by using the peculiar velocities from the simulations and once by turning it off. 
%
%In all case, we fit the \lya\ absorbers with Voigt profiles using {\sc viper} \citep[][]{gaikwad2017b}  and calculate the longitudinal two-point, three-point and reduced three-point correlation using the method described earlier. We do this for $N_{\rm HI}>10^{12.5}$cm$^{-2}$ { absorbers}. 
%
In the left, middle and right panels of Fig.~\ref{Corr_sim}, we plot 
%The observed as well as simulated 
two-point, three-point and reduced three-point correlations respectively
as a function of $r_\parallel$.  
Results from MBII simulations (with and without $b_{\rm turb}$) and Sherwood simulations are presented in the top and bottom rows respectively. {  The errors in the correlations are { one-sided poissonian uncertainty for the data-data pairs} computed over 8000 sightlines in case of the MBII simulation and 4000 sightlines for each of the Sherwood simulations.}
%Top row presents the results for MPII simulations and the bottom row present results for the Sherwood simulations.
%
%It is apparent that 
For both simulations, unlike in observations, there is no suppression
%The first thing that is apparent from this figure is that the 
of two- or three-point correlation in the first bin (i.e $r_\parallel < $ 1 pMpc). 
%\Anand{Is this due to SNR? Soumak: Maybe or maybe due to intrinsically lower b?}
%tends to be higher than the observed value. 
%There is no suppression in the two-point correlation in the first bin. 
%The same trend is noticed even in the case of Sherwood simulations albeit with less excess. 
%
In the case of MBII simulations, the predicted two- and three-point functions roughly follow the observations for $r_\parallel>1$ pMpc.  {In the $r_\parallel=1-2$ pMpc bin, the two- and three-point correlation are found to be $1.65^{+0.13}_{-0.13}$ and $4.76^{+1.98}_{-1.67}$, respectively. In case of simulations, they are found to be $1.88^{+0.02}_{-0.02}$ and $3.83^{+0.19}_{-0.18}$, respectively. }
%{ \color{magenta} PG: I think instead of quoting $\chi^2$ for the two bins, we should state the maximum difference in terms of $\sigma$. $\chi^2$ can be used if we have sufficiently large number of bins e.g., 10 or higher.}
%
When we include $b_{\rm turb}$,  the predicted correlations are higher for $r_\parallel<4$ pMpc. 
{ 
As noted in Fig.~\ref{NHI_b_sim}, in these models there is a reduction in the number of low \NHI\ { absorbers}.
%This is probably related to some of the low-\NHI\ { absorbers} will be excluded by  $>4\sigma$ level requirement. 
Thus we may be probing clustering among the relatively high \NHI\ { absorbers} even when we use the same \NHI\ cut-off when we include $b_{\rm turb}$. This exercise, clearly illustrates that simple addition of a constant $b_{\rm turb}$ in quadrature to thermal $b$ will not provide correct solution to the missing sub-grid physics.} 
%\Anand{If this is true then when we use \NHI$>$13.3 we should not find any diference. Is this true?  Only except the first point, they match. }
%as we use the measured \NHI\ distribution to get the random pairs and triplets.

%For $r_\parallel>1$ pMpc the two-point correlation function obtained for different simulations agree well with each other, though it seems slightly weaker in Sherwood simulations. The simulations with a SNR of 50 are also roughly consistent with the observed values. 
{ In the case of Sherwood simulations, the predicted two-point and three-point correlations are also roughly consistent with the observations as well as with the MBII simulation. In the $r_\parallel=1-2$ pMpc bin, the two- and three-point correlation are found to be $1.84^{+0.04}_{-0.04}$ and $4.38^{+0.57}_{-0.53}$, respectively for the Wind+AGN case.}
%lower than those predicted by MBII simulations ({  The two-point correlation is lower by about $\sim$27\% at the scale of 1-2 pMpc and decreases to about $\sim$9\% at the scale of 2-4 pMpc (for Wind+AGN case). The three-point correlation is lower by about $\sim$44\% at the smallest scale of 1-2 pMpc and decreases to about $\sim$36\% at the scale of 2-4 pMpc}).
%They are also slightly lower than the observed distributions but within 2$\sigma$.  }
In the case of two-point correlations, the difference between the predictions of the  three Sherwood simulations are { small (within $\sim$13\%)}. 
In the case of three-point correlation the dispersion between the three models is slightly higher ({  within $\sim$17\%}) but still consistent within errors.  %\Anand{Even for the last two lines we can give some quantitative number. For example, you can give spread in percentage.}
{Therefore, it appears that various feedback process included in the Sherwood simulations produce little difference to the predicted $\xi$ and $\zeta$ at various scales}.
%Sherwood simulations also show similar trend with no strong suppression in the first bin and roughly following the observed distributions at large scales. 
%
%{  Why does High SNR match better with observation?}

{\it In summary, the simulations considered here roughly reproduce the observed profile of $\xi$ and $\zeta$ at scales greater than 1 pMpc . 
{This could just be the mere reflection of the large scale \lya\ distribution being consistent with matter distribution in the $\Lambda$CDM models. The difference between the two simulations can come from slightly different cosmological parameters used (see initial paragraphs of section~\ref{sec:simulations} for details). }
However, the models fail to reproduce the large suppression we notice for $r_\parallel<1$ pMpc in the observations. } The suppression in the first bin could originate from: (i) difference in the SNR between simulations and observations; (ii) differences in density field at small scales like presence of excess smoothing in the observed data { (as suggested by $b$-distribution)} and (iii) differences in the line fitting routine used.  Note the simulated spectra have higher SNR and typically individual components have lower $b$-values. This helps in the component decomposition of the blended profile and this could provide higher measured correlations at the lowest velocity bin. 
We confirm that the difference is not due to our automatic Voigt profile routine {\sc viper} ({  which we use for fitting simulated spectra}) as our fits to the observational data produce consistent results to what we get from the line list of \citet{danforth2016} ({  which we have used for calculating observed correlations}).
{Our analysis of simulated data with SNR = 12 did not show large suppression in correlation function at $r<1$ pMpc.}
Therefore, it is possible that the simulations do not capture the density distribution at small scales {(i.e missing sub-grid physics)}. This could also be the reason for the simulated $b$ distribution being different from the observed one as seen in Fig~\ref{NHI_b_sim}.
%
%
%However, for the intermediate scales (i.e 1 to 4 pMpc) the $\xi(r)$ and $\zeta(r)$  from simulations roughly follow the observed profile. 
%

%We also show the results from simulations with sub-grid micro-turbulence added to the line width parameter in Fig~\ref{Corr_sim}. As expected in the first bin, where the effect of exclusion due to blending is high, we do find both $\zeta$ and $\xi$ show a decrease in the case of models with turbulence. However  the difference between the predictions of these two models is very minor 
%in other $r_\parallel$ bins.

\subsubsection{Reduced three-point correlation function (Q):}

\citet[][]{maitra2020}, in their $z\sim 2$ simulations, have found that the reduced three-point correlation function is less sensitive to the astrophysical parameters compared to $\xi$ or $\zeta$. 
In panels (c) and (f) of Fig.~\ref{Corr_sim}, we plot the Q values predicted in simulations as a function of $r_\parallel$. {In the observations,  we do not see any evidence for the scale dependent Q due to large measurement errors.
%value thanks to large errors in the Q measurements.
%
However,
it is evident that the simulated Q values increase with increasing scale { in the case of MBII simulation. Similar trend is also seen for Sherwood simulations, but with large errorbars.} 
%consistent with what we measure from observations (also shown in Fig.~\ref{Corr_sim}). 
%
%Interestingly, the observed and predicted Q values agree with each other very well for $r_\parallel<$ 1 pMpc despite having differences for $\xi$ and $\zeta$. 
%
In the case of MBII, Q values tend to be lower in the model with $b_{\rm turb}$ (i.e Q = $0.61^{+0.04}_{-0.04}$) is
lower than that without $b_{\rm turb}$ (i.e Q = $0.71^{+0.04}_{-0.04}$)
%{  ( Q(with $b_{\rm turb}$) = $0.62\pm 0.05$ and Q (without $b_{\rm turb}$) = $0.72\pm 0.06$ 
at $r_{\parallel}=1-2$ pMpc bin.}

We also notice Q values measured in MBII are slightly lower than what is found in the case of Sherwood simulations {(Q for Sherwood simulation having Wind+AGN feedback is $0.88^{+0.12}_{-0.12}$ at $r_{\parallel}=1-2$ pMpc bin)}. 
%\Anand{It will be good to quantify this in the two bins you discuss for the correlations.}
%
%The measured Q values are nearly constant in the first three distance bins albeit with large errors. 
%For $r_\parallel$ = 1-2 pMpc the measured {Q$=xxx\pm yyy$} and the same for 2-4 pMpc bin is {$xxx\pm yyy$}. 
%The Q values also match with observations at large scales. However, observed Q has large errors at large $r_\parallel$.
%We notice that the actual value of Q at large distances (i.e $r_{\parallel} = 2- 4$ pMpc) measured in MPII are less than those found for Sherwood simulations.
However, thanks to large errors,  Q values predicted by different simulations are consistent with observed values  {($0.95^{+0.39}_{-0.38}$) at $r_{\parallel}=1-2$ pMpc bin}.
%and both simulations produce  slightly higher Q values than the mean found in observations. However observed Q has  large errors. 
%
%We have found Q for the \lya\ absorbers in the observed sightlines to be $2.2\pm 0.8$ at the scale, $r_{\parallel}$ = 1-2 pMpc.
%
%The dependence on scale and $N_{\rm HI}$ is found to be weak. 
%
%The simulations, however, predict a strong increase in Q with increasing scale. It is also found to decrease with increasing $N_{\rm HI}$ threshold above $N_{\rm HI}>10^{12.75}$cm$^{-2}$. The observed and simulated Q match only at the scale of 1-2 pMpc. MBII simulation predicts a Q value of $\sim 1.8$ at this scale for $N_{\rm HI}>10^{12.5}$cm$^{-2}$ { absorbers} while Sherwood predicts Q values of 2.1 and 2.7 with and without feedback respectively. Also, Sherwood seems to have a slightly steeper radial profile of Q in comparison to MBII. The Q value in the Sherwood simulation at the 0.5-1 pMpc bin seems to be much smaller in comparison to the MBII simulation. Discrepancies at such scales can be ascertained to gas physics and not the cosmology of the simulations involved.
We need to keep in mind that MBII and Sherwood simulations use slightly different cosmologies (i.e $\Omega_m$ and $h$ differ by $\sim$10\% and $\sigma_8$ is consistent within 2\%). 
%We  do see Q predicted in Sherwood simulations are slightly higher than those from MBII simulations.
We notice that Q also depends on the feedback process included in the simulations (see panel (f) in Fig.~\ref{Corr_sim}) . The 80-512 Sherwood simulation produces slightly large Q values compared to those include feedback effects. {\it However, these differences are smaller than the difference in Q we see between different scales in simulations.}

\subsubsection{Effect of peculiar velocities:}
In this section, { using MBII simulations,} we study the effects of the line of sight peculiar velocities on the measured correlation functions.
For this case, we switch off all the peculiar velocities (considering only the cosmological expansion and thermal broadening effects) while generating the mock spectra. We then fit Voigt profiles to these spectra to identify individual components for our clustering analysis. 
%We use MBII simulations only for this exercise.
%
%We plot this for two cases: one with non-thermal turbulence $b_{\rm turb}$ added to the line-width parameter $b$ and one without $b_{\rm  turb}$. We find that $b_{\rm  turb}$ has a minor effect of reducing correlations in the first radial bin only (0.5-1 pMpc) . We find the scales for positive two-point and three-point correlations to be similar to observations. However, the radial profile of the correlations are very different from observations. In simulations, we do not see the small scale suppression due to exclusion that is seen in observations. The reduced three-point correlation Q in simulation seems to match with observation at the scale of 1-2pMpc. We find that Q has a strong scale dependence and seems to increase with scale, which was not the case in observation. 
%
%Next, we explore the effect of redshift space distortions on two-point , three-point and reduced three-point correlation 
In the top panels of Fig.~\ref{Corr_v}, 
we present the ratio of measurements without and with the inclusion of line of sight velocity fields for MBII simulations.

%First of all 
The presence of line of sight velocity field is found increase both $\xi$ and $\zeta$ for length scales of our interest. 
%The effect is  stronger at the lowest velocity bin. 
%
%Interestingly, we find that redshift space distortions have a significant impact on the magnitude of both two-point and three-point correlations at both the $N_{\rm HI}$ thresholds. In both the cases, redshift space distortions seem to amplify the signal siginificantly at smaller scales. 
At the scale of 1.0-2.0 pMpc, two-point and three-point correlations are amplified by a factor of $\sim 1.7$ {  and 2.1, respectively} by the line of sight velocity field. {This could imply the \lya\ absorbers are part of converging flows}.
%\Anand{Soumak put exact value for the above two numbers}. 
On the other hand, presence of line of sight velocity field makes Q weaker at smaller scales ({  $r_{\parallel}<2$pMpc}).
%{ , although with a relatively lesser magnitude}. 
At the scale of 1.0-2.0 pMpc, redshift space distortion reduced the intrinsic Q value by a factor of $\sim 0.3$ ({ though with large errorbars}). { We confirm similar trends in Sherwood simulations also (not shown here). Due to large errors we did not probe the effect of peculiar velocity on Q at large scales (i.e $r>4$ pMpc).} 

The discussions presented here suggests that the actual matter clustering is {  smaller (roughly 60\% and 50\% in the case of two- and three-point correlations, respectively)} than what we measure in the redshift space without accounting for the peculiar velocities. {  The actual Q values are slightly larger at smaller scales. This trend is similar to what is found in the higher order clustering of galaxies. As shown in Fig.4 of \citet{mcbride2011a}, the projected measurements of Q are found to be slightly larger than the redshift space measurements at smaller scales. At larger scales, they roughly follow each other. \citet[][]{Marin2008} reports similar trends based on the real space vs redshift space distribution of galaxies from cosmological simulations.}
%{We also find the scale dependence of Q  remains even when we consider no peculiar velocities. Therefore, we conclude that the scale dependent Q is not an artefact of peculiar velocities.}
%\end{comment}
%
%\Anand{Quote the ratio at different scales over which we measure correlations in the observed data. We should use this to interpret the observations after removing the RSD.}

\subsubsection{ Dependence on $N_{\rm HI}$}

Next we study the $N_{\rm HI}$ dependence of simulated correlation statistics. In panels d-f in Fig.~\ref{Corr_v}, we plot two-point($\xi$), three-point ($\zeta$) and reduced three-point (Q) correlation functions measured  at $r_\parallel$ = 1-2 pMpc as a function of $N_{\rm HI}$ thresholds. In panel (d) and (e), we find that the two- and three-point correlations steadily increases with increasing $N_{\rm HI}$ threshold. 
While the trend seen is similar to what we see in observations we see the increase in $\xi$ and $\zeta$ with \NHI\ threshold is slightly slower %{  (although consistent within errorbars)} 
as compared to what is seen in the case of observations. {This could be related to the lack of high \NHI\ absorbers in simulations compared to observed distribution.} 
%({  see Fig.~\ref{Corr2}}).

%We also show results with and without $b_{\rm turb}$. 
%
{The measured $\xi$ and $\zeta$ values for low \NHI\ threshold are higher when we include $b_{\rm turb}$. However, the results are consistent for higher \NHI\ thresholds for models with and without $b_{\rm turb}$ included. As discussed before inclusion of $b_{\rm turb}$ reduces the number of low \NHI\ systems detected, which cluster weakly. Therefore, for low \NHI\ thresholds we get enhancement in $\xi$ and $\zeta$. However, \NHI\ distribution is not affected at high \NHI\ thresholds by $b_{\rm turb}$ (see Fig~\ref{NHI_b_sim}). Therefore, $\xi$ and $\zeta$ values are similar at high \NHI\ thresholds. This  also suggests that fixing the shortcomings of the simulation (i.e sub-grid physics) using a constant $b_{\rm turb}$ term is not the correct solution.
}
%The measured $\xi$ values in this case are slightly higher at low \NHI\ and consistent at high \NHI\  with models without $b_{\rm turb}$. This is consistent with we missing some low \NHI\ component due to incompleteness.
%This could be related to the lack of high column density systems in the simulation we notice in Fig.~\ref{NHI_b_sim}.
%

%In panel (e) of Fig.~\ref{Corr_v}, we find that the three-point correlation increasing with increasing \NHI\ threshold. This trend is again consistent with what we saw in our observations. As in the case of $\xi$, the increase of $\zeta$ with \NHI\ is slower than what we see in observations. This could be related to the number of high \NHI\ { absorbers} being deficient in our observations. In particular, as we go to higher \NHI, the number of absorbers we see per simulated sightline becomes very small. 
%Therefore larger simulation boxes are needed to explore this in more detail.
%We also notice that inclusion of $b_{\rm turb}$ produce very minor effects on $\zeta$.

The dependence of Q on \NHI\ threshold is shown in panel (f) of Fig.~\ref{Corr_v}. The Q remains nearly constant and  consistent value with observations. As expected the inclusion of $b_{\rm turb}$ reduces the Q values in the low \NHI\ range.

%Same trend is seen for Q in panel f. This is contrary to what is found in observations, where the two-point and three-point correlation function steadily increases with $N_{\rm HI}$ and Q remains constant.

\subsubsection{Dependence on $b$-parameter}

In observations, we found the two- and three-point correlation function is independent of $b$-threshold for low threshold values (see Fig.~\ref{Corr2}). We did find the suppression when we consider systems with $b>40$ \kms (i.e high-$b$ absorbers).
When we consider high- and low-$b$ sub-samples (see Fig.~\ref{Corr_b}) three-point correlation was not detected in the case of high-$b$ sub-sample.
In panels (g)-(i) in Fig.~\ref{Corr_v}, we plot two-point, three-point and reduced three-point correlation functions at a scale of 1-2 pMpc as a function of $b$ thresholds. We show the results from MBII simulations with and without addition of $b_{\rm turb}$. It is evident from this figure that unlike observations both two- and three-point correlation functions show monotonic increasing with increasing $b$-threshold even for the high-$b$ absorbers. We also notice that Q has very weak dependence on $b$ threshold.

{ When we include $b_{\rm turb}$ as expected the $\xi$ and $\zeta$ are independent of limiting $b$ when it is less than 20 \kms. However shows monotonous increase with increasing $b$ afterwards. In particular detection of positive $\zeta$ for $b>40$ \kms\ is intriguing. We believe the difference is dominated by two possible effects (i) higher SNR used in our simulations and (ii) the differences in the $b$-distribution between simulations and observations noted before. That is a given b-threshold will pick typically high \NHI\ systems in the simulations compared to observation (see Fig~\ref{NHI_b_sim2}).}

%Inclusion of $b_{\rm turb}$ reduces the amplitude of both two- and three-point correlations (at large scales) for large $b$-thresholds but unlike observations the trend is monotonous.
%
%In panel (g), we see that the two-point correlation steadily increases with $b$, contrary to observations, where we see a dip above 40 \kms. Inclusion of $b_{\rm turb}$, reduces $\xi$ at a given $b$ threshold. However the trend remains same.
%
%In panel (h), the three-point correlation increases up to $b>20$\kms and above that, it starts to fall down. This trend in three-point correlation is similar to what is seen in observations.
%When we include $b_{\rm turb}$, the peak in $\zeta$ occurs at a higher $b$-threshold.
%
% Similar trend is seen in Q in panel (i) .
% 

\begin{figure*}
 \includegraphics[viewport=0 7 300 285,width=7.2cm, clip=true]{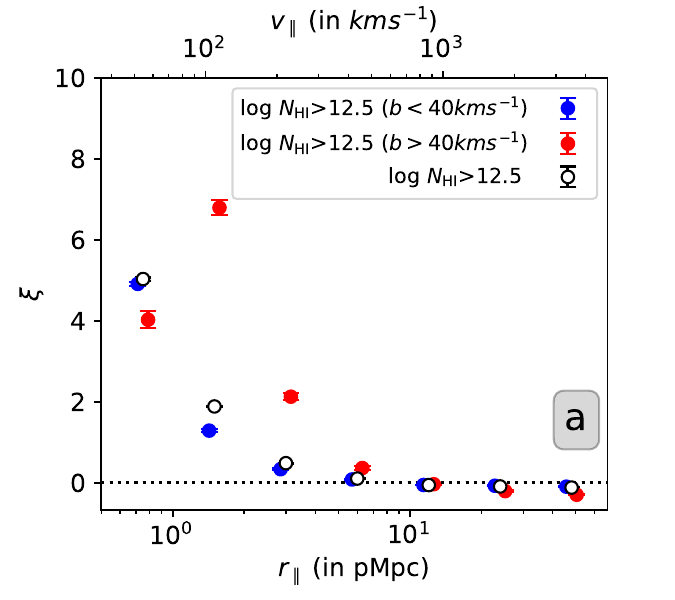}%
	\includegraphics[viewport=0 7 300 285,width=7.2cm, clip=true]{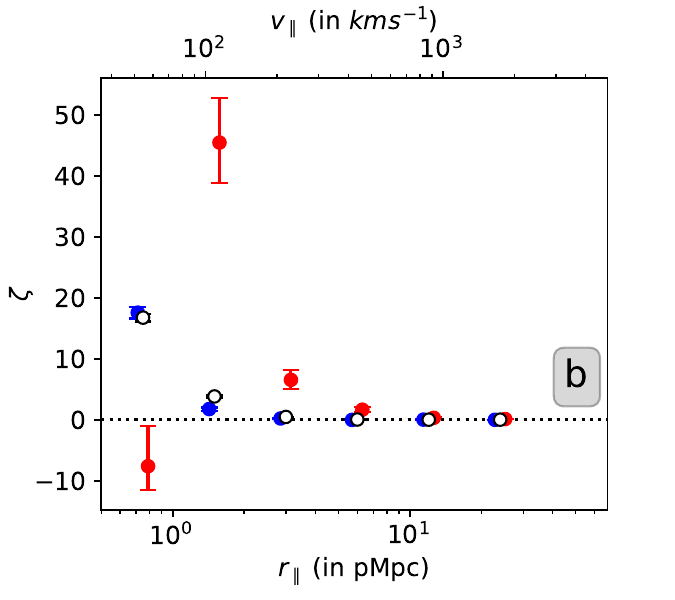}%

	\caption{Two-point (panel a) and three-point (panel b) correlations of \lya\ absorbers in MBII simulation for full sample, high-$b$ ($b>40$ \kms) and low-$b$ ($b<$ 40 \kms) sub-samples. 
	%The errors represent 68\% confidence interval about the mean value of correlations calculated for different sightlines. Results for the full sample is also shown for comparison. 
	Unlike in the case of observations (see Fig.~\ref{Corr_b}) high-$b$ { absorbers} show stronger two- and three-point correlations at large scales.}
\label{Corr_b_sim}
\end{figure*}

%Clearly the trend seen for $b$-dependence is very different from what we have found from observations. 
To investigate this further in Fig~\ref{Corr_b_sim}, we plot the radial dependence of two- and three-point correlation functions for low- and high-$b$ sub-samples from our simulations (similar to Fig~\ref{Corr_b} for observations).
{For $r>1$ pMpc, two and three-point correlation function of high-$b$ systems are found to be larger than those of  low-$b$ systems. We notice that when we consider SNR = 12 simulations, as in observations we do not detect positive three point correlations for $r<4$ pMpc. However we notice that these models have much larger two-point correlation at r = 1-2 pMpc bin. Therefore we conclude that while low SNR in the observations have a role to play in suppressing the three-point correlation in the case of high-$b$ absorbers this alone can not explain the increase in $\xi$ and $\zeta$ we find as a function of $b$-threshold.}

\section{Main results \& discussions:}

We report the detection of longitudinal (redshift space) three-point correlation function ($\zeta$) of $z\le0.48$ \lya\ absorbers using HST-COS spectra of 82 QSOs compiled by \citet{danforth2016}. We study the dependence of $\zeta$ on \NHI, $b$-parameter, $z$ and presence of metals. We also correlate the \lya\ absorbers at $z<0.2$ contributing to the three-point correlation function with the galaxy distribution for 41 of these sightlines.
We compute \lya\ clustering using four different simulation boxes at $z\sim0.1$ and
compare them with observations.
%explore whether the observed trends are also reproduced in these simulations. 
%In particular, we quantify the effect of redshift space distortion on the measured two- and three-point correlation functions. 
Here we discuss the main results of our study.
%

%\begin{enumerate}
\vskip 0.1in
\par \noindent{\bf Clustering properties of low-z IGM:}
 %We report the detection of longitudinal (i.e redshift space) three-point correlation function from the low-$z$ \lya\ forest. 
 {  
 \citet{danforth2016} has shown that the low-$z$ \lya\ absorbers cluster among themselves up to a scale of $\sim$10 pMpc with a positive two-point correlation.
 %that is consistent with the measurements of \citet[][]{danforth2016}. 
 Using the same data,
 we report for the first time the detection of a non-zero probability excess in \lya\ triplets ($\rm PE_3$) up to 8 pMpc, with the strongest detection ($\rm PE_3=8.8^{+2.0}_{-1.7}$ for { absorbers} having \NHI$\ge 10^{12.5}$ cm$^{-2}$) coming from the $r_\parallel = 1-2$ pMpc bin with equal arm configuration at a significance level of $\sim 5\sigma$.
 In the case of 1:2 arm ratio configuration, we find $\rm PE_3 = 2.1^{+0.5}_{-0.4}$ (4$\sigma$ level) for $r_{\parallel,1} = 1-2$ pMpc and $r_{\parallel,2} = 2-4$ pMpc. Using the triplet probability excess we obtain positive longitudinal three-point correlation ($\zeta$) at scales below 4 pMpc. The strongest detection is at the scale of 1-2 pMpc with equal arm configuration ($\zeta=4.8^{+2.0}_{-1.7}$) for \NHI$\ge 10^{12.5}$ cm$^{-2}$). 
 {\it These are the first reported measurements of three-point correlation function from the \lya\ forest observations and using Voigt profile decomposed components.} We do not detect any three-point correlation for 1:2 arm ratio configurations.
The measured amplitudes of \lya\ three-point correlation is at least an order of magnitude weaker than that measured for galaxies
($\zeta\sim 200$ for $M_r<-19$ galaxies at $r\sim 1$ pMpc and $\theta=0$ for $r_1:r_2=1:2$ configuration; see \citet{guo2016}).
%
%\citep[see figure 1 of][for r$_1$ = 1 Mpc and $\theta=0^\circ$]{guo2016}.
}
 %
 %\Anand{Update the two point, three point clustering values and Q. Also write about the scales here itself. Compare them with the values from galaxies. This discussion can be on the scale as well. For $theta = 0$ deg and $theta = 90$ deg. Mention that at high column density threshold it gets closer to the galaxy values.}
%
{%We measure the two-point correlation function ($\xi$) for different scales probed in this study and show them to be consistent with the measurements of
%of $\xi$ reported by 
%\citet[][]{danforth2016}. 
%two-point correlation function together with the three-point correlation function was 
Using our measured $\xi$ and $\zeta$ we obtain the reduced three-point correlation function Q ($0.95^{+0.39}_{-0.38}$) for $r_\parallel = 1-2$ pMpc bin and equal arm configuration . This is similar to the Q values measured in the case of low-$z$ galaxies \citep[see figure 4 of][for $\theta =0$ for $r_1:r_2=1:2$ configurations]{mcbride2011a}.
%However, our measurement errors are large and it will allow $Q = 1.3$ within 1$\sigma$ level.
}

{ Both two- and three-point correlations are suppressed for $r_\parallel <$ 1.0 pMpc 
{  (or $r_\parallel <$72\kms\ at $z\sim0.1$).} 
%
%\Anand{What did people say about this suppression in the early paper. It will be good to cite these papers. You may find them from my talks on low-z clustering I gave some time back.}.
This suppression is related to the efficiency at which 
%we will be able to perform
Voigt profile decomposition can be performed at these scales. This depends on the matter distribution at small scales, instrumental resolution, {  line blending}, SNR of the spectra used \citep[see][]{danforth2016}. Measuring $\zeta$ and $\xi$ in the transverse direction using closely spaced quasar sightlines will be important to have further insights into this issue.

For $r_\parallel>1$ pMpc, both two- and three-point correlations show a decreasing trend with increasing scale. {  However, we do not notice any scale dependence of Q for $r_\parallel > 1$ pMpc range in the observations, thanks to large measurement errors.
%Below this, in the region of exclusion, the Q value is suppressed. 
In the case of galaxies Q values are measured over much larger scales (i.e 3-20 pMpc) and found to be scale independent when peculiar velocity effects are taken care of \citep{mcbride2011b}. }
 }

 %   \item{
 %   In this work, we have used low-redshift HST-COS spectra to study higher order clustering properties of \lya\ absorbers or "{ absorbers}" in the redshift space. As a sanity check, we have match the logitudinal two-point correlation function of these { absorbers} with previous studies based on HST-COS spectra. We report the first observational signatures of non-gaussianity in \lya\ forest with a $3.1\sigma$ detection of longitudinal three-point correlation amplitude of $10.3\pm 3.3$ at the scales of 1-2 pMpc for $N_{\rm HI}>10^{12.5}$ \lya\ { absorbers}. The reduced three-point correlation at this scale is found to be $2.2\pm 0.8$. We detect three-point correlation upto a length scale of 3 pMpc beyond which it is found to be 0. Within these scales, we find the reduced three-point correlation to be constant within the errobars. The three-point correlation power is suppressed at scale below 1 pMpc due to exclusion effect.
 %   }
   
   \vskip 0.1in
    \par\noindent
    {\bf Dependence of \NHI:} {  As we discussed in the introduction, it is known that the $N_{\rm HI}$ of an absorption feature is related to the underlying baryonic (and dark matter) overdensities. At low-$z$, it is also known
that high \NHI\ systems ($N_{\rm HI}>10^{14}$cm$^{-2}$)  strongly cluster around galaxies while the weaker absorbers are distributed more randomly or associated with low density IGM or galaxy voids \citep{Penton2002,Tejos2014}. It is also known that the stronger $N_{\rm HI}$ absorbers have larger two-point correlation compared to low \NHI\ absorbers \citep[see][for example]{Penton2002,danforth2016}. { In this work, we find that like two-point correlation the three-point correlation function of the \lya\ absorbers also shows an increasing tendency with  increasing $N_{\rm HI}$ thresholds for a given scale probed}. Compared to $\zeta=4.8^{+2.0}_{-1.7}$ for $N_{\rm HI}>10^{12.5}$cm$^{-2}$ absorbers, we obtain $\zeta=34.2^{+19.4}_{-14.1}$ for $N_{\rm HI}>10^{13.5}$cm$^{-2}$ absorbers. }
    %It is expected that with stronger absorbers, one will get even stronger three-point correlation similar to what one gets in the case of clustering in galaxies 
     On the other hand, the reduced three-point correlation Q is relatively independent of the $N_{\rm HI}$ thresholds. 
    %However, it is interesting to note that in high-$z$ simulations, we found Q measured in the transverse direction to increase with increasing  \NHI\ threshold \citep[see figure 5 of][]{maitra2020}. It will be interesting to measure the \NHI\ dependence of Q using longitudinal correlations at high-$z$.
  %  \Anand{We need to cite the low-z papers where they claim clustering depends on \NHI. Some discussion should come here.}

    \vskip 0.1in
    \par\noindent{\bf Dependence on b-parameter:}
    { 
    %We also study the dependence of clustering on the  $b$-parameter. 
    {  While the low-$b$ \lya\ absorbers ($b<40$ \kms) probe the cool photoionized gas, some fraction of the high-$b$ \lya\ absorbers ($b>40$ \kms) could arise from the highly ionized warm hot intergalactic gas or WHIM \citep[][]{Richter2006,tepper2012}. Atleast 20\% of the total baryonic content of the Universe is located in highly ionized WHIM and the cool photoionized gas comprises of about 30\% of the baryons \citep{Lehner2007}. A $b$-dependent clustering study essentially allows us to probe matter distribution as one transitions from one gaseous phase to another. } 
    We find the two-point correlation (probed over 1-2 pMpc) to remain nearly constant for low-$b$ thresholds but shows a decreasing trend when we consider absorbers with $b>40$ \kms.
    This trend is also seen in the case of three-point correlation albeit with a much sharper decrease. %{  Soumak: Comparison between observation and simulation? }
    %In this case,
    %increases with $b$ thresholds at the scale of 1-2pMpc upto $b>40$\kms\ above which it starts decreasing. In case of three-point correlation, it increases upto $b>30$\kms\ and then starts decreasing. The decrease at $b>40$\kms\ is not significant and 
    %there is a sharp decrease in three-point correlation.  
    When we consider high- and low-$b$ sub-samples based on $b$-parameter 
    %(i.e high-$b$ systems with $b>40$ \kms and low-$b$ systems with $b<40$ \kms) 
    the measured radial profiles of two- and three-point correlations are found to be very different. While low-$b$ sub-sample shows positive three-point correlation for  $r_\parallel<$ 4 pMpc no correlation is detected for high-$b$ sub-sample over the same scale.
    The lack of three-point correlation among high-$b$ systems could be related to biases involved in detected three such absorbers in smaller velocity separations using automatic Voigt profile fits. Therefore, to draw any useful physical conclusions it will be important to confirm this result with high SNR data. 
    { Just to quantify this, we considered the number of high-b ($b>40$ \kms) triplets seen in 8000 sightlines (i.e $ dz\sim 290$ which is $\sim$ 15 times the redshift path of the observed sample used here)  of simulated data for two SNR values (i.e 12 and 50). In the case of SNR of 50, we identify 4 and 88 high-b triplets in the first two r bins (i.e 0.5-1.0 and 1.0-2.0 pMpc). However, for SNR of 12, we do not find any triplet. This clearly demonstrates that, while having a larger path length is always useful for detecting the signal at a higher significant level, better SNR is crucial for the identification of the high-$b$ triplets. 
 }

    }
    %The low $b$ dependence of two-point and three-point correlation can be attributed to the existence $N_{\rm HI}-b$ correlation at small $b$ values. The increase in correlation amplitudes with $b$ thresholds simply reflects the increase with $N_{\rm HI}$ thresholds at small $b$ values.
    %\Anand{I think in the beginning of this section you should review what is the origin of BLA and suggested from the literature study. You may look at Lehner's paper and paper by Richter et al., 2006 and Tepper-Garcia et al., 2012. Then how do you fit this into the clustering property you see.}

 \vskip 0.1in
 \par\noindent
    {\bf Metal components and clustering:}
  {  \citet{danforth2016} have shown that { metal bearing} \lya\ absorbers cluster strongly compared to those without detectable metals. Therefore, we investigated whether the three-point correlation we see comes mainly from these { metal bearing \lya} components. For this we identified \lya\ absorption with associated 
    \CIV, \OVI\ and \SiIII\ absorption (with rest equivalent width more than 30 m\AA).
    We studied the \lya\ clustering excluding these components with metal detection. This exercise confirmed that our clustering measurements are not dominated by the { metal bearing \lya\ absorbers}. As metal line systems are sparse it is not possible to probe the three-point correlation between these systems using the present data.
    }
    %dependence of  %possibility of 
    %two-point and three-point correlation on the presence of metals. 
    %at low redshifts to originate mainly from metal enriched regions. We based our study on .
   % Only a small fraction of the \lya\ absorbers are found to be associated with these metal lines. Also only 40\% of the identified triplets have associated metals.
    %We find that the \lya\ correlation we measure  are also contributed by absorbers without detectable metal lines. 
    %{  This is consistent with the findings in \citet{shen2010} that majority of the metals are locked in stars (80-90\%) at low-$z$ as opposed to high-$z$, where larger proportions of metals are present in the IGM due to more efficient wind escape.
    %do not contribute significantly to the overall two-point and three-point correlation of \lya\ absorbers
    %}
    %\Anand{Again: there is no discussions here. See what people say about where from these metals orginate and try to connect that withthe clustering properties we see...}
    
 \vskip 0.1in
 \par\noindent{\bf Connection to Galaxy distribution:}
    {Since the distance scale probed in our study corresponds to velocity ranges consistent with what one expects in galaxy halos it is also important to see the contribution of CGM absorption to the measured clustering. We use the  catalog of $z<0.2$ galaxies around the 41 quasars available in the literature \citep{keeney2018,prochaska2011} to study the relationship between \lya\ triplets and galaxy distribution.} 
    {We find majority of \lya\ triplet systems at $z<0.2$ to have nearby  galaxies. For the chosen triplet configurations ($r_1=r_2=0.5-1$ pMpc, $1-2$ pMpc, $2-4$ pMpc and $r_1=0.5-1$ pMpc, $r_2=1-2$ pMpc), we find that 88\% of the triplets have at least one nearby galaxy within a velocity separation of 500 \kms. The impact parameters of these galaxies range from 62-3854 pkpc (median of 405 pkpc).  Therefore, a good fraction of triplets originate from impact parameters that are inconsistent with them being associated to a single galaxy. { We also find that the impact parameters of isolated \lya\ absorption is statistically  larger than that of the doublets and triplets. This suggest that multiple component \lya\ absorbers are more closely related to galaxies than the isolated absorbers.
    All these are consistent with the clustered \lya\ regions being closer to the galaxy distributions.
    } We also find that $b>40$ \kms BLA { absorbers} occur more frequently for triplet systems ($\sim 85$\%) in comparison to individual absorbers ($\sim 31.9$\%) .
    As we mentioned before,
    \citet{Wakker2015} found that the \lya\ absorption with 
\NHI $\geq10^{13}$ cm$^{-2}$  have the filament impact parameter less than 2.1 pMpc and all BLAs are found to be located within 400 pkpc to the filament axis and all the absorbers showing multiple velocity components are located within 1 pMpc to the filament axis. 
\citet{Tejos2016} have also found the BLAs have high detection rate close to the filament connecting cluster pairs in the redshift range $0.1<z<0.5$.
Thus our results are consistent with clustered regions originating from filament like structures. Establishing such a connection will allow us to interpret the redshift evolution of $\zeta$ in terms of evolution on large scale structures. { However, we need more sensitive galaxy observations over a wider projected scale compared to what has been used in this study.}
%\citep[as also found by][]{Burchett2020}}.

 %   All these are in line with the observations by \citet{Wakker2015} that most \lya\ absorbers originate from the filamentary structures with BLAs and multiple absorbers having low impact parameters to the filament axis compared to the low-$z$ isolated absorbers.
    }

\vskip 0.1in    
\par\noindent{\bf Clustering in the simulated data:}    
    The four different hydrodynamical simulations considered here roughly reproduce the scale dependence of of $\xi$, $\zeta$ and Q for $r_\parallel >1$ pMpc (within 2$\sigma$ level). {This is could just be the reflection of matter distribution at large scale being consistent with $\Lambda$CDM predictions.}
    At the smallest scale, the simulations (performed at SNR $\sim$ 50) do not show suppression as seen in the data. {  The simulations also show a similar dependence of $\xi$, $\zeta$ and Q on \NHI\ as the observations.} However, the rate of increase of $\zeta$ and $\xi$ with \NHI\ is found to be lower in simulations. We also note that different feedback processes have little effect on the observed clustering.
    Unlike observations, simulations show a monotonic increase in $\zeta$ and $\xi$ with $b$-parameter.  This differences could be linked to the fact that none of the simulations correctly reproduce the observed distribution of $b$-parameter and \NHI\ distribution at high \NHI\ range.
    While low SNR of the observed data may explain lack of three-point correlation among high-$b$ systems, this alone can not explain the $b$-dependent trend found in observations.
  %  We also find that \lya\ absorption from the metal polluted regions cluster strongly compare to the absorption from less polluted regions.

      { Using our simulations we find that the line of sight peculiar velocities tend to amplify the observed $\xi$ and $\zeta$ by a factor of $\sim$ 1.7 and 2.1, respectively, compared to the real space clustering. The effect on Q value is smaller. The Q values are seemingly suppressed by $\sim
      $30\% within $r_{\parallel}=2$pMpc with large errorbars.} {  This is similar to the findings in \citet{mcbride2011a} for clustering of SDSS galaxies, wherein the projected measurements of Q are found to be slightly larger than the redshift space measurements at smaller scales. Similar results are also reported based on galaxy distribution in real space vs redshift space from cosmological simulations in 
    \citet{Marin2008}}.

\vskip 0.1in
\noindent{\bf Future directions:} {It is important to accurately measure the three point correlation and Q preferable over larger scales. This will allow us to measure the linear and non-linear bias parameter for the \lya\ forest. With the existing archival data it is possible to improve our measurement by more than 50\%.
The scale dependence of Q (shown by simulations) is interesting. 
As we discussed in the introduction, low-$z$ \lya\ may originate from CGM, ICM, IGM or from the interface region between them. If different populations trace different scales that can lead to a scale dependent Q values. Non-linear clustering studies also predict scale dependent Q values. Accurate measurement of $Q$ over  different scales is therefore very important.}

{In the simulation front, it is important to have large simulations with appropriate sub-grid physics to first match the observed \NHI\ and $b$-distributions before trying to match the observed clustering properties of the \lya\ absorbers. It is also important to calibrate the simulations to correctly produce the properties of metal line absorbers to be able to probe the influence of feedback on the clustering properties of low-$z$ absorbers.
Such a constrained simulation will allow us to establish a connection between large scale structure and the \lya\ absorbers. }

%\Anand{Add some ref here}.
%Alternatively, as we are measuring the clustering along the line of sight, presence of peculiar velocities can artificially change $\xi(r)$ and $\zeta(r)$.

%\Anand{We can say we need more data to accurately measure the $\zeta$ at different scale and different \NHI\ thresholds. You may mention about the requied number of absorbers. You may also mention about the importance of measuring clustering in the transverse direction. }
%\end{enumerate}

%Measuring $\zeta$, $\xi$ and Q over a large redshift range is important to gain more insight into the evolution of structures traced by the \lya\ absorption.

\section*{Acknowledgments}
{
We acknowledge the use of High performance computing facilities PERSEUS and PEGASUS at IUCAA.
We thank Tiziana Di Matteo and Rupert Croft for the MBII simulation, J. Bolton for the Sherwood simulations and Kandaswamy Subramanian, Tirthankar Roy Choudhury and Aseem Pranjape  for discussions. { We are also grateful to the anonymous referee for his/her valuable comments and inputs regarding the manuscript.} Support by ERC  Advanced  Grant  320596 `The Emergence of Structure During the Epoch of
reionization' is gratefully acknowledged. PG acknowledges the support of
the UK Science and Technology Facilities Council (STFC) .
%We thank J. Bolton for making the Sherwood simulation
%suite available to our work.
NK acknowledges the support of the Ramanujan 
Fellowship of DST, India
%\footnote{Awarded by the 
%Department of Science and Technology, Government of India} 
and  the IUCAA
%\footnote{Inter University Centre for Astronomy and 
%Astrophysics, Pune, India} 
associateship programme.
The Sherwood simulations were performed using the Curie supercomputer at the
Tre Grand Centre de Calcul (TGCC), and the DiRAC Data Analytic system at the
University of Cambridge, operated by the University of Cambridge High
Performance Computing Service on behalf of the STFC DiRAC HPC Facility
(www.dirac.ac.uk) . This was funded by BIS National E-infrastructure
capital grant (ST/K001590/1), STFC capital grants ST/H008861/1 and
ST/H00887X/1, and STFC DiRAC Operations grant ST/K00333X/1. DiRAC is part of
the National E-Infrastructure. The MBII simulation was run on the Cray XT5 supercomputer-- Kraken -- at the National Institute for Computational Sciences, supported by the National Science Foundation (NSF) PetaApps program, OCI-0749212.
}

\section*{Data Availability}
{HST-COS data products of \citet{danforth2016} used in this study can be accessed from https://archive.stsci.edu/prepds/igm/. %Tabulated correlation measurements will be shared on request to the corresponding author. 
}

%%%%%%%%%%%%%%%%%%%%%%%%%%%%%%%%%%%%%%%%%%%%%%%%%%



\bibliographystyle{mnras}
\bibliography{main} % if your bibtex file is called example.bib

\bsp	% typesetting comment
\label{lastpage}
\end{document}